\documentclass[12pt, b5paper,twoside]{tesi_upf}

\usepackage[latin1]{inputenc}
\usepackage{caption}

\usepackage[english,catalan]{babel}
\usepackage[cam,b5,center,frame]{crop}
\usepackage{braket}
\usepackage{graphicx}
\usepackage{hyperref}

\usepackage{times}

\usepackage{upgreek}

\usepackage{makeidx}
\makeindex

\addto\captionscatalan
  {\renewcommand{\contentsname}{\Large \sffamily Sumari}}

\usepackage{amsmath}
\usepackage{amssymb}
\usepackage{amsfonts}

\newcommand{\fullpagefigure}[4][.76\textheight]{%
  \clearpage
  \thispagestyle{empty}%
  \noindent
  \begin{minipage}[c][\textheight][c]{\columnwidth}
  \centering
  \vspace*{\fill}
  \includegraphics[width=#1\columnwidth,keepaspectratio]{#2}\par
  \vspace{0.5\baselineskip}
  \captionof{figure}{#3}\label{#4}
  \vspace*{\fill}
  \end{minipage}
  \clearpage
}

\title{Twisted Multilayer Graphene}
\subtitle{ Superperiodicity and quasicrystals}
\author{Pedro Alc\'azar Guerrero}
\thyear{2026}
\department{Institut Catal\`a de Nanoci\`encia i Nanotecnologia}
\supervisor{Stephan Roche, Aron Cummings, Sergio O. Valenzuela}

\makeatletter
\newlength{\margintemp}
\makeatother
\begin{document}

\frontmatter

\maketitle

\cleardoublepage


\noindent A Luc\'ia

\cleardoublepage


\noindent {\Large \sffamily \textbf{Agradecimientos}}.
\\
\newline
\newline
Esta tesis representa un largo trabajo de cuatro a\~nos en el que much\'isima gente ha jugado un papel fundamental. Es un proceso que no habr\'ia sido posible sin el apoyo de un gran n\'umero de personas. Quiero dedicar aqu\'i un espacio para acordarme de todos ellos y darles las gracias.

Quiero comenzar agradeci\'endole a mis padres, Pedro y Loli, que hayan estado aqu\'i para acompa\~narme durante estos cuatro a\~nos. La realizaci\'on de esta tesis, como la de la mayor\'ia que he visto, ha tenido momentos muy felices en los que ellos han estado para compartir y alegrarse conmigo, pero tambi\'en ha tenido momentos muy frustrantes en los que han estado all\'i para ayudarme. Sin vosotros no habr\'ia sido posible en absoluto. Igualmente quiero agradecer a mi hermana Luc\'ia, la persona m\'as importante de mi vida, no puedo estar m\'as orgulloso de tener una hermana como t\'u y sin ti esto no habr\'ia sido posible. Muy especialmente quiero agradecer tambi\'en a mi abuelo Pedro, que se ha preocupado durante todo el proceso, a mi abuelo Manuel, que fue una fuerte inspiraci\'on para m\'i y a mi abuela Eulalia, que no pudo verla acabar y estoy seguro de que no podr\'ia estar m\'as feliz. Y a todos mis primos, que m\'as que solo mis primos son mis amigos, no sab\'eis lo afortunado que me siento.

Quiero agradecer tambi\'en a mis amigos de Madrid por aguantarme durante todo este tiempo: In\'es, Javi, Elena y Mois\'es, Pablo, Pablo Getafe, Jaime, Fer, Asier, Gonzalo. A todos mis amigos de Valdeca\~nas: Nano, Pablo, Rub\'en, Daniela, Mar\'ia, Paloma y Paula. Y especialmente a todos los que han conseguido que Barcelona se sienta un hogar: Joaqu\'in, Jorge, Jaime, Patri, Catalina, Sara, Fede, Roberta, Santiago, Dorye, Thomas, Onurc, Chema, Mireia y Aitana.

Si hoy estoy terminando esta tesis y si decid\'i dedicarme a la F\'isica, es en gran parte gracias al trabajo de incre\'ibles divulgadores. Por eso quiero dedicar algunas l\'ineas para acordarme de quienes construyeron esta pasi\'on. Quiero acordarme de los nombres de Javier Santaolalla, Jos\'e Luis Crespo, Ana Morales y  Eduardo S\'aenz de Cabez\'on y extender el agradecimiento a toda la gente que se dedica a la labor incansable de hacer divulgaci\'on cient\'ifica.

Pero si bien ellos fueron una de las razones de mi pasi\'on por la f\'isica, a\'un m\'as importante han sido aquellos mentores y profesores que me han acompa\~nado. Mucho m\'as importantes han sido para m\'i Juan Ignacio Beltr\'an, Mari Carmen Mu\~noz de Pablo y Juan Ram\'on Mu\~noz de Nova, espero que sep\'ais que hab\'eis sido incre\'iblemente importantes para m\'i y quiero agradeceros por seguir contando conmigo incluso despu\'es de terminar la universidad.

Por \'ultimo y de forma muy especial quiero agradecer a Stephan Roche por haberme tenido en su grupo, haberme guiado durante la tesis y haber cuidado, y seguir cuidando, mi desarrollo profesional. Thanks to Aron for teaching and helping me, but also for his advice, for taking care of me, and for listening and helping me decide on a career. And thanks to all the members of the Theoretical and Computational Nanoscience Group. Equally, I want to thank ICN2 for their support, and all the international collaborators without whose work this would not have been possible. I also want to acknowledge grant PCI2021-122035-2A-2 funded
by MCIN/AEI/10.13039/501100011033 and European
Union ``NextGenerationEU/PRTR'' and the support from Departament de Recerca i Universitats de la Generalitat de Catalunya.

\cleardoublepage


\selectlanguage{english}
\section*{\Large \sffamily Abstract}


This thesis investigates how superperiodicity, quasiperiodicity, and disorder shape electronic and spin transport in graphene-based systems, with an emphasis on experimentally relevant length scales and realistic atomistic modeling. Using large-scale real-space quantum-transport methods, it first establishes controlled transport fingerprints that distinguish conventional Bloch propagation in periodic structures from the anomalous dynamics induced by quasiperiodic modulations. Building on this framework, the thesis analyzes magic-angle twisted bilayer graphene and shows that, within a finite disorder window where flat-band features remain robust, moderate Anderson disorder can counterintuitively enhance the mean free path. This disorder-induced delocalization is further linked to changes in the quantum metric extracted from optical conductivity, revealing a direct connection between transport, electronic geometry, and the real-space extent of the underlying states. The study then turns to graphene quasicrystal approximants and hybrid multilayer stacks, identifying sub-ballistic transport and self-similar localization patterns as signatures of quasicrystalline order, while also demonstrating their strong fragility against disorder and interlayer proximity effects. Finally, the thesis addresses spin transport in suspended monolayer graphene, showing that atomic-scale corrugations generate short-range fluctuating Rashba fields that can limit spin lifetimes to the nanosecond range even when charge transport remains close to ballistic. Taken together, these results provide a unified picture of how geometry, disorder, and structural complexity govern transport phenomena in twisted and corrugated graphene systems.

\selectlanguage{catalan}
\vspace*{\fill}
\section*{\Large \sffamily  Resum}
Aquesta tesi investiga com la superperiodicitat, la quasiperiodicitat i el desordre modelen el transport electr\`onic i d'esp\'in en sistemes basats en graf\`e, amb un \`emfasi especial en les escales de longitud rellevants experimentalment i en una modelitzaci\'o atom\'istica realista. Mitjan\c{c}ant m\`etodes de transport qu\`antic en espai real a gran escala, s'hi estableixen primer empremtes de transport controlades que permeten distingir la propagaci\'o convencional de tipus Bloch en estructures peri\`odiques de la din\`amica an\`omala indu\"ida per modulacions quasiperi\`odiques. A partir d'aquest marc, la tesi analitza el graf\`e bicapa girat a l'angle m\`agic i mostra que, dins d'una finestra finita de desordre en qu\`e les caracter\'istiques de les bandes planes es mantenen robustes, un desordre d'Anderson moderat pot augmentar de manera contraintu\"itiva el cam\'i lliure mitj\`a. Aquesta deslocalitzaci\'o indu\"ida pel desordre es relaciona, a m\'es, amb canvis en la m\`etrica qu\`antica obtinguda a partir de la conductivitat \`optica, cosa que revela una connexi\'o directa entre el transport, la geometria electr\`onica i l'extensi\'o en espai real dels estats subjacents. L'estudi aborda despr\'es aproximants de quasicristalls de graf\`e i piles h\'ibrides multicapa, i identifica el transport subbal\'istic i els patrons d'auto-similitud en la localitzaci\'o com a signatures de l'ordre quasicristal\textperiodcentered l\'i, alhora que en demostra la gran fragilitat davant del desordre i dels efectes de proximitat entre capes. Finalment, la tesi estudia el transport d'esp\'in en graf\`e monocapa susp\`es i mostra que les corrugacions a escala at\`omica generen camps de Rashba fluctuants i de curt abast que poden limitar la vida d'esp\'in a l'escala dels nanosegons, fins i tot quan el transport de c\`arrega es mant\'e proper al r\`egim bal\'istic. En conjunt, aquests resultats ofereixen una visi\'o unificada de com la geometria, el desordre i la complexitat estructural governen els fen\`omens de transport en sistemes de graf\`e girat i corrugat.

\vspace*{\fill}

\selectlanguage{english}
\cleardoublepage

{\bf Preface.} This thesis is the result of extensive computational research on the effects of super- and quasiperiodicities in graphene. This thesis is intended for the readers with basic knowledge of tight-binding theory and condensed matter physics who want a deeper insight into the effects of these long-range modulations. We will try to introduce all the necessary concepts and establish general effects of super- and quasiperiodicities in graphene. This thesis is organized as follows: 

\paragraph{Chapter 1: Introduction.}
We introduce the motivation for the thesis as well as the current scientific context of moir\'e physics and twistronics.

\paragraph{Chapter 2: Fundamentals.}
We introduce the electronic structure of graphene and multilayer graphene, from tight-binding to low-energy continuum descriptions, and we set notation for reciprocal-space concepts (Dirac valleys, mini-Brillouin zones, and interlayer scattering channels).
We also provide the minimal background on twisted multilayers as a route to superperiodicity, and introduce the quasicrystalline $30^\circ$ limit as a canonical example of aperiodic order.

\paragraph{Chapter 3: Computational methods.}
We present the quantum-transport framework used throughout the thesis: Kubo-type formulations of diffusion, mean-squared displacement, and mean free path; diffusion coefficient for dealing with transport in disordered systems; and the linear scaling KPM techniques that enable spectral and time-evolution calculations at million- to billion-atom scale.
We also summarize auxiliary methods used in specific chapters (electrostatic modeling and mean-field interaction models).

\paragraph{Chapter 4: Superperiodic and quasiperiodic modulations as controlled models.}
We first study wave-packet dynamics in a superperiodic one-dimensional chain to identify how a finite superlattice length introduces characteristic time/length scales in the diffusion coefficient. We then generalize to graphene under externally imposed superperiodic potentials with different symmetries, and finally to a quasiperiodic (quasicrystalline) modulation. This chapter establishes a set of transport ``fingerprints'' differentiating periodic systems, strictly periodic superlattices  from quasiperiodic modulations. These controlled results serve as a reference point for interpreting the more complex moir\'e and quasicrystalline behavior in twisted multilayers.

\paragraph{Chapter 5: Transport in twisted multilayer graphene: disorder and structural complexity.}
We apply the methodology to realistic tight-binding models of magic-angle twisted bilayer graphene and quantify how the Anderson-type disorder reshapes both the density of states and transport length scales.
A key result is that, in a finite window of disorder strengths where flat-band features survive, disorder can increase the mean free path by broadening the flat bands and reducing the scattering rate: a disorder-induced delocalization mechanism.
We then connect this transport evolution to the disorder dependence of the quantum metric obtained from optical conductivity via the SWM sum rule, providing a geometric interpretation in terms of the real-space extent of the electronic ground state.
Finally, we extend the analysis to the graphene quasicrystal (via periodic approximants) providing the first result for its sub-ballistic exponent and to a trilayer stack combining a magic-angle interface with a quasicrystalline layer.

\paragraph{Chapter 6: Magnetism in twisted bilayer graphene.}
Building on the transport/localization perspective, we review the experimental landscape of magnetic and topological phases in MATBLG and summarize representative microscopic modeling routes.
This sets the first steps for the simultaneous consideration of magnetism and disorder in MATBLG. To date, this chapter represents a work in progress as we have found it not possible with the available computational resources to replicate previous results in the literature with our formalism.

\paragraph{Chapter 7: Spin transport in corrugated monolayer graphene.}
We investigate the microscopic limit of spin transport in suspended graphene by combining atomistic corrugation profiles with a tight-binding description including curvature-induced SOC terms.
We show how atomic-scale corrugations generate short-range fluctuating Rashba fields that can dominate spin relaxation, while leaving charge transport comparatively close to the ballistic limit.
This provides a concrete mechanism that can reconcile the long-standing gap between early idealized predictions and experimentally observed spin lifetimes.

\paragraph{Chapter 8: Conclusions and outlook.}We review the main results of the thesis and propose future research lines.

\paragraph{Scope of the thesis} The goal of this thesis is to understand the role of superperiodicity in transport and its interplay with disorder, and to extract general trends common among different super- and quasiperiodic systems. We will review how this superperiodicity and disorder interplay and how this interplay might affect the localization, thereby inferring its possible effects on correlated phases emerging in these systems.

\paragraph{What this thesis does not attempt}
\label{sec:non_goals}

To keep the scope controlled, the thesis does not aim to provide a full microscopic theory of interactions in MATBLG and their interplay with disorder.
Instead, it focuses on robust and computationally tractable single-particle quantities that strongly constrain any interacting description: localization properties, transport length scales, and their evolution under disorder and structural complexity. Yet, we propose this as an interesting continuation of the work presented here, and include a discussion of the possible challenges. Likewise, while proximity effects between different layers and materials are a broad and important theme in graphene heterostructures, here they are used mainly as a means of testing the robustness of spectral and transport signatures (and are introduced in greater detail
only when they become essential to the spin-transport discussion).

\paragraph{Original contributions of this thesis}
\label{sec:non_goals}

The goal of this thesis is to present a coherent story starting from superperiodicities and quasiperiodicities in the linear chain towards the study of superperiodic and quasicrystalline twisted bilayer graphene. The discussion and connection between them in chapters 4-6 is original and sustained in simulations supported by results in the literature. Among the main contributions of this thesis to published articles, which are discussed in their dedicated sections, we can find the following publications:

\begin{itemize}
    \item \cite{PhysRevLett.134.126301} describes how the disorder affects the localized states of magic angle twisted bilayer graphene, finding the surprising fact that for increasing disorder, the flat-band states delocalize. We can find the main results of this article in Sec. \ref{Sec:MATBLG_paper}.
    
    \item \cite{22s2-vpm5} shows the quasicrystalline transport behavior of the QCTBLG and how these states are disrupted by disorder. It also proposes the creation of a trilayer with a quasicrystal and a magic angle. The article shows that the magic-angle flat-band states remain robust upon addition of the third layer, while showing the fragillity of the quasicrystalline states in the $30^\circ$ twisted interface. These results can be found in Secs. \ref{Sec: Graphene Quasicrystal} and \ref{Sec: trilayer}.
    
    \item \cite{Cummings_2025} explores an upper limit of spin relaxation in freestanding graphene. It analyzes how thermal corrugations induce spin-orbit coupling, thereby placing limits on the spin-relaxation lengths. It links these results to the presence of a random spin-orbit Rashba field, comparing the results with previous analytical models in the literature. The results of this article can be found in Chapter \ref{chapter:spin}.
\end{itemize} 

Also, results regarding Sec. \ref{sec:qdot_array} and \ref{sec:mean-field-simulations} represent ongoing work, still not published, and are current research lines.

\cleardoublepage

\tableofcontents

\listoffigures
\addcontentsline{toc}{chapter}{List of figures}

\listoftables
\addcontentsline{toc}{chapter}{List of tables}

\mainmatter

\chapter{Introduction}
\label{chapter:introduction}

\sloppy

Graphene has become a paradigmatic platform for quantum transport in two dimensions. Since its first isolation, its combination of ultrahigh mobility, mechanical robustness, and an electronic spectrum governed at low energy by massless Dirac fermions has enabled both fundamental discoveries and device concepts that were difficult to realize in conventional semiconductors \cite{GeimNovoGraphene,RevModPhys.81.109}. Beyond monolayers, stacking graphene sheets provides an additional and unusually powerful control parameter: the relative geometry between layers. A small lattice mismatch or a relative twist angle generates long-wavelength interference patterns (moir\'e superlattices), which reshape the electronic spectrum and the nature of electronic states. In the last decade, this geometric control has evolved from a  qualitative observation into a quantitative route to engineering band structures and stabilizing new electronic regimes giving rise to the field of twistronics.

A central theme of this thesis is that {geometry can act as an  effective potential with its own characteristic length scales}, producing superlattice-driven localization and transport anomalies even in the absence of conventional disorder. In twisted graphene multilayers, this idea appears in two complementary limits: (i) \textbf{superperiodic} structures, where the moir\'e pattern is commensurate and defines an enlarged unit cell, and (ii) \textbf{quasiperiodic/quasicrystalline} structures, where no finite unit cell exists but long-range order remains. Both limits can be accessed experimentally, and have clear theoretical signatures.
How do long-range moir\'e pattern modulations compete with disorder?
And how do these single-particle effects connect to the stability of interaction-driven phases? We will address these questions and clarify the impact of disorder on the physics emerging from these long-range geometry stacks.

\section{From moir\'e flat bands to fragile quasicrystalline states}
\label{sec:motivation}

\subsection{Flat-band moir\'e systems and the role of disorder}
Magic-angle twisted bilayer graphene (MATBLG) is one of the most prominent examples of moir\'e band engineering.
Near the so-called magic angle, the low-energy moir\'e bands become extremely narrow and the wave functions acquire strong real-space structure (with enhanced weight in AA-stacked regions), dramatically increasing the relative importance of electron-electron interactions \cite{doi:10.1073/pnas.1108174108,Cao2018,Cao2018CI,Balents2020}.
Experimentally, the resulting phase diagrams display correlated insulators and superconductivity as a function of carrier density, and in many devices also exhibit interaction-driven symmetry breaking and topological responses \cite{doi:10.1126/science.aaw3780,doi:10.1126/science.aay5533,0256b11272db4e4fb10d4e4bb162df16}.

At the same time, a robust and predictive understanding of these phases requires a careful baseline: real samples are never perfectly clean or perfectly uniform.
Electrostatic inhomogeneity, strain, twist-angle variations, and local defects are all present to some degree and can reshape transport and localization. Because correlations in flat-band systems are amplified by reduced kinetic energy and enhanced localization, any mechanism that modifies the single-electron localization properties is expected to impact the propensity toward correlated phases.
This thesis is therefore motivated by a practical and conceptual need: to quantify the competition between geometry-induced localization and disorder in twisted graphene multilayers using large-scale microscopic simulations of transport and spectral properties.
This analysis will be performed to establish a first realistic picture of how disorder affects flat-band states whose localization underlies correlation effects in MATBLG \cite{Balents2020}.

\subsection{Quasicrystals: long-range order without periodicity}

A second major motivation of this thesis comes from the quasicrystalline limit of twisted graphene. In contrast to periodic crystals, quasicrystals display long-range structural order without translational invariance, so Bloch's theorem no longer provides the natural starting point for describing their electronic states. This absence of periodicity has profound consequences for transport: electronic spectra often develop pseudogaps, eigenstates can become critical rather than simply extended or exponentially localized, and wave-packet dynamics may deviate from the standard ballistic-to-diffusive picture familiar from ordinary metals. These ideas were established in the quasicrystal literature well before graphene quasicrystals became available experimentally, and they provide the conceptual backdrop for the present work.

In particular, real-space Kubo-Greenwood studies showed that quasiperiodic systems can exhibit propagation modes qualitatively different from periodic ones, including regimes where the conductivity remains nearly unchanged or can even increase with disorder when the Fermi energy lies near a pseudogap \cite{Roche1997ConductivityOQ}. Complementary reviews and ab initio analyzes further emphasized that transport in quasicrystals and related approximants often falls outside the standard semiclassical Bloch-Boltzmann framework, because small band velocities and anomalous quantum diffusion become central ingredients of the conductivity itself \cite{10.1063/1.531914,MaciaQC}.

This general perspective is especially relevant to twisted graphene at a relative angle of $30^\circ$. In that limit, the two graphene layers form a dodecagonal quasicrystal with twelvefold rotational symmetry but no translational periodicity, giving rise to unconventional electronic states and resonant spectral features that are absent in ordinary moir\'e superlattices \cite{PhysRevB.99.165430}. Experiments and theory have shown that this graphene quasicrystal hosts characteristic high-energy resonances and distinctive real-space patterns associated with quasicrystalline order. From the broader quasicrystal literature, one also expects such states to display anomalous transport fingerprints, including sub-ballistic spreading and unconventional optical conductivity \cite{Roche_Fujiwara_1998}.  Indeed, the generalized Drude picture developed for quasicrystals predicts that anomalous diffusion can strongly suppress the usual metallic Drude peak and, in sufficiently non-Boltzmann regimes, even replace it by a dip while allowing the DC conductivity to increase with disorder \cite{PhysRevLett.85.1290}. These phenomena make quasicrystals especially attractive for this thesis: they offer a paradigmatic setting in which geometry alone generates localization tendencies and transport anomalies, but they also raise a central question about robustness. As will be explored later in this work, quasicrystalline resonances in twisted graphene can be highly fragile to disorder and to proximity coupling with additional layers, so understanding both their emergence and their breakdown is essential for assessing their physical observability and their possible coexistence with flat-band moir\'e physics in more complex graphene stacks.



\subsection{Spin transport as a complementary probe of structural complexity}
Finally, graphene is also one of the most relevant materials in spintronics. Its weak intrinsic spin-orbit coupling (SOC) suggests long spin lifetimes and long spin diffusion lengths, yet experiments have historically reported spin lifetimes far shorter than early theoretical expectations.
Even in very clean devices, observed lifetimes remain on the order of nanoseconds rather than microseconds, indicating that subtle mechanisms may dominate spin relaxation.
One candidate mechanism is the presence of thermally induced atomic-scale corrugations that generate fluctuating, curvature-induced SOC fields. A later part of this thesis addresses this question by connecting atomistic corrugation profiles with quantitative predictions for spin relaxation in suspended graphene
\cite{Cummings_2025}.

\chapter{Fundamentals of transport in multilayer graphene}
This chapter presents the conceptual and model background required for the rest of the thesis. We first review the essential electronic structure of graphene and graphene multilayers. We begin with the lattice and reciprocal-space description of monolayer graphene and then introduce the tight-binding and effective low-energy models that will be used throughout the thesis.

Building on this basis, we discuss multilayer stackings, interlayer coupling, and twisted graphene structures as a platform where geometry acts as a long-wavelength modulation of the electronic states. We also introduce the quasicrystalline $30^\circ$ limit, which will serve later as a key example of aperiodic order in graphene-based systems. In this way, the chapter establishes both the notation and the physical intuition needed to understand the transport results presented in the following chapters.


\section{Electronic models for graphene}

\subsection{Graphene structure}

\begin{figure}[h]
  \centering
  \includegraphics[width=\linewidth]{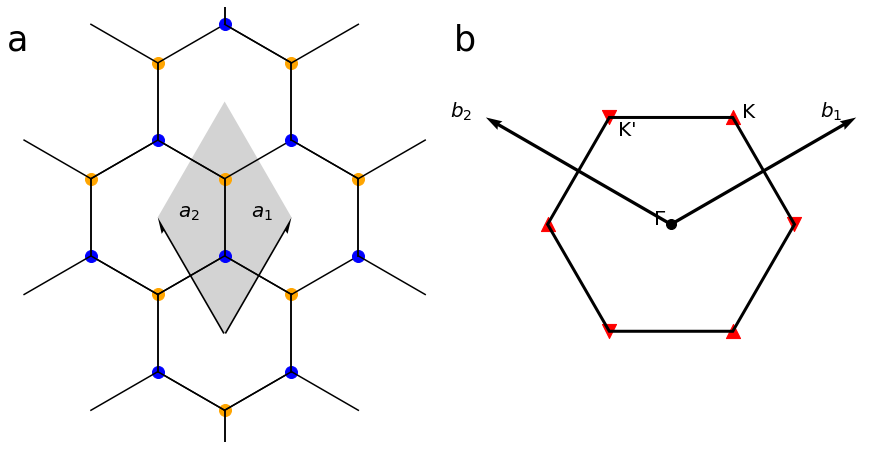}
    \caption{\textbf{a.} Crystalline structure of graphene where the unit cell is shadowed and the colors correspond to the A (blue) and B (orange) sublattices. \textbf{b.} First Brillouin zone of graphene where the equivalent K points are marked as triangles pointing up and the K' with triangles pointing down.}
    \label{fig:graphenestruct}
\end{figure}

Graphene is a 2D material formed by a monolayer of carbon atoms disposed in a hexagonal lattice as shown in Fig. \ref{fig:graphenestruct}a). We can define the lattice with its lattice vectors.

\begin{equation*}
    \begin{array}{cc}
         \mathbf{a_1}=\left( \frac{\sqrt 3}{2}a_{cc},\frac{{3}}{2}a_{cc}\right)&
         \mathbf{a_2}=\left( \frac{-\sqrt 3}{2}a_{cc},\frac{{3}}{2}a_{cc}\right) . \\
         
    \end{array}
\end{equation*}
Here $a_{cc}$ is the distance between two nearest neighbors. Thus, we can find the position of any carbon atom in the lattice by doing $\mathbf{r}=n_{1} \mathbf{a_1}+ n_{2} \mathbf{a_2}+\mathbf{\delta_{i}}$ being $n_{1,2}$ two integers and $\delta_{i}$ the displacement vector for the different sublattices given by:

\begin{equation*}
    \begin{array}{cc}
         \mathbf\delta_A=\frac{1}{3}\left(\mathbf a_1+\mathbf a_2\right)&
         
         \mathbf\delta_B=\frac{2}{3}\left(\mathbf a_1+\mathbf a_2\right) .\\
         
    \end{array}
\end{equation*}

As the lattice is periodic, we can also define its first Brillouin zone by the vectors \cite{Ashcroft:102652}
\begin{equation}\label{eq: reciprocal-lattice-vectors-graphene}
    \begin{array}{cc}
         \mathbf{b_1}=\frac{2\pi}{a_{cc}}\left(\frac{1}{\sqrt{3}},\frac{1}{3}\right) &\mathbf{b_2}=\frac{2\pi}{a_{cc}}\left(-\frac{1}{\sqrt{3}},\frac{1}{3}\right).
    \end{array}
\end{equation}
 We can see the shape of this first Brillouin zone in Fig. \ref{fig:graphenestruct} b, where the high-symmetry $k$ points $K$ and $K'$ are explicitly marked. These points have coordinates

\begin{equation}
    \begin{array}{cc}
         K=\frac{(\mathbf{b_1}-\mathbf{b_2})}{3}&
         
         K'=-\frac{(\mathbf{b_1}-\mathbf{b_2})}{3}. \\
         
    \end{array}
\end{equation}
 As we will see later, among other reasons, these two high-symmetry points are especially important because all low energy properties occur in their surroundings, where electrons in graphene behave as massless Dirac fermions \cite{Novoselov2005}. We call the surroundings of these points the $K$ and $K'$ Dirac cones.
\subsection{Tight-binding models}\label{Sec:Tight-Binding-graphene}

 A general shape for the Hamiltonian of a material can be read as \cite{Amano2025,Ashcroft:102652}

\begin{equation}\label{eq:full material equation}
    H=T_{nuclei}+T_{electrons}+V_{el-el}+V_{el-nuc} + V_{nuc-nuc} + V_{external}
\end{equation}
where $T_{electrons,nuclei}$ is the kinetic energy of electrons and nuclei respectively, and $V_{el-el,el-nuc,nuc-nuc}$ are the interactions between each pair mentioned. Frequently we can assume $T_{nuclei}$ and $V_{nuc-nuc}$ to be negligible for the electronic modeling due to the huge difference in mass and speed from electrons and nuclei. This approximation is called the Born-Oppenheimer approximation \cite{BornOppenheimer1927}. Thus, we can write (\ref{eq:full material equation}) as 

\begin{equation}\label{eq: born oppenhaimer}
    H=T_{electrons}+V_{el-el}+V_{el-nuc} + V_{external}=H_{TB}+V_{el-el}+V_{external}.
\end{equation}

For graphene at zero temperature, we will go further in our approximation and rely on the tight-binding models \cite{Ashcroft:102652,PhysRev.94.1498}. For that purpose, we will merge $T_{electrons}$ and $H_{el-nuc}$ into one term assuming the electron can only settle in the atomic positions of the lattice and tunnel from one to another. This results in the $H_{TB}$ term of the equation. For now, we will only consider contributions of $H_{TB}$, but we will consider the other two terms later in the context of disorder $V_{external}$ and correlated systems ($V_{el-el}$).

In the literature, graphene tight-binding model is frequently considered to use mainly its $p_z$ orbitals for the description of bands that are close to the charge neutrality point \cite{RevModPhys.81.109,PERES20061559}. Given the symmetries of these orbitals, in a lattice as described above, a general tight-binding model of graphene $p_z$ orbitals can be written as \cite{PERES20061559}

\begin{equation}
    H_{TB}=\sum_{i,j}t(|\mathbf{r_j}-\mathbf{r_i}|)c_i c_j^\dagger
    \label{tight binding}\quad ,
\end{equation}
where $c_i,c_i^\dagger$ are the annihilation and creation operators. Here, $\mathbf{r_i}$ is the position of each lattice site and $t(|{r_i}-{r_j}|)$ the overlap function.

There are two considerations that are frequently made. First, we will take advantage of lattice periodicity by taking the Fourier expansion of and expressing $c_i,c_i^\dagger$ in the reciprocal space. Thus, we make $c^{(\dagger)}_i=\sum_k c_{i,k}^{(\dagger)}e^{\pm i\mathbf{k}\cdot \mathbf{r}}$ where the plus (minus) sign is chosen for the creation (annihilation) operators \cite{fetter-walleka}.

\begin{figure}[h]
  \centering
  \includegraphics[width=\linewidth]{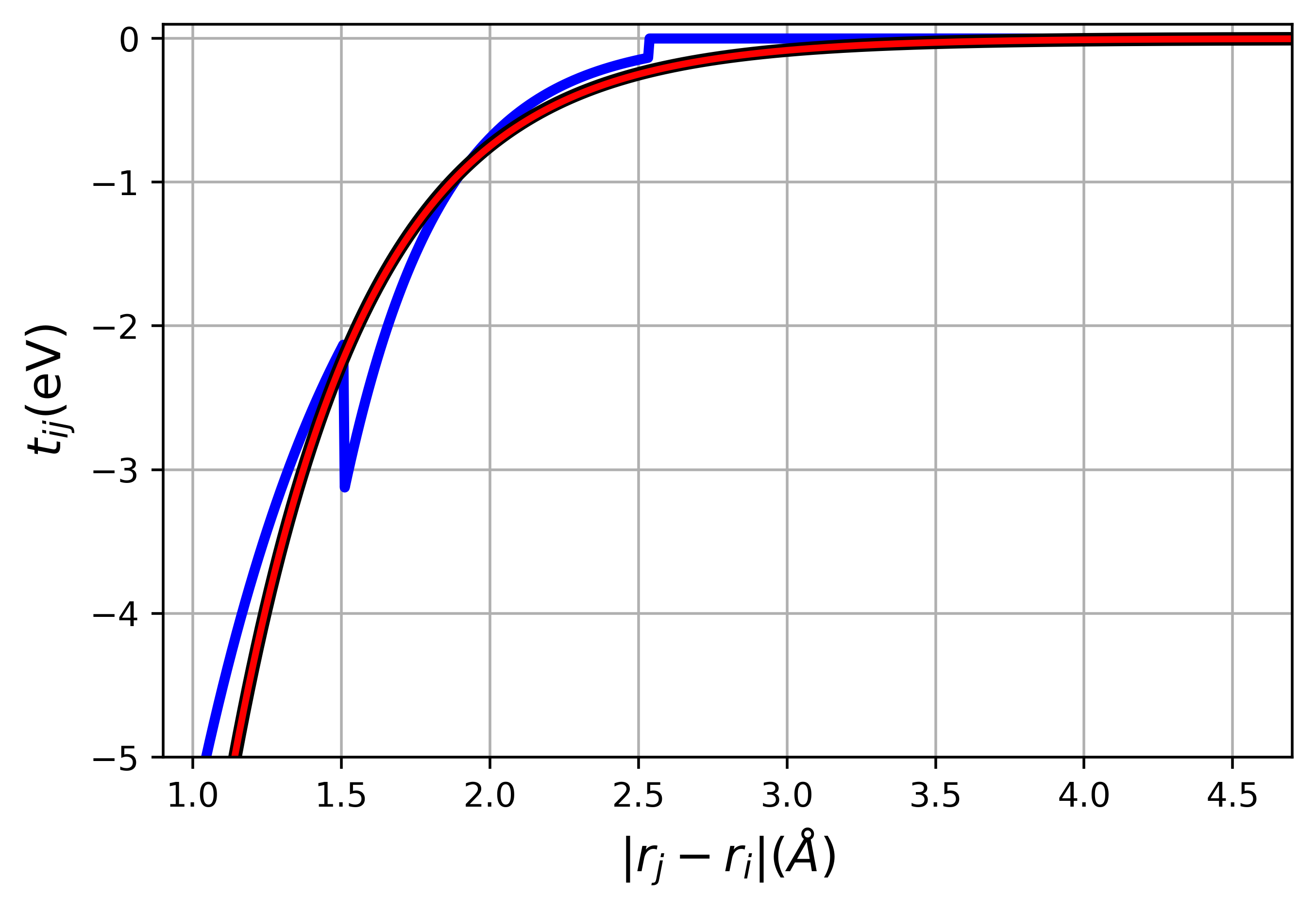}
    \caption{Decay of the different overlap functions used for this integral during this thesis. In black we observe the model displayed in \eqref{eq:charlier_model}, in red \eqref{eq:lewenkopf_model} and in blue \eqref{eq:dubois model} explained in chapters \ref{Chapter: Transport in twisted multilayer graphene}, \ref{chapter:magnetism} and \ref{chapter:spin} respectively. }
    \label{fig:overlapdecay}
\end{figure}

The second concerns the nature of the overlap function. As we can see in Fig. \ref{fig:overlapdecay}, this function decays (in absolute value) for long enough distances. With that in mind, we can assume that there exists a $r_{max}$ such that $t(r)=0,\quad\forall r>r_{max}$. Different values of $r_{max}$ are going to be used during this text, let us start by choosing a simple model. If we choose a value of $r_{max}$ small enough for the overlap function to become $t(|\mathbf{r_j}-\mathbf{r_i}|) = t$ for the coupling between nearest neighbors $i,j$ and 0 elsewhere. Note that no dependence on the distance is present here as all nearest neighbors in pristine graphene are equidistant.  Thus, equation (\ref{tight binding}) becomes

\begin{equation}
    H_ {TB}=\sum_{\langle i,j\rangle;k}te^{i \mathbf{k}\cdot(\mathbf{r_j}-\mathbf{r_i})}c_{i,k} c_{j,k}^\dagger\quad ,
\label{nearestneighborsgraphene}
\end{equation}
where the sum in $k$ is taken over the first Brillouin zone.

\begin{figure}[h]
  \centering
  \includegraphics[width=\linewidth]{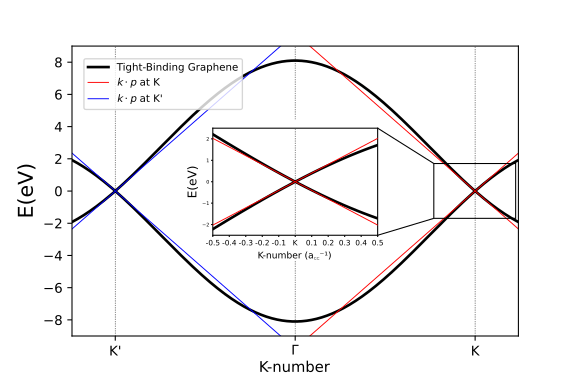}
    \caption{Band structure computed from a nearest-neighbor approximation of equation (\ref{tight binding}). The red and blue curves show the $K\cdot p$ approximation in equation (\ref{KdotP}) that remains valid for low energies.}
    \label{fig:bandsgraphene}
\end{figure}
By choosing a path and diagonalizing this Hamiltonian we can generate the graphene band structure such as in Fig. \ref{fig:bandsgraphene} where we can clearly see the linear dispersion around the $K$ and $K'$ points. In their three dimensional version, the band structure forms two cones around these points. These cones cross the Fermi level only in the measure-zero set of points located at $K$ and $K'$, presenting a linear conic dispersion in their surroundings. These two regions in the $k$-space are called the Dirac cones and are the origin of most of the exotic electronic physics of graphene \cite{Foa_Torres_Roche_Charlier_2020}.

\subsubsection{The Fermi velocity}

There is a vast amount of information we can extract from the band structure of a material. We will start with the Fermi velocity. With the Hamiltonian expanded in the $k$ space, the velocity of each state can be written as \cite{Kittle}

\begin{equation}\label{eq:fermi velocity}
v=\frac{1}{\hbar}\nabla_k E(k).
\end{equation}
Thus, the Fermi velocity is represented by the slope of the bands. Note that for perfectly flat bands, $v_F=0$. Later on, we will delve into this as the origin of correlations in the case of magic-angle twisted bilayer graphene.

\subsection{$k\cdot p$ model}
\label{Section:kp}
In computing the electronic properties of graphene, we will frequently be interested in its low-energy electronic properties. As we can see in Fig. \ref{fig:bandsgraphene}, the low-energy states are only present in the Dirac cones in the vicinity of $K$ and $K'$, so a description of this region will often be enough to analyze transport in experimentally feasible devices \cite{RevModPhys.81.109,MarconciniKP}.

Taking the expansion of equation (\ref{nearestneighborsgraphene}) around the $K$ and $K'$ point we get
\begin{equation}
    \label{KdotP}
    H_\tau=v\tau(\mathbf{\sigma}\cdot \mathbf{\tilde{k}}),
\end{equation}
where $\mathbf{\sigma}$ is the Pauli matrix vector defined as $\mathbf{\sigma}=\left(\sigma_x,\sigma_y,\sigma_z\right)$ with $\sigma_x$, $\sigma_y$ and $\sigma_z$ the Pauli matrices defined as
\begin{equation}
\label{Pauli}
\begin{array}{ccc}
     \sigma_x=\left(\begin{array}{cc}
     0&1  \\
     1&0 
\end{array}\right)&
\sigma_y=\left(\begin{array}{cc}
     0&-i  \\
     i&0 
\end{array}\right)&
\sigma_z=\left(\begin{array}{cc}
     1&0  \\
     0&-1 
\end{array}\right)
\end{array}  .
\end{equation}
It is easy to see that $v(t_{ij})=v(t_{ij})=\frac{3a_{cc}}{2\hbar}t$ is the slope of the band dispersion obtained from \eqref{KdotP}, and thus, from \eqref{eq:fermi velocity}, we can connect it with the Fermi velocity of graphene 
for a nearest-neighbor Hamiltonian as the one in (\ref{nearestneighborsgraphene}). $\tau$ is $\pm 1$ for the $K$ and $K'$ valleys, respectively, and we will use the subscript $\tau$ to refer to the different valleys. Note that here we have made $\tilde{k}=k-K_{\tau}$. From now on in this text, we will use the same notation $k$ for $\tilde{k}$ when we refer to the $k\cdot p$ models.

\section{The proximity effect}

The model on \eqref{KdotP} can be seen as a valid minimal model for free-standing graphene. However, graphene is highly sensitive to its surrounding environment, and its electronic properties can be modified when placed in close contact with other two-dimensional materials. This phenomenon is commonly referred to as the proximity effect. It arises from weak interlayer coupling which preserves the intrinsic structure of each layer while enabling the transfer of symmetry-breaking fields, electrostatic potentials, or spin-orbit interactions across the interface. Unlike chemical functionalization, proximity does not require charge transfer or bonding, and therefore provides a controlled route to engineer graphene's band structure without degrading its crystalline quality \cite{PhysRevB.92.155403}.

The proximity effect makes graphene highly versatile. In chapter\ \ref{chapter:spin} we will further introduce the implications of this effect on spin properties as well as the new terms emerging in the Hamiltonian. A simple example is the induction of sublattice-asymmetric potentials, such as those generated when graphene is placed on top of hexagonal boron nitride (hBN), leading to a small mass term and to the opening of a band gap at the Dirac point \cite{PhysRevB.76.073103,pmid23686343}, turning graphene from a highly conductive material into a semiconductor. Similarly, layered environments that break inversion symmetry can generate Rashba-type spin-orbit coupling (SOC), while materials with strong intrinsic SOC can imprint additional spin-valley coupled terms onto graphene. We will briefly discuss the effects of this proximity effect on the Hamiltonian in Sec.\ \ref{subsec:SOC_graphene} in the context of spin dynamics and later use these emergent terms to model the corrugations of graphene itself.

The relevance of this effect is key to understanding the recent interest of the scientific community in van der Waals materials \cite{Geim2013,Sierra2021}, that is, stackings of 2D materials bound by their out-of-plane van der Waals forces. This family of materials is a first motivation for understanding in detail the superperiodicities in graphene, as they naturally emerge in the stacking of different materials as a result of differences in their geometries. The correct understanding of superperiodicities is needed for the correct modeling of such layered structures.

It is also worth mentioning that proximity interactions are also relevant in multilayer graphene even when all layers are themselves graphene sheets. In this case, each layer can be illustrated as a perturbing environment for the others, modifying the effective Hamiltonian and allowing interlayer tunneling. When the layers are perfectly aligned, this manifests as the well-known differences between AA and AB (or ABC) phases of multilayer graphene. When a relative twist angle is introduced, the proximity landscape becomes spatially modulated, giving rise to moir\'e patterns in which local stacking order varies over tens of nanometers. This provides the main motivation for our work, which we introduce carefully in the next sections.

\section{Multilayer Graphene}\label{Sec: Multilayer AA, AB, ABC graphene}
\begin{figure}[h]
    \centering
    \includegraphics[width=1\linewidth]{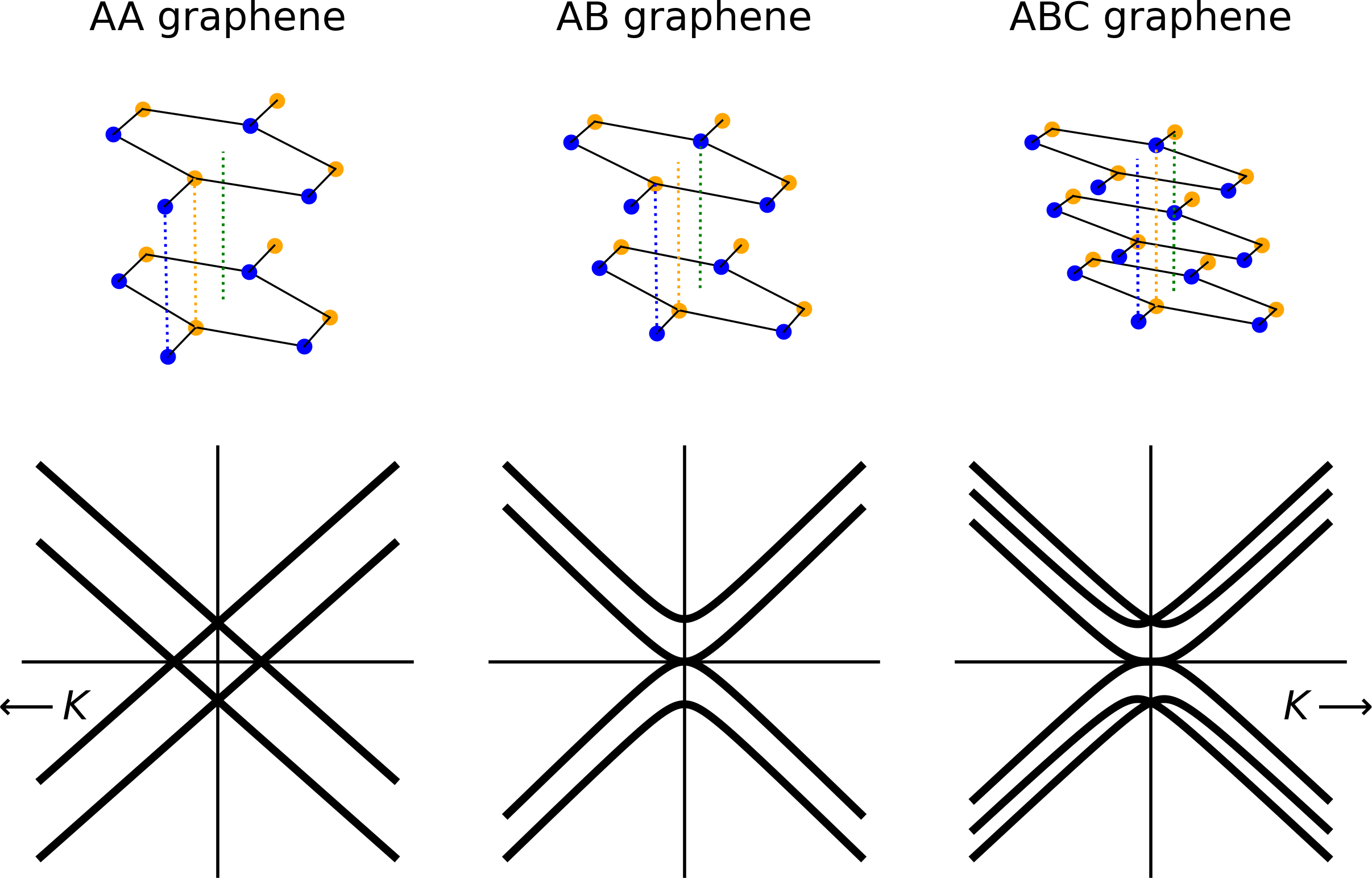}
    \caption{Different possible stackings of graphene and their corresponding band structures. The blue, orange and green vertical lines in the top figures are a guide to the eye to understand the matching conditions for the bottom layer A, B lattice sites and the center of the hexagons respectively.}
    \label{fig:stackings}
\end{figure}

The proximity between two different layers of graphene can also have an impact on the electronic properties. This is the case of multilayer graphene, where three possible stacking configurations are found. We can see the AA, AB and ABC stackings of the multilayer graphene represented in Fig. \ref{fig:stackings} with their respective band structures.

Among the three structures shown in Fig.\ \ref{fig:stackings}, AB-stacked bilayer graphene (often called Bernal stacking) is the thermodynamically most stable configuration \cite{PhysRevB.71.235415}, corresponding to the stacking of bulk graphite \cite{RevModPhys.81.109}. We can see its structure in the middle of Fig.\ \ref{fig:stackings}, which can be visualized as placing the $B$ atoms of the top layer directly above the $A$ atoms of the bottom layer, while the remaining $B$ sites are located  above the centers of the hexagons. This geometry maximizes the number of near-vertical interlayer bonds and leads to a strong dimerization between the $A$--$B$ pair of sites. At low energies, the corresponding band structure is characterized by two approximately parabolic bands that present a crossing at the $K/K'$ points, describing a massive gapless semimetal with a small gap overlap in the absence of external fields \cite{McCann_2013}.  When a transverse electric displacement field is applied, the inversion symmetry between the two layers breaks and a tunable band gap opens at the charge neutrality point \cite{McCann_2013}. We will use this feature in chapter\ \ref{chapter:superperiodicities} to build analogous configurations to the twisted moir\'e graphenes. 

On the left-hand side of Fig.\ \ref{fig:stackings} we find the AA-stacked bilayer graphene. It is obtained when both sublattices are aligned on top of each other, so that every $A$ ($B$) atom in the top layer lies directly above an $A$ ($B$) atom in the bottom layer. This stacking is only metastable, but has been realized in epitaxial and quasi-freestanding samples \cite{10.3389/fnano.2023.1333127}. Such metastability will have an impact on the relaxation of our structures, as we will see later in chapter\ \ref{Chapter: Transport in twisted multilayer graphene}. In the simplest tight-binding description, the interlayer coupling hybridizes the two Dirac cones into a pair of cones shifted in energy by the interlayer hopping amplitude. As a result, AA bilayer graphene behaves as a semimetal whose low-energy spectrum can be viewed as a superposition of two monolayer-like Dirac dispersions, with an increased density of states with respect to its monolayer and Bernal counterparts \cite{PhysRevB.90.155415,LAREF2020163755}. This suggests that when AA and AB stackings are combined in the same stacking, as will be the case of the twisted bilayer graphene, electrons will tend to localize in the regions of space with AA stacking.

Finally, on the right-hand side of Fig.\ \ref{fig:stackings}, we can find ABC graphene. This stacking (also referred to as rhombohedral stacking) appears only for three or more layers. It can be constructed by starting from an AB bilayer and placing the third layer so that its $B$ sites lie above the $A$ sites of the second layer, with each additional layer shifted in the same direction. In this geometry, only two sites (one on each outermost layer) remain undimerized, and they host the low-energy electronic states. For an ABC trilayer, the effective Hamiltonian near the valleys $K/K'$ produces a cubic dispersion, which is much flatter than the linear or parabolic spectra of mono- and bilayer graphene \cite{PhysRevB.80.165409}. As the number of layers increases, the surface bands become increasingly flat and develop a very large density of states at the Fermi level, thus producing correlated phenomena and superconductivity \cite{abcsuper,Nery2020,PhysRevB.108.144504}.
 In this text only AA and AB stackings are going to be relevant for our discussion, but we consider it important to mention ABC graphene here as another example of a relevant graphenic flat-band system.

\section{Twisted Multilayer Graphene: superperiodicity and quasiperiodicity}\label{Sec: super-quasiperiodicity in tblg}

\begin{figure}[h]
    \centering
    \includegraphics[width=\linewidth]{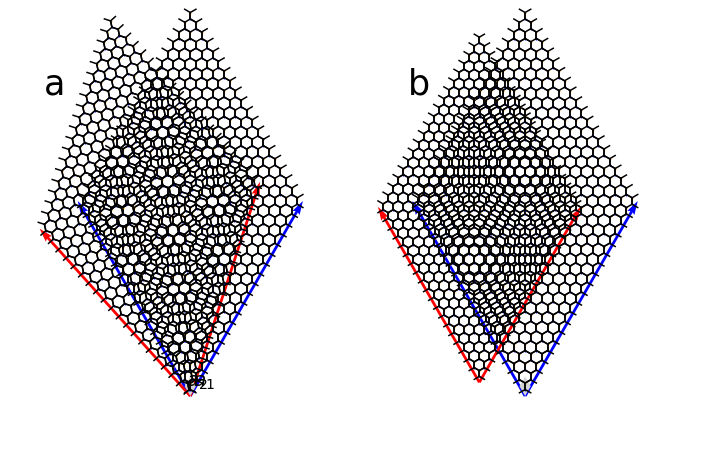}
    \caption{Two cases of moir\'e patterns. In panel a: two identical hexagonal (graphene) structures are rotated $12^\circ$. In panel B the unit cells are parallel but there is a mismatch between the lattice constant in red and blue unit cells. }
    \label{fig:moire structure}
\end{figure}

In addition to stable stackings in graphene, we can stabilize novel systems for multilayer graphene by stacking layers with differences in their geometry. These differences can be inequivalences in the structure of the layers (usually the lattice constant or the induced strain) and the differences in their relative positions (i.e., the presence of a twist angle between them) generating a moir\'e pattern similar to those of Fig. \ref{fig:moire structure} where every atom of the layers settles in a different dielectric environment due to the other. In a bilayer structure as the ones in Fig.\ \ref{fig:moire structure}, this will generate some zones with more AA-like character (the lighter ones) and some with more AB (the darker ones).

Depending on the matching between the lattice vectors of the layers, we can have two different cases:
\begin{itemize}
    \item If there exist two non-parallel vectors $\mathbf{A_1},\mathbf{A_2}$ such that they are linear combinations at the same time of each layer basis vectors, then 
    \begin{equation}  \label{eq:matching conditions}
    \begin{array}{ccccccc}
         \mathbf{A_1}&=&\alpha_{11}\mathbf{a_1}+\alpha_{12}\mathbf{a_2}&=&\beta_{11}\mathbf{b_1}+\beta_{12}\mathbf{b_2} &=&\dots  \\
         \mathbf{A_2}&=&\alpha_{21}\mathbf{a_1}+\alpha_{22}\mathbf{a_2}&=&\beta_{21}\mathbf{b_1}+\beta_{22}\mathbf{b_2} &=&\dots
    \end{array}
    \end{equation}
    with all the coefficients real, we will say that the system is superperiodic. This condition implies that we can build a larger unit cell of lattice vectors $\mathbf A_1, \mathbf A_2$, so that if we translate the whole set of positions $\lbrace \mathbf r_i\rbrace$ to $\lbrace \mathbf r_i+\mathbf{A_n}\rbrace$ the Hamiltonian will remain invariant. This is easy to prove: for any $\mathbf r_i$ lattice site located in layer $\ell$ with unit vectors $\mathbf l_1$, $\mathbf l_2$, by the translation invariance of layer $\ell$, must exist and belong to the lattice another site $\mathbf r_i+\mathbf A_n = \mathbf{r_i}+ \ell_{n1}\mathbf l_1+ \ell_{n2}\mathbf l_2 $ with integers $\ell_{n1}, \ell_{n2}$.
    
     This condition is sufficient for the existence of superperiodicity and a way to generate very long-range unit cells. 
     For example, in the case of multilayer graphene, these superperiodicities are often on the scale of tens of nanometers, about 100 times larger than the usual graphene unit cell.
    
    \item If such vectors do not exist, or they would be extremely large \cite{Uri2023}, we will call the system quasicrystalline because there is no possible unit cell for it. A quasicrystal is a material that presents some order but has no real-space periodicity. This distinguishes it from amorphous structures (which have no order) and from crystals (with a commensurate unit cell). This is the case because all of the layers come from periodic crystals. 
\end{itemize}
 
In both cases, we can build models similar to the one in Sec.\ \ref{Section:kp} where we approximate a certain energy or $k$-space region. In this chapter, we will build two different models that are very popular in the literature: The Bistritzer-MacDonald and the Koshino-Moon models for the twisted bilayer graphene.

\subsubsection{Continuum models for moir\'e: The mini-Brillouin zone and Umklapp scattering}

To understand electronic states of twisted bilayer graphene, it is important to introduce first the idea of Umklapp scattering. Assuming the  condition (\ref{eq:matching conditions}) is met, there must exist a common reciprocal space of unit vectors $\tilde b_1,\tilde b_2$ linear combination of both reciprocal space basis. From here one could (and we will do in future chapters) build full tight-binding models for the unit cell and expand into this reduced unit cell. In a simpler approach, to gain better control and insight about our Hamiltonian, it is often useful to reduce our problem to a continuous form. This is the case of the Bistritzer-MacDonald (BM) models \cite{doi:10.1073/pnas.1108174108,mcdonaldDynamics}, which understand the coupling between the two graphenes as the coupling between their Dirac cones.

The general idea behind the BM models is the expansion of the interlayer terms in the Hamiltonian in reciprocal space. In practice, the $k$-space version of Fig.\ \ref{fig:overlapdecay} follows a similar decay with $|k|$ (Fig. 2 in \cite{doi:10.1073/pnas.1108174108}). This will allow us to take into account only the interactions between points in the reciprocal lattice of the layers that are close in the $k$-space. Then, we can expand both Hamiltonians into their reciprocal spaces and generate the $k\cdot p$ models for each cone. Inside the layers, the states in different points in the $K$-space are orthogonal, but due to the inclusion of the interatomic potential, interactions between the $k$ and $\tilde k$ spaces of the Hamiltonians are allowed
\begin{equation}
    H=H_{bot}(k)+H_{top}(\tilde k)+T(|\tilde k-k|),
\end{equation}
where $H_{bot}(k)$ and $H_{top}(\tilde k)$ are the Hamiltonians of the bottom and top layers in their reciprocal spaces $k$ and $\tilde k$ and $T(|\tilde k-k|)$ is the Fourier version of the interlayer potential. This term allows for the existence of non-zero scattering outside the first Brillouin zone (Umklapp scattering) of each graphene.

\begin{figure}[h]
    \centering
    \includegraphics[width=.7\linewidth]{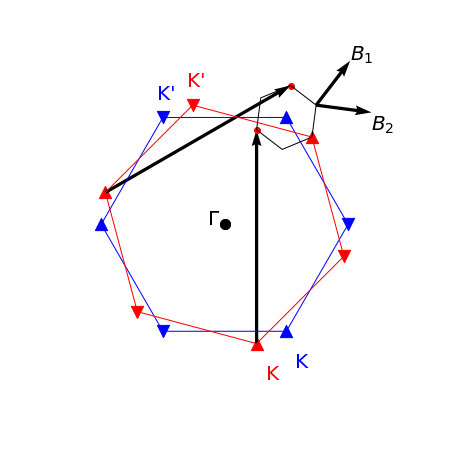}
    \caption{Mismatch between the first Brillouin zone of the top (red) and bottom (blue) layers. Black arrows mark the folding of the Dirac points of the top layer that generate Umklapp processes with a specific $K$ point of the bottom layer. In black we can see the mBZ and their basis vectors.}
    \label{fig:miniBrillouinZone}
\end{figure}
To establish a distance between the $k$ points $|\tilde k-k|$, we can use the equivalence condition between the two reciprocal spaces
\begin{equation}\label{condition BM}
    k_{bot}+G_{bot}= k_{top}+ G_{top}
\end{equation} for $G_{top, bot}$ linear combinations of the reciprocal space basis of each of the layers.

This $T(|\tilde k-k|)$ allows multiple processes to mix different $k$-points between layers (Umklapp processes), but often a truncated version of this model is enough to get accurate results \cite{lam2022theoryconstructingeffectivemodels}. We will consider only the interactions between the $K$ point of the bottom layer and the closest $K$ point of the top layer. Three processes between the three equivalent versions of these cones are present in the first Brillouin zone of graphene. The geometry of these three equivalent processes can be seen as the red dots of the mBZ in Fig.\ \ref{fig:miniBrillouinZone} where each $K$ point in the top layer presents a similar interaction with its closest $K$ point of the bottom layer.

For the $K$ cone of graphene, the full effective continuous model reads as
\begin{equation}\label{eq:Bistritzer}
    H=\left[\begin{array}{cccc}
         H_{bot}(\mathbf { k})& T_1&T_2&T_3  \\
         T_1^\dagger&H_{top}(\mathbf {\hat k_1})&0&0\\ 
         T_2^\dagger&0&H_{top}(\mathbf {\hat k_2})&0\\ 
         T_3^\dagger&0&0&H_{top}(\mathbf {\hat k_3})\\ 
    \end{array}\right],
\end{equation}
with 
$$
\begin{array}{c}
T_1=w(|\mathbf s|)\left(\begin{array}{cc}
     1&1  \\
     1&1 
\end{array}\right)\quad 
T_2=w(|\mathbf s|)\left(\begin{array}{cc}
     1&e^{-2i\pi/3}  \\
     e^{2i\pi/3}&1
\end{array}\right)\\\\
T_3=w(|\mathbf s|)\left(\begin{array}{cc}
     1&e^{2i\pi/3}  \\
     e^{-2i\pi/3}&1 
\end{array}\right).\qquad 
       \\
      
\end{array}$$

Here $H_{top}(\mathbf{\hat k}_i)$ represents the $k\cdot p$ Hamiltonian \eqref{KdotP} developed from all the red points in the mBZ of Fig.\ \ref{fig:miniBrillouinZone}. These red dots are located at points 

\begin{equation}
    G_i=K_{bot} + s_i,
\end{equation}
 where $K_{bot}$ is the Dirac cone of the bottom layer and $s_i$ are the vectors emerging from the condition \eqref{condition BM} such that
\begin{equation}
    \begin{array}{ccc}
         s_1&=&K_{top}-K_{bot}  \\
         s_2&=& K_{top}-K_{bot}+G_{bot,1}-G_{top,1} \\
         s_3&=& K_{top}-K_{bot}+G_{bot,1}-G_{bot,2}-G_{top,1}+G_{top,2} \\
    \end{array},
\end{equation}
are the folding of the top layer Dirac points onto the mBZ, where in this equation $K_{top, bot}$ represent the Dirac points of top and bottom layers respectively.

Therefore, $\mathbf{\hat k_i}-\mathbf{k}=\mathbf s_i$ is the transformation between two equivalent points in the $k$ space given by \eqref{condition BM}. For small angles, this distance can be geometrically approximated to have magnitude $|\mathbf{s}_i|=|\mathbf s|\approx2 K \sin(\theta/2)$. The phase in $T_1,T_2,T_3$ is extracted from the hexagonal symmetry of graphene and $\omega$ is the strength in the reciprocal space of the interlayer coupling, frequently considered to be $\omega(|\mathbf s|)\approx110\,\mathrm{meV}$.

 This model can be understood as a reciprocal space version of a tight-binding problem similarly to the one described in \eqref{tight binding}. That allows us to choose a unit cell in the reciprocal space for such vector. For that, a valid choice can be $B_1=b_{bot,1}-b_{top,1}+2b_{bot,2}-2b_{top,2},\,B_2=b_{top,1}-b_{bot,1}$. In Fig.\ \ref{fig:miniBrillouinZone} we can see this mini-Brillouin zone (mBZ) in black where the Brillouin zones of the top and bottom graphenes are displayed in blue and red respectively. 

 Even if the model presented here is only valid for small angles of twisted bilayer graphene, certain generalizations have been made both for incommensurate angles \cite{Koshino_2015} and for other materials such as graphene on hBN \cite{PhysRevB.90.155406}. We will focus on a specific generalization for the quasicrystalline angle. As we increase the twist angle, the interactions between the $K$ point of one of the layers and the $K'$ point of the other can become more important. The limit case occurs at $30^\circ$ where the $K$ point of the bottom layer is equidistant from the $K$ and $K'$ points of the top layer.

\begin{figure}[h!]
    \centering
    \includegraphics[width=0.7\linewidth]{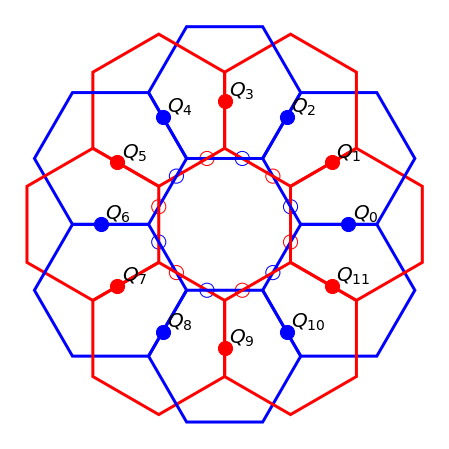}
    \caption{First Brillouin zone of the bottom (blue) and top (red) graphene layers. We can explicitly see the position of the twelve points $Q_i$ as the solid dots. Their reductions to the first Brillouin zone can be seen as the inner ring of empty dots where we can explicitly see they are equidistant to the Dirac cones of both layers.}
    \label{fig:FBZ_Koshino_Moon}
\end{figure}

 To solve this, Mikito Koshino and Pikyung Moon (KM) proposed a generalization for the BM model \cite{PhysRevB.99.165430} into a twelve-band model that couples the twelve Dirac cones present in the two layers by twelve high-symmetry points in the reciprocal space $Q_i = |b_1|\pi/\sqrt{3}\left(\cos(i\pi/6),-\sin(i\pi/6)\right)$ with $|b_1|$ the reciprocal lattice vector defined as in \eqref{eq: reciprocal-lattice-vectors-graphene}. These points are located at the same distance from both Dirac cones due to the periodicity of the system, as can be seen by the empty dots in Fig.\ \ref{fig:FBZ_Koshino_Moon}, where we can see the projections in the unit cell of each $Q_i$.

They propose that instead of hybridizing with the closest available $K$, the model in the 30$^\circ$ twisted bilayer graphene can be described by resonant Umklapp processes in a ring where every Dirac cone of each layer is coupled to the $K$ and $K'$ points of the other layer by those $Q_i$, and thus, the full Hamiltonian reads

\begin{equation}\label{eq:Koshino-Moon}
    H=\left(\begin{array}{ccccccc}
         H_0(\mathbf{\hat k}_0)&T_K&&&&&T_K^\dagger  \\
         T_K^\dagger&H_1(\mathbf{\hat k}_1)&T_K\\
         &T_K^\dagger&H_2(\mathbf{\hat k}_2)&T_K\\
         &&\ddots&\ddots&\ddots\\
         &&&&T_K^\dagger&H_{11}(\mathbf{\hat k}_{11})&T_K\\
         T_K&&&&&T_K^\dagger&H_{12}(\mathbf{\hat k}_{12})
    \end{array}\right)
\end{equation}
where $H_i$, in a similar way to the BM model, is the Hamiltonian in the inner space of each Dirac cone, with each of the $\mathbf{\hat k}_i=(R(-7i\pi/6)k+Q_0)$ with $R(\theta)$ the two-dimensional rotation matrix is defined as $R(\theta)=e^{-i\theta\sigma_y}$. $T_K$ are the overlap matrices, analogous to $T_i$ in the original BM, where $$T_K=\omega(|\mathbf s|)\left(\begin{array}{cc}
     e^{-2\pi i /3}&1  \\
     1& e^{2\pi i /3}
\end{array}\right),$$ for this angle $\omega(|\mathbf s|)\approx\omega(|2 K \sin(15^\circ)|)\approx 157\ \mathrm{meV}$.

We will see that this model produces a highly symmetric Hamiltonian with a twelvefold rotational symmetry, which is the symmetry of the quasicrystal twisted bilayer graphene. Later on in this thesis we will see the impacts of this high order symmetry in the states of the dodecagonal graphene.

Here we have presented two very popular continuous models that are frequently used to approximate the electronic properties of low angle and quasicrystalline bilayers of graphene. We can see the explicit bands obtained by these models in Figs.\ \ref{fig:BM} and \ref{fig:KM}. We will use them to review some important results in low angle and quasicrystalline twisted bilayers of graphene. But first, it is important to note that not all the states in these superperiodic systems will be forced to be super- or quasi- periodic, and to know if they are, it is important to talk about their projections.

\subsubsection{Layer-Projected states}

We have already explained that the interaction between layers and Umklapp processes can be relevant for states of superperiodic systems and generate exotic features, but the existence of superperiodicity is not going to be a sufficient condition for Umklapp processes to take relevance. It is important to discriminate for the purposes of the study of super-/quasi- periodicity when talking about twisted multilayers of graphene, which states present this superperiodic nature and which don't. Thus, it will be important to include these definitions.

\begin{itemize}
    \item We will call \textbf{intralayer} states those that belong only to one layer, this is $\psi=P_L\psi$ where $P_L$ is the projector onto that layer subspace. These states must present all the symmetries of the separate layer in which they live and generally do not present the long-range behavior of the multilayer. We can see these states as the colored parts of Figs. \ref{fig:BM} and \ref{fig:KM}.

    \item We will call \textbf{interlayer} states those that belong to two or more layers. Such states must respect the symmetries of both layers and present the symmetries of the whole set of layers; thus these states must be intrinsically super-/quasi- periodic and we will consider this as a criterion for super- and quasiperiodicity in moir\'e. Those states correspond to the gray sections of Figs. \ref{fig:BM} and \ref{fig:KM} where the state does not correspond to a projection in any of the layers.
\end{itemize}

%
With that, we have a criterion to distinguish in a twisted bilayer graphene which states can be understood as states of each separate layer and which ones are intrinsic to the complex geometry of the moir\'e. Later in  chapter\ \ref{Chapter: Transport in twisted multilayer graphene} we will see that this also has an impact on the behavior of transport for these multilayers.

\subsection{The magic-angle twisted bilayer graphene}\label{Sec:MATBLG_intro}

As discussed in the previous section, introducing a relative twist angle between two graphene layers generates a long-wavelength moir\'e superlattice and a corresponding mini-Brillouin zone, where interlayer Umklapp processes strongly modify the electronic structure.
Among all possible twist angles, a special role is played by the so-called
magic-angles, where the Dirac cones of twisted bilayer
graphene become remarkably flat, leading to strongly enhanced interaction effects. There have been several angles predicted to be ``magic'' \cite{doi:10.1073/pnas.1108174108} but only one has been experimentally found to date \cite{Cao2018} around $1.05^\circ$, so we will refer to it as the magic-angle twisted bilayer graphene.

\begin{figure}[t]
    \centering
    \includegraphics[width=\linewidth]{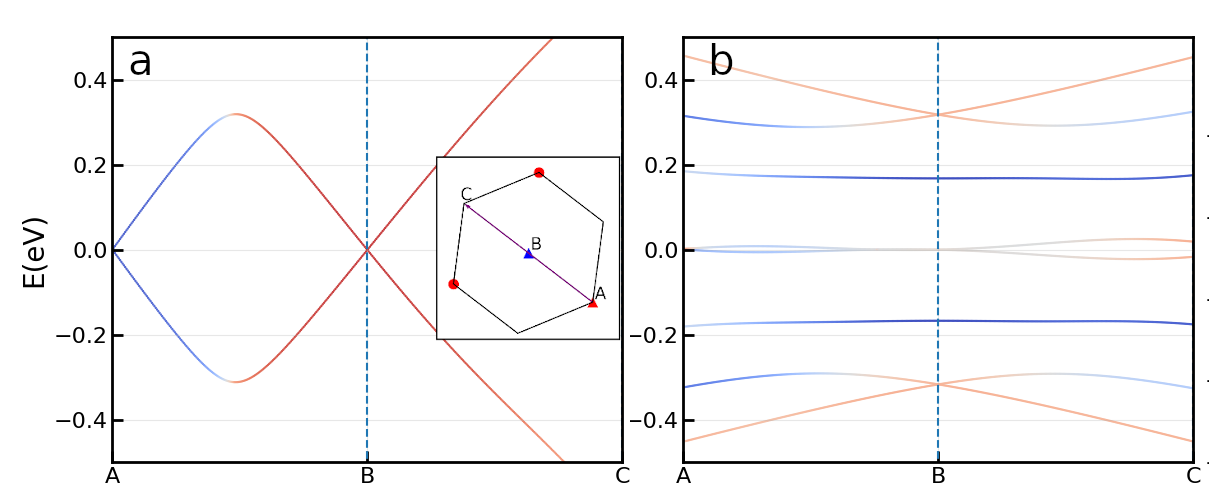}
    \caption{Band structure of two different angles for the twisted bilayer graphene computed using the Bistritzer-MacDonald models \cite{doi:10.1073/pnas.1108174108} for angles of $5^\circ$ (left) and $1.05^\circ$ (right), following \eqref{eq:Bistritzer}.
    Inset of panel a shows the $K$-path used where the mBZ corresponds to the black hexagon in Fig.\ \ref{fig:miniBrillouinZone}. Red (blue) colors in the main panels represent projections onto the top (bottom) layer.}
    \label{fig:BM}
\end{figure}

From a continuum-model perspective, the physics of small-angle twisted bilayer
graphene can be understood in terms of two Dirac cones, one from each layer,
coupled in momentum space by a vector of magnitude
$k_\theta \approx 2K \sin(\theta/2)$. As the angle is reduced, these two Dirac cones get closer, which pull the interlayer states lower in energy until the limit case of the magic-angles. At these angles the interaction between the layers and the proximity in reciprocal space between their Dirac cones generate the interlayer flat band at low energy which calls for electron localization and correlations. We can see explicitly this behavior in the transition between panel a of Fig.\ \ref{fig:BM} at $5^\circ$  with two well-defined Dirac cones for the top and bottom layers, to panel b at $1.05^\circ$.

In real space, the small twist angle produces a moir\'e unit cell of the order of $\sim 14$~nm at the magic angle. Within this enlarged unit cell, the local stacking varies smoothly between AA, AB/BA and intermediate configurations. As shown by both theory and experiments, the flat-band wave functions are strongly enhanced in the AA regions, where the two layers are almost perfectly aligned, while being suppressed in AB/BA domains. This has been experimentally confirmed by STM measurements of the local occupations of MATBLG \cite{STM-Magic-angle}.

As we discussed previously, the flatness of the bands at the magic angle can be directly related to the
general expression for the band velocity in \eqref{eq:fermi velocity}.
Around charge neutrality, the bandwidth of the active moir\'e bands is reduced
to only a few meV, so that the corresponding Fermi velocity is
strongly suppressed compared to that of monolayer graphene.
 This suppression of kinetic energy is a key ingredient for its exotic correlated nature and will be revisited in detail in chapter\ \ \ref{Chapter: Transport in twisted multilayer graphene} for realistic
magic-angle structures.

\subsubsection{Flat bands and electronic correlations}

The strong reduction of the kinetic energy scale elevates the relative importance of electron-electron interactions.
In the tight-binding description introduced in Sec.\ \ref{Sec:Tight-Binding-graphene}, the electronic Hamiltonian can be read as
\begin{equation*}
 H = H_{\mathrm{TB}} + H_{\mathrm{el-el}} ,
\end{equation*}
where $H_{\mathrm{TB}}$ describes the single-particle tight-binding part and
$H_{\mathrm{el-el}}$ accounts for Coulomb interactions.
For monolayer graphene or for weakly perturbed systems, it is often a good approximation to neglect $H_{\mathrm{el-el}}$ at the level of band structure and transport. In contrast, in magic-angle twisted bilayer graphene (MATBLG), the flatness of the active bands implies that the kinetic energy per electron becomes
comparable to, or even smaller than, the characteristic Coulomb energy scale.

A simple model for these interactions in the moir\'e lattice are
\begin{equation}
 H_{\mathrm{el-el}} =
 \frac{1}{2}
 \sum_{i\neq j} V_{ij}\,\hat{n}_i \hat{n}_j
 \qquad
 V_{ij} = \frac{e^2}{4\pi\varepsilon_0\varepsilon_r
 |\mathbf{r}_i-\mathbf{r}_j|},
 \label{eq:first coulomb appearence}
\end{equation}
where $\hat{n}_i$ is the electronic density operator at the lattice site $i$,
$\varepsilon_r$ is an effective dielectric constant accounting for the screening
from nearby layers and substrate, and $\mathbf{r}_i$ denote the positions of
localized Wannier states associated with the flat bands.
For this reason, MATBLG is a paradigmatic example of a correlated electron system \cite{Balents2020}: the ratio between interaction strength and bandwidth is large, and the ground state cannot be understood within a simple band picture.

Experimentally, this enhanced role of correlations is reflected in the  appearance of interaction-driven phases when the flat bands are partially filled. Transport measurements have revealed transitions between different correlated insulating states at certain integer fillings of the flat bands and superconducting phases in their vicinity \cite{Cao2018}. In their study Cao et al. found not only a superconducting state for low temperatures but also a phase diagram similar to those of the high-temperature superconductors. The exact mechanism for the superconductivity of MATBLG remains an open question to date and thus many studies on its electron-electron interactions have been conducted \cite{Balents2020}. 
In later studies other correlated phases have been found, such as Chern insulating states \cite{Nuckolls2020} or optical cascades \cite{efetov-cascades,KrishnaKumar2025, nhs2-mbhr}. These correlated phases are highly sensitive to experimental conditions such as strain, twist-angle inhomogeneity, and disorder, underscoring the delicate balance between localization, kinetic energy and Coulomb interactions in MATBLG \cite{ciepielewski2024transporteffectstwistangledisorder,PhysRevLett.134.126301}. This provides a fundamental motivation for more realistic studies of such effects against experimental conditions. To have a first realistic approximation to the effects of disorder on the flat-band states producing these interactions is the main purpose of this thesis.

\subsection{The quasicrystal twisted bilayer graphene}\label{sec:quasicrystal intro}

As we mentioned at the beginning of Sec.\ \ref{Sec: super-quasiperiodicity in tblg} generic misalignments between several graphene layers often lead to incommensurate stacking geometries, an effect that is even increased when more than two layers are involved \cite{Uri2023}. Among all such possibilities, the most prominent and experimentally established example in the bilayer case is the so-called graphene quasicrystal, obtained for a relative twist angle of $30^\circ$ between two graphene sheets \cite{doi:10.1073/pnas.1720865115,doi:10.1126/science.aar8412}. This structure consists of a two-dimensional quasicrystal that preserves the massless Dirac character of the carriers.

\begin{figure}[]
    \centering
    \includegraphics[width=\linewidth]{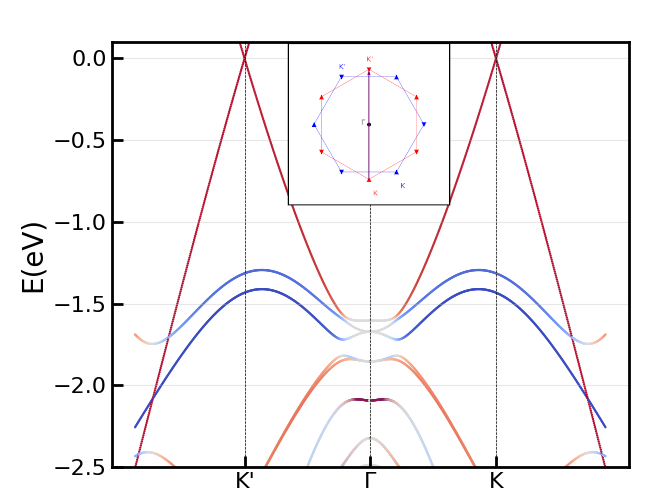}
    \caption{Band structure of the quasicrystal twisted bilayer graphene computed with the effective Koshino-Moon model \cite{PhysRevB.99.165430} following equation\ \eqref{eq:Koshino-Moon}. The inset shows the $K$-path for the Koshino-Moon model where the Brillouin zone of bottom and top graphene corresponds to the central hexagons of Fig.\ \ref{fig:FBZ_Koshino_Moon}. Red (blue) colors in the main panels represent projections onto the top (bottom) layer.}
    \label{fig:KM}
\end{figure}

As the structure of $30^\circ$ twisted bilayer graphene is quasicrystalline, neither a unit cell nor a superperiodicity can be found. Unlike other incommensurate twisted bilayers of graphene whose structure can also be considered quasicrystalline, here the structure defines two of the main interesting opportunities that quasicrystals offer \cite{MaciaQC}:

\begin{itemize}
    \item A well-defined 12-fold (dodecagonal) \textbf{rotational symmetry}, which is forbidden for regular Bravais lattices \cite{Kittle}.
    \item A \textbf{fractal} structure, where stacking in different regions can be mapped with inflations of the Stampfli pattern \cite{10.3389/frcrb.2024.1496179,stampfli1986dodecagonal}, a 12-fold rotationally symmetric self-similar fractal pattern invented by Peter Stampfli in 1986, inflations of this pattern, we will introduce later in this text this fractal pattern and its inflations.
\end{itemize}

This is reflected in diffraction and ARPES measurements, which reveal sharp Bragg peaks and multiple replicas of the Dirac cones arranged with 12-fold symmetry in momentum space, a rotational order that is forbidden in ordinary crystals \cite{doi:10.1073/pnas.1720865115,doi:10.1126/science.aar8412}.

From an electronic point of view, we have already seen a description of the continuous model in \eqref{eq:Koshino-Moon}. As a result of the lack of periodicity, instead of a single mBZ, we find the relevant states generated by a hierarchy of Umklapp processes between 12 Dirac cones located at equivalent positions given the symmetries of the lattice. In Fig.\ \ref{fig:KM} we can explicitly see those states, where we can distinguish two sectors. First, we can see the two Dirac cones of the bottom graphene layer, where we find the intralayer states; later we will see that the electrons in these regions behave as a combination of two monolayer graphenes. Second, at higher energies one finds a set of interlayer states; we will focus here on the $\alpha$, $\beta$, and $\gamma$ states. In these states, the multiple Umklapp processes become simultaneously resonant and produce flat bands, giving rise to a spiky density of states, as we will see later on in Sec.\ \ref{Sec: Graphene Quasicrystal}.

In real space, high-energy resonant states display characteristic quasicrystalline features. Tight-binding calculations and ARPES/STM experiments find that the local density of states at these resonance energies presents the same high order symmetry as the structure. In addition, studies on local descriptions of states find different resonant peaks following inflations of the Stampfli pattern \cite{10.3389/frcrb.2024.1496179,stampfli1986dodecagonal}. In chapters\ \ref{chapter:superperiodicities} and \ref{Chapter: Transport in twisted multilayer graphene} we will see the impact of these fractal wave functions have on transport, producing high resistivity materials due to the hierarchical electron scattering.

Another important feature of 30$^\circ$-tBLG that will be relevant for chapter\ \ref{Chapter: Transport in twisted multilayer graphene}, is the energy separation between these interlayer quasicrystalline states and the low-energy Dirac sector. The quasicrystalline resonances appear at relatively high energies, typically in the window $|E|\sim 1.5$-$2$ eV. This separation of energy scales makes the graphene quasicrystal a natural counterpart to MATBLG. In MATBLG, interlayer Umklapp scattering at small angles produces flat interlayer bands and strong correlations at low energies, while in 30$^\circ$-tBLG, the strongest hybridization is shifted to high energies. In later chapters, we will see the effects of this contrast by combining MATBLG with quasicrystalline layers, using periodic approximants of 30$^\circ$-tBLG to study how flat-band physics and quasicrystalline resonances hybridize in realistic large-scale heterostructures.

\subsubsection{The Stampfli pattern and fractality}
During this section, we have mentioned several times the Stampfli pattern as the geometrical disposition of the $30^\circ$ twisted bilayer graphene structure. Here, we will introduce this fractal pattern.

The Stampfli pattern is a self-similar dodecagonal quasiperiodic tiling introduced by Peter Stampfli in 1986 \cite{stampfli1986dodecagonal}. In the version relevant here, it is built from three elementary tiles: equilateral triangles, squares, and rhombi with an acute angle of $30^\circ$. A convenient construction starts with a 12-fold rosette, or equivalently from a central dodecagonal cluster, and applies an inflation-substitution rule with scale factor $2+\sqrt{3}$. In each iteration, the pattern is first enlarged by this factor, and each enlarged tile is then replaced by a symmetric arrangement of smaller triangles, squares, and rhombi. Repeating this procedure generates a hierarchy of similar motifs at increasing length scales. Because the inflation factor is irrational, the structure has a long-range orientational order and an exact twelvefold symmetry but no translational periodicity. In this sense, the three primitive tiles define the local building blocks of the pattern, while successive inflations define its higher-order geometry. This hierarchical construction makes the Stampfli pattern a natural geometric framework for describing the real-space organization of quasicrystalline states in $30^\circ$ twisted bilayer graphene.

We can see this pattern described in Fig.\ \ref{fig:stampfli}, and matched to the $30^\circ$ twisted bilayer graphene.

\fullpagefigure[1]{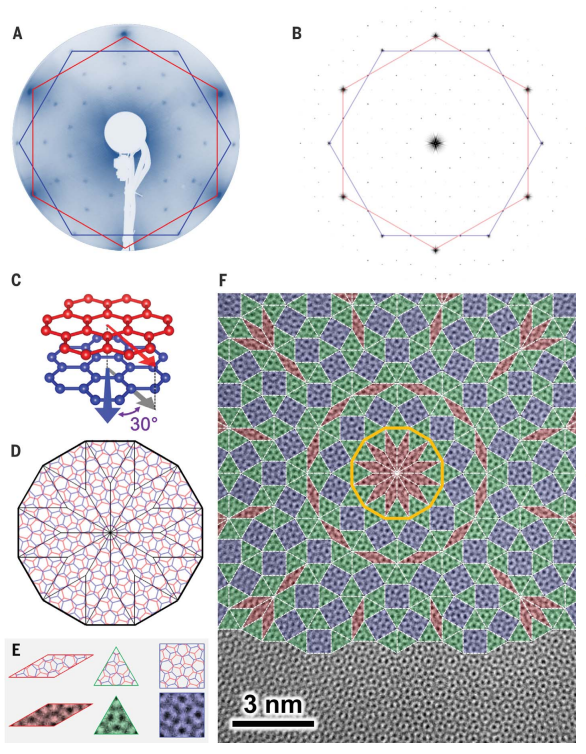}
{(A) A low-energy electron diffraction  pattern of graphene quasicrystal. (B) A Fourier-transformed pattern of graphene quasicrystal and the first Brillouin zone of each layer. (C and D) An atomic structure model of twisted bilayer graphene with 30$^\circ$. (E) Atomic structures and TEM images of Stampfli tiles [rhombuses (red), equilateral triangles (green), and squares (blue)]. (F) A false-colored TEM image of graphene quasicrystal mapped with 12-fold Stampfli-inflation tiling. From \cite{doi:10.1126/science.aar8412}. Reprinted with permission from AAAS.}{fig:stampfli}


    

\chapter{Computational methods}\label{chapter:computational}
This chapter presents the computational framework used throughout the thesis to model electronic structure and quantum transport in graphene-based systems with very large superperiodic length scales, and in some cases with no unit cell at all. Because these systems range from large moir\'e superlattices to genuinely quasiperiodic structures, the methodology must remain accurate at the atomistic level while also scaling to samples containing millions of orbitals.

For this reason, the chapter develops the real-space Kubo formalism used to compute spectral and transport observables, including the density of states, diffusion coefficient, mean-squared displacement, conductivity, and mean free path. It also introduces the linear-scaling kernel polynomial methods \cite{FAN20211} that make these calculations feasible, together with the auxiliary self-consistent schemes used later in the thesis to model externally generated electrostatic potentials and mean-field interaction effects. Altogether, this chapter establishes the numerical tools that connect realistic atomistic Hamiltonians to experimentally relevant transport and localization metrics.

    

\section{Basics of quantum transport}\label{Section: quantum transport}

\subsection{Kubo Formalism and the Chester-Thellung formula}

All our transport calculations are going to be made under the Kubo formalism. This will allow us to obtain a precise calculation of the conductivity for zero-temperature systems under the single-electron approximation. We start with the Kubo-Bastin conductivity at a certain temperature $T$ and chemical potential $\mu$ \cite{PStreda_1982,KUBO}

\begin{equation}\label{eq:kuboBastin}
    \sigma(\mu,T)=-\frac{ \hbar e^2}{\Omega}\int \mathrm d E'\left[-\frac{\partial f(E'-\mu)}{\partial E'}\right]\times \mathrm{Tr}\left[\delta\left(E'-\hat H\right)\hat V\,\mathrm {ImG^+}\left(E'\right)\hat V\right].
\end{equation}

Here $f$ is the Fermi-Dirac distribution \cite{Ashcroft:102652}, $\Omega$ is the volume of the phase space, $e$ and $\hbar$ are the electric charge of the electron and the reduced Planck constant respectively. $\hat V$ is the velocity operator that can be computed from the Heisenberg equation of motion \cite{fetter-walleka} as 

\begin{equation}\label{eq:velocityOperator}
    \hat V=-\frac{i}{\hbar}\left[\hat H,\hat R\right].
\end{equation}

The $\mathrm{G^+ (E')}$ is the retarded Green's function. For it, we may choose a regularization resulting from an adiabatic transformation to a steady state of energy $E'$ \cite{FAN20211}, thus making
\begin{equation}\label{eq:green_functions}
    \mathrm {ImG^+}\left(E'\right)=-\pi\delta(E'-\hat H)=-\pi\lim_{\tau\to\infty}\int_{-\tau}^\tau\frac{\mathrm{dt}}{2\pi\hbar}e^{i(E'-\hat H)t/\hbar}.
\end{equation}
With that, changing the derivative as $-\partial f (E'-\mu)/\partial E'=\partial f (E'-\mu)/\partial \mu$ and expanding the functions over the Hamiltonian onto their eigenstates with $f(H)=\sum_nf(E_n)|n\rangle\langle n|$ with $F(E)$ the function; we can integrate over $E'$ to obtain the Chester-Thellung formula \cite{GVChester_1959,GVChester_1961}

\begin{equation}\label{eq:Chester-tellung}
    \sigma(\mu,T)=\lim_{\tau\to\infty}\frac{e^2}{2\Omega}\int_0^\tau \mathrm d t \mathrm{Tr}\left[\frac{\partial f(\hat H-\mu)}{\partial \mu}\left\lbrace\hat V(t),\hat V(0)\right\rbrace\right].
\end{equation}
This formula, in its zero temperature version, is going to be the basis for our transport calculations.


\subsection{The mean-squared displacement }

Equation (\ref{eq:Chester-tellung}) has a temperature dependence via the Fermi function $f(\hat H -\mu)$. For simplicity, as we will only work with systems at zero temperature, we will take the limit $T\to0$ at this point. With that we have
\begin{equation}\label{eq:Chester tellung 0 temperature}
    \sigma (\mu)=\lim_{\tau\to\infty}\frac{e^2}{2\Omega}\int_0^\tau \mathrm d t \mathrm{Tr}\left[\delta(\hat H-\mu)\left\lbrace\hat V(t),\hat V(0)\right\rbrace\right].
\end{equation}

From there, we can simplify this formula if we define the equilibrium density of states as
\begin{equation}\label{eq:equilibrium dos}
    \hat \rho_{eq}(\mu)= \frac{1}{\Omega}\frac{\delta(\hat H - \mu)}{\mathrm{Tr} \left[\delta(\hat H - \mu)\right]},
\end{equation}
where $\mathrm{Tr} (\delta(\hat H - \mu))$ corresponds to the total occupancy at a chemical potential $\mu$; and use it to define the mean-squared displacement
\begin{equation}\label{eq:msd}
    \Delta X^2(\mu,t)=\mathrm{Tr}\left[ \hat\rho_{eq}(\mu)(\hat X(t)-\hat X(0))^2 \right] .
\end{equation}

Using its second derivative and (\ref{eq:equilibrium dos}), inserting them into (\ref{eq:Chester tellung 0 temperature}) we arrive at the relation
\begin{equation}
    \label{eq:msd conductivity}
    \sigma (\mu)= \lim_{t\to \infty} e^2\rho (\mu)\frac{1}{2}\frac{\mathrm d \Delta X^2(\mu,t)}{\mathrm dt}.
\end{equation}

To compute transport during this thesis, we will mainly exploit this relation between the mean squared displacement and the conductivity making use of the diffusion coefficient as described in the next section.

\subsection{The diffusion coefficient}

It will be important for us, then, to introduce the diffusion coefficient \cite{10.1143/PTP.20.948} as:

\begin{equation}\label{eq:diffusion coefficient definition}
    D(E,t)=\frac{1}{2}\frac{\mathrm d \Delta X^2(\mu,t)}{\mathrm dt}.
\end{equation}
Here we can express conductivity as $\sigma(\mu)=\lim_{t\to\infty} e^2\rho(\mu)D(\mu,t)$ taking advantage of \eqref{eq:msd conductivity}.

It is important to note that this quantity, as well as mean squared displacement and electronic spreading are not just mathematical tools, but contain relevant information of electronic transport; later on in this thesis we will make further connections between such quantities and relevant characteristic parameters for the system.
In the case of the diffusion coefficient, it is important to see how it can be related to a time dependent conductivity \cite{FAN20211}, in particular 
\begin{equation}\label{eq:conductivity_diffusion_coeff}
    \sigma(\mu,t)= e^2\rho(\mu)D(\mu,t),
\end{equation}
where far from being a simple definition to simplify our equations, during this thesis, it will prove to be related to transport at different length scales for the system.

This diffusion coefficient is going to be the main quantity shown for transport in this thesis, and therefore it is interesting to take a closer look at how it evolves in the different transport regimes. We will dedicate the next section to studying how $D(E,t)$ evolves in disordered media, taking as a reference the case of a linear chain.

\subsection{Transport in disordered media}


We will cover here the main regimes of transport in materials for different strengths of disorder. For that purpose, we will take the disorder defined as Anderson disorder \cite{PhysRev.109.1492} and connect it to the evolution of the wave packet.

For a tight-binding system defined as (\ref{tight binding}) we add an onsite potential with the form
\begin{equation}
    \label{eq: Anderson disorder}
    V_{anderson}=\sum_i \omega_ic_ic_i^\dagger,\quad\omega_i\in\left[-W,W\right],
\end{equation}
where $\omega_i$ are random real numbers contained in the interval $[-W,W]$ with homogeneous distribution. To that $W$ we will call the Anderson strength. Depending on the strength of this disorder, eventually we will fall into one of the following transport regimes: ballistic, diffusive, or localized.

\subsubsection{Ballistic regime}\label{sec: ballistic regime}

Let us start with a system without disorder. Thus, any electron in the system must behave as a Bloch wave with constant Fermi velocity, we say that we are in a ballistic regime, where there is no scattering between different points of the first Brillouin zone. From (\ref{eq:fermi velocity}), (\ref{eq:Chester-tellung}) and (\ref{eq:msd conductivity}), it follows that the mean squared displacement behaves in a ballistic regime as 
\begin{equation}
    \Delta X(\mu) ^2 = v_f(\mu)^2 t^2.
\end{equation}.

Thus,  taking the derivative, the diffusion coefficient 
\begin{equation} \label{eq: diffusivity ballistic}
    D(\mu,t)=v_f^2(\mu) t
\end{equation}
must have a linear behavior with slope that is equivalent to the average Fermi velocity at this chemical potential. This $v_f(E)$ is equivalent to the expected Fermi velocity, as defined in (\ref{eq:fermi velocity}) over the entire Fermi contour.

\subsubsection{Diffusive regime}

If we add some disorder, we will enable the scattering inside the first Brillouin zone. 
Instead of a linear evolution with Fermi velocity, the average velocity of the wave packet (extracted as $v(E)=\partial_t D(E,t)$) is reduced with time. Eventually, the system will arrive at the so-called diffusive regime. 

In this regime, the main source for transport is the spreading of the wave packet, and thus the diffusion coefficient reaches a constant value, and thus using (\ref{eq:Chester tellung 0 temperature}),(\ref{eq:msd conductivity}) and the expansion of the velocity operator (\ref{eq:fermi velocity}) we find 
\begin{equation}
    D(\mu,t)=D(\mu)=\lim_{t\to \infty}\frac{1}{4}\frac{\mathrm d}{\mathrm d t}\left( \Delta X^2+\Delta Y^2\right)=\lim_{t\to \infty}\frac{1}{2}\frac{\mathrm d}{\mathrm d t} \Delta X^2
\end{equation}
 for 2D transport, where the last equivalence only holds for homogeneous transport with $\Delta X ^2 = \Delta Y^2$.
We will call this value of $D(\mu)$ the diffusion constant.

Let us go deeper into the meaning of the term $Tr\left[\rho_{eq}\left\lbrace\hat V(0),\hat V(t)\right\rbrace\right]$ in equation (\ref{eq:Chester tellung 0 temperature}). This quantity is proportional to the velocity autocorrelation. It is easy to see that it reaches a value of $2v_f(\mu)^2$ for pure ballistic regimes and of 0 for pure diffusive regimes. Now we can assume that this transition has an exponential shape, such as
\begin{equation}
    Tr\left[\rho_{eq}\left\lbrace\hat V(0),\hat V(t)\right\rbrace\right]=v_f(\mu)^2e^{-{t/\tau_p}},
\end{equation}
where $\tau_p$ is the momentum relaxation time. Then, from equations (\ref{eq:Chester tellung 0 temperature}), (\ref{eq:msd conductivity}), (\ref{eq:diffusion coefficient definition}), we have
\begin{equation}\label{eq:deffinition_momentum_scattering_time}
    D(\mu)=\lim_{t\to \infty}D(\mu,t)=\frac{1}{2}v_f(\mu)^2\tau_p.
\end{equation}
From here, we can also define the mean free path
\begin{equation}
    \label{eq:deffinition mfp}
    \ell (E)=v_F(E)\tau_p(E),
\end{equation}. 
where we recover the semiclassical values for conductivity in diffusive systems. Using scaling arguments, we can also find the localization length to be dependent on the mean free path.


\subsubsection{Localized regime}

When the disorder is too large, there is such a large amount of scattering that the electron remains confined under a certain localization length $\xi(\mu)$. In those systems, we expect the mean-squared displacement to become time independent, and thus we achieve
\begin{equation}
    \lim_{t\to\infty}D(E,t)=0.
\end{equation}
following an exponential decay as $D\propto e^{-L/\xi(E)}$.
This is the so-called strong localization regime.

Nevertheless, there is the possibility of a different kind of localization. If the localization length is much larger than the mean free path, $\xi>>\ell$, then the diffusion coefficient decays smoothly as a function of time, and thus $\lim_{t\to\infty} D(E,t)=0$, this is the weak localization regime (WL), and making use of similar scaling arguments as in equation (\ref{eq:deffinition_momentum_scattering_time}), we can express the diffusion coefficient to decay as \cite{FAN20211}
\begin{equation}\label{eq:weak-localization}
    D\propto -log(\frac{L}{\ell}).
\end{equation}

In general, the localization length scales of the system are mediated by the mean free path, so we would expect that the longer the mean free path, the longer the localization length.

\subsection{The Anderson scaling theory of localization} \label{sec:Anderson-scaling-theory-of-localization}

\begin{figure}
    \centering
    \includegraphics[width=\linewidth]{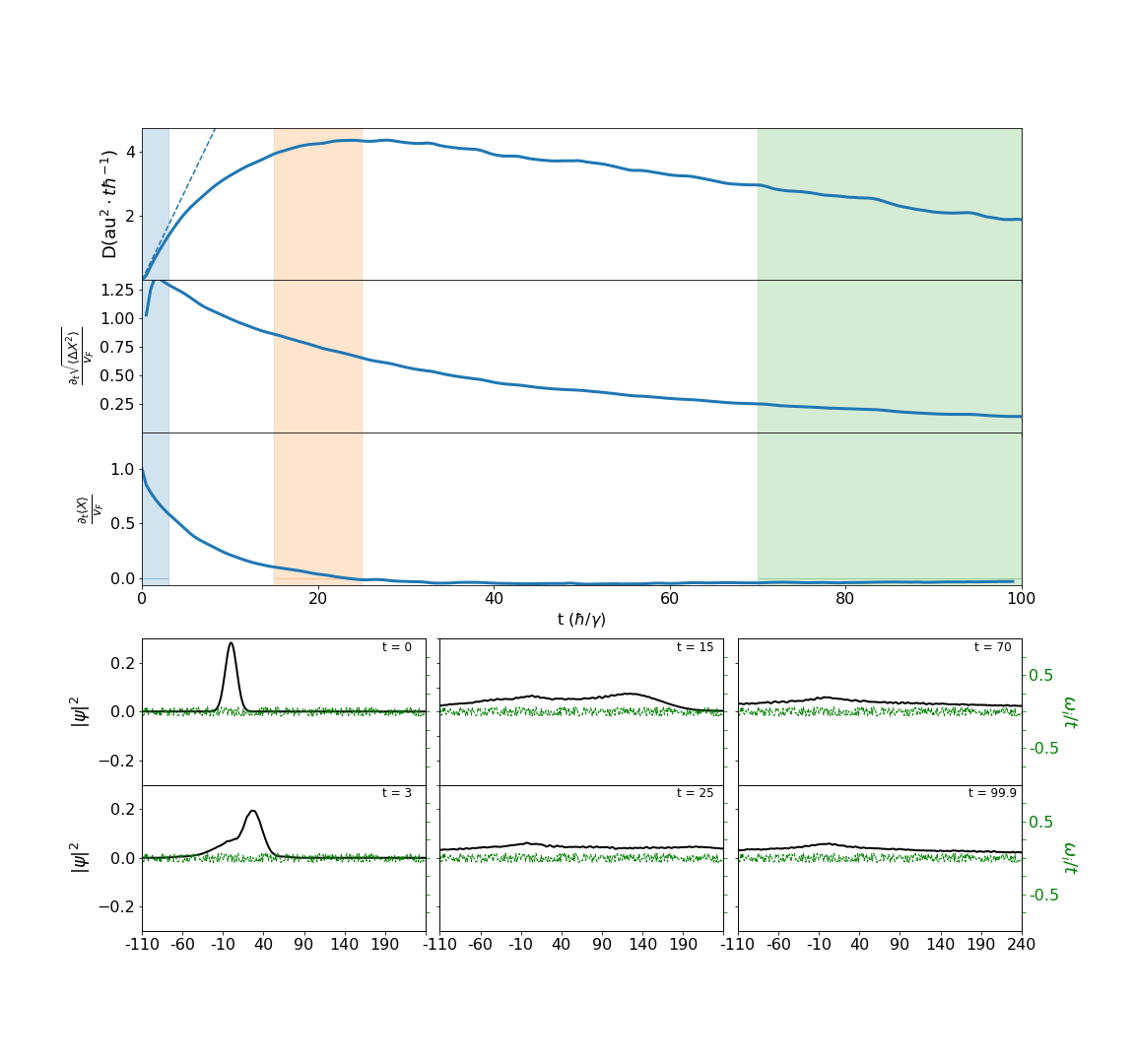}
    \caption{Transport of a Gaussian pulse in a linear chain defined as in (\ref{eq:linear chain}) with Anderson disorder $W=t/8$. In the top panel we can see the diffusion coefficient (top), the time derivative of the electron spread (middle) and the time derivative of the expected position (bottom) as a function of time. The dashed line in the top panel is a guide to the eye representing $v^2 t$ (dashed) expected for the ballistic case. The three shadowed regions are contained in the ballistic(blue), diffusive (orange) and localized (green) regimes. The bottom panels are snapshots for the times starting and finishing the marked regions.}
    \label{fig:anderson_scaling}
\end{figure}

At this point, and with a semiclassical quantity as the mean free path connected both with the diffusive and the localized regimes, we can extend our prior results. As predicted in \cite{Mott01041961,10.1143/PTPS.53.77,Licciardello_1975}, any crystal in 1D or 2D under the effect of a certain amount of disorder is going to go through these three regimes, ballistic, diffusive, and localized. We have an illustrative example for that in Fig.\ \ref{fig:anderson_scaling}, where we can see the evolution of a Gaussian wave packet in a linear chain, where we can see the three different aforementioned regimes. We will make further emphasis on this model later in Sec.\ \ref{sec:linear-chain-superp}.

In the blue region in the top panels of Fig.\ \ref{fig:anderson_scaling}, we can see this linear behavior of $D$ and make a direct comparison with the function (\ref{eq: diffusivity ballistic}). In the middle and bottom panel we can see that both the velocity of the electron spreading $\partial_t\sqrt{\langle\Delta X^2\rangle}$ and the velocity of the expected position  $\partial_t \langle X\rangle$ present values close to the Fermi velocity of the pristine system. In the two left plots of the bottom panel, we can see that the wave packet in this regime is displacing more than spreading.

For the orange region, the system is reaching the diffusive regime, and thus the diffusion coefficient is reaching a plateau where $\partial_t D(t)=0$. In this regime, we can see that the velocity of the spreading (middle plot of the top panel) keeps a finite value, meanwhile the velocity of the position expected value (bottom plot of the top panel) is exponentially suppressed. In the middle column of the bottom panel, we can see that the electrons are indeed mainly spreading.

From then on, the system is localizing. We can see explicitly in the suppression of the spreading velocity in the green region of Fig.\ \ref{fig:anderson_scaling}, and by the fact that the real-space profile (right column of the bottom panel), remains mainly unmodified, predicting the wave function does not escape a certain region in the space.

\section{The kernel polynomial method}\label{Sec: KPM-intro}

In the previous section, we can see that most of the quantities in which we are interested are related to the density of states $\rho_{eq}$ or with a $\delta(H-E)$. This is a singular function whose necessity of regularization we have already mentioned. Frequently, regularization for this function is made by inversion of the Hamiltonian \cite{economou}. However, the cubic scaling of the computational complexity of matrix inversion with time makes it impossible for us to invert our matrices in a reasonable amount of time \cite{RevModPhys.78.275}. Several procedures can be used to reduce this scaling, such as the Lanczos recursive method \cite{RHaydock_1972,RHaydock_1975,Lanczosbook} or methods based on Green's functions \cite{NguyenGreen}. We will make use of the kernel polynomial Method (KPM) where an expansion in Chebyshev polynomials is used to approximate this delta. Yet, in Appendix\ \ref{Appendix: green's function}, we also provide an explanation of the Green's function method.

\subsection{The Chebyshev polynomials}

The Chebyshev polynomials \cite{FAN20211} are a basis of orthonormal functions defined in the interval [-1,1] that can be computed recursively as 
\begin{equation}\label{eq:chebyhev polinomia definition}
    \begin{array}{c}
    T_0(x)=1\\
    T_1(x)=x\\
    T_{n}(x)=2xT_{n-1}(x)-T_{n-2}.
    \end{array}
\end{equation}
 In this new basis, any function can be expanded as \cite{10.1063/1.448136}:
\begin{equation}\label{eq:Chebyshev expansion}
\begin{array}{c}
    f(x)=\sum_{n=0}^{\infty} f_m(x)T_m(x) \\
    \\
    f_m(x)=(2-\delta_{m0})\int_{-1}^1\frac{f(x)T_{m}(x)}{\sqrt{1-x^2}}\mathrm d x.
\end{array}
\end{equation}

We want to take advantage of this expression and the recursion of Chebyshev polynomials to compute $\delta(H-E)$. To do that, we need to rescale the Hamiltonian to the domain of Chebyshev polynomials. Here, we can generate a new reduced Hamiltonian
\begin{equation}\label{eq: reduced hamiltonian}
    \tilde H=\frac{2(\hat H - E_{avg})}{\Delta E},
\end{equation}
with $E_{avg}$ and $\Delta E$ being the band center and bandwidth, respectively. This Hamiltonian is defined in the domain of Chebyshev polynomials $\tilde E=\frac{2( E - E_{avg})}{\Delta E}$. So, making use of the scaling law of the delta
\begin{equation}
    \label{eq: reduced energies}
    \delta (\hat H- E)=\frac{2}{\Delta E}\delta (\tilde H-\tilde E)
\end{equation}
and introducing it into (\ref{eq:Chebyshev expansion}) we find 
\begin{equation}\label{eq:Chebyshev expansion delta}
\begin{array}{c}
    \delta(\tilde H - \tilde E)=\sum_{n=0}^{\infty} \Gamma_m(\tilde E)T_m(\tilde H) \\
    \\
    \Gamma_m(\tilde E)=\left(2-\delta_{m0}\right)\frac{T_{m}(\tilde E)}{\sqrt{1-\tilde E^2}}.
\end{array}
\end{equation}

Note that $\Gamma_m (\tilde H)$ is analytical and will be key to computing our properties making use of the recursion (\ref{eq:chebyhev polinomia definition}) but before that there is an important factor to consider.

In real simulations, we will not be able to make the infinite sum. In practice, we will compute the sum for different number of moments $M$, and converge $M$ to be sufficiently large to model the properties of the system well.

\subsection{The kernel choice}

At this point, we are almost ready to make our calculations. Before that, it is important to note that in the expansion made in \eqref{eq:Chebyshev expansion delta}, we are expanding in Chebyshev polynomials a non-differentiable function. This will result in Gibbs oscillations \cite{RevModPhys.78.275}.

To avoid that, we will include an extra term $g_m$ as a damping factor in convolution with our $\delta$. Like this we will convolute our solution with smooth statistical functions reducing thus the noise. In the Chebyshev expansion, this is as easy as including a new term in the expansion (\ref{eq:Chebyshev expansion delta})
\begin{equation}\label{eq: delta kernel function}
    \delta(\tilde H - \tilde E)=\sum_{n=0}^{\infty} \Gamma_m(\tilde H)g_m T_m(\tilde H). \\
\end{equation}

Here several choices of the kernel can be made. We will make use of the Jackson kernel \cite{RevModPhys.78.275} defined as
\begin{equation}\label{eq: Jackson Kernel}
    g_m=\frac{(M+1-m)\cos\left( \frac{\pi m}{M+1} \right)+\sin\left(\frac{\pi m}{M+1}\right)\cot(\frac{\pi}{M+1})}{M+1}.
\end{equation}
This kernel choice has been shown to approximate the delta and to be an optimal choice for previous work with graphene \cite{FAN20211,SILVER1996115}. This damping term can be understood as the Chebyshev expansion of a Gaussian function, and the broadening will be determined by the number of moments, extracted from $\tilde \eta = 1/M$. Once we recover the original energy spectra of our calculations, the broadening will be 
\begin{equation}\label{eq:Jackson-kernel-broadening}
    \eta=\pi\cdot\Delta E/M.
\end{equation}
This will lower bound the energy resolution we can achieve with the kernel polynomial method for a certain amount of moments.


\subsection{The stochastic trace approximation}\label{sec:stochastic trace}

Even if the Chebyshev expansion makes our scaling better $O(N^2)$, it is not enough to achieve the linear scaling promised by KPM. For that, it is also necessary to approximate the trace with the stochastic phase approximation. Let us call $N$ the dimension of the system and $N_{rv}$ the number of random vectors. We choose a set of random numbers $\varphi_{\alpha,n}\in[0,2\pi)$ for $\alpha\in [0,N)$ and $n\in[0,N_{rv})$.

We can prepare a state of the following shape, in which we use our random complex phase for each of the elements of the basis.
\begin{equation}\label{eq: random phase state definition}
    |\psi_{n}\rangle=\sum_{\alpha=0}^N\frac{e^{i\varphi_{\alpha,n}}}{\sqrt{N}}|\varphi_\alpha\rangle,
\end{equation}
where we have called $|\varphi_\alpha\rangle$ each vector of the basis.

If we take the expected value of an operator in this state, we have
\begin{equation}
\begin{array}{c}
    \frac{1}{N_{rv}}\sum_{n=0}^{N_{rv}}\langle \psi_n |A| \psi_n\rangle=\frac{1}{N_{rv}N}\sum_{n=0}^{N_{rv}}\sum_{\alpha,\beta}A_{\alpha\beta}e^{i(\varphi_{\alpha,n}-\varphi_{\beta,n})}=\\\\
     =\frac{1}{N}Tr\left[A\right]+\frac{1}{N_{rv}N}\sum_{n=0}^{N_{rv}}\sum_{\alpha\neq\beta}A_{\alpha\beta}e^{i(\varphi_{\alpha,n}-\varphi_{\beta,n})}.
\end{array}
\end{equation}

If we assume $A$ to be a dense operator, the second term is a sum over a random distribution that averages to zero (because of the complex phases)  where the last term decays as $\Delta A \propto \frac{1}{\sqrt{NN_{rv}}}$. This is especially easy to see if we approximate $A_{ij}$ to be uniformly distributed, where the central limit theorem is applicable.

In practice, the number of random vectors needed to compute the trace is relatively small $N_{rv}\ll N$. This takes the method from $\mathcal O(N^2)$ to $\mathcal O(N)$.


\subsection{Spectral calculations} \label{Section: Spectral calculations}

With that, we have all the pieces that we need to do our simulations.
We can divide the simulations we are going to face during this thesis into \textbf{spectral} and \textbf{time-evolution} calculations.

The \textbf{spectral} calculations are those of properties that depend only on the density of states. In particular, we have properties with shape 
\begin{equation}
    \label{spectral operator}
    \mathrm{Tr}\left[A \delta (H-E) B\right],
\end{equation}
being $A$ and $B$ sparse operators. Note that spectral calculations reduce to a density of states for the case $A=B=\mathbb I$ being $\mathbb I$ the identity operator.

To compute them, we will follow the following recursive algorithm:

\begin{enumerate}

    \item Generate a random state $|\phi\rangle$ as in (\ref{eq: random phase state definition}). With that, we can express the trace as $\mathrm{Tr}\left[A \delta (H-E) B\right]\approx \langle\phi|A\delta(H-E)B|\phi\rangle$.

    \item Multiply the operators $A,B$, having $|\phi_A\rangle=A|\phi\rangle$ and $|\phi_B\rangle=B|\phi\rangle$.
    
    \item Compute the reduced Hamiltonian using (\ref{eq: reduced hamiltonian}).

    \item Compute the bracket with the first two moments $T_m(\tilde H)$ of the expansion (\ref{eq: delta kernel function})
    \begin{equation}
    \begin{array}{c}
        \langle T_0\rangle_{AB}=\langle\phi_A|\phi_B\rangle\\
        \langle T_1\rangle_{AB}=\langle\phi_A|\tilde H|\phi_B\rangle.\\
    \end{array}
    \end{equation}
    We will store the values of $|\phi_{B,m}\rangle=T_m|\phi_B\rangle$.

    \item Iteratively solve moments up to $M$ with
    \begin{equation}
        \langle T_m\rangle_{AB}=\langle \phi_A|\left[2\tilde H|\phi_{B,m-1}\rangle-|\phi_{B,m-2}\rangle\right].
    \end{equation}

    \item For each target energy $E$, we find its reduced version as $\tilde E=2(E-E_{avg})/\Delta E$. 
    \item With the value $\tilde E$ we make the sum in (\ref{eq: delta kernel function}) for $\langle A\delta(E-H)B\rangle=\frac{\Delta E}{2N}\sum_m \Gamma_m g_m \langle T_m\rangle_{AB}$.

    \item Average several random phase iterations repeating from step 2.
\end{enumerate}

This algorithm results in a linear scaling complexity algorithm to compute the spectral properties of materials, allowing us to cover our million-atoms-sized systems. 

An interesting remark to make is that the role of the random trace approximation here serves as a tool to efficiently compute the trace, but in practice, we will often want to reduce the space where we will compute that trace. This will happen in the case of local or projected densities of states. We can allow $\ket{\phi}$ to be non-zero only in a finite set of orbitals without loss of generality and repeat this algorithm. If we do so, a larger sample of random phases is going to be needed to allow proper convergence of the states.

In the specific case of the local densities of states we can reach even larger precision as no random phase will be needed to be involved. In particular, we can use a single $|\varphi_i\rangle$ and repeat this process for $\ket{\phi}=\ket{\varphi_i}$:
\begin{equation}\label{eq:LDOS}
    LDoS(E,i)=\bra{\varphi_i} \delta (H-E) \ket{\varphi_i}.
\end{equation}
This quantity is going to be useful to understand the complex real-space projections of the electronic states of moir\'e and superperiodic systems that we advanced in previous sections.

\subsection{Time-evolution calculations}\label{Sec:Time evolution}

We will also need to compute the evolution of operators with time. These are calculations on operators that depend on time in the form
\begin{equation}\label{eq:lsquant:time-evolution}
    \mathrm{Tr}\left[\hat U(t)\delta(H-E)A\hat U(t)\right].
\end{equation}
If we think in the Schr\"odinger formalism of quantum mechanics, where instead of time evolving the operators we time evolve our wave functions, we see a similar procedure as in Sec.\ \ref{Section: Spectral calculations} can be done, where we replace $|\phi\rangle$ by $|\phi(t)\rangle$. Only a method to efficiently compute the time evolution of the wave function is going to be needed.

For that, we will allow our wave function to evolve in a determined number of time steps by choosing a discretization of time in a sufficiently small grid $\Delta t$. Here, a typical value for our calculations is $\Delta t = 1\ \mathrm{fs}$. With that, for each time in our grid $t$, we can compute the next time step $t+\Delta t$ as  $|\phi(t+\Delta t)\rangle=\hat U(\Delta t)|\phi(t)\rangle$.

It remains to understand the nature of $U(t)$. The evolution operator in quantum mechanics can also be treated as a dense function, and thus it will be convenient for our formalism to expand it onto the Chebyshev domain. We can do
\begin{equation}\label{eq: Expansion of time evolution}
\begin{array}{c}
    \hat U(\Delta t)=\sum_{m}\Upsilon_{m}(\Delta t)T_m(\tilde H)  \\\\
     \Upsilon_{m}=(2-\delta_{m0})(-i)^{m}J_{m}(\frac{\Delta E\Delta t}{2\hbar}),
\end{array}
\end{equation}
where $J_m(x)$ are the Bessel functions of the first kind. In practice, for the computation of $\ket{\phi(t+\Delta t)}$, we will take the sum until the absolute value of $\Upsilon_m$ is smaller than some tolerance, making the higher order contributions to $\hat U(\Delta t)$ negligible. For the calculations in this thesis we have always chosen this tolerance as $10^{-15}$.

The procedure then follows in a similar way to that in Sec.\ \ref{Section: Spectral calculations} where, given a wave function $|\phi(t)\rangle$, we compute $|\phi(t+\Delta t)\rangle$ as:

\begin{enumerate}

    \item We compute the reduced Hamiltonian from equation (\ref{eq: reduced hamiltonian}).
    \item We compute (similarly to step 4 of Sec.\ \ref{Section: Spectral calculations})
    \begin{equation}
    \begin{array}{c}
        |\phi_0\rangle=|\phi\rangle\\
        |\phi_1\rangle=\tilde H|\phi\rangle.\\
    \end{array}
    \end{equation}
    \item We iteratively solve the coefficients (Step 4 of Sec.\ \ref{Section: Spectral calculations})
    \begin{equation}
        |\phi_{m'}\rangle=2\tilde H|\phi_{m'-1}\rangle-|\phi_{m-2}\rangle.
    \end{equation}

    \item We make the sum in (\ref{eq: Expansion of time evolution}) with
    $|\phi(\Delta t)\rangle=\sum_{m'}\Upsilon_{m'}(\Delta t)|\phi_{m'}\rangle$.

\end{enumerate}

A combination of this procedure to compute $|\phi(t)\rangle$ at each time and the procedure of the previous section to perform the explicit calculation $\bra{\phi(t)}\delta(H-E)A\ket{\phi(t)}$ will be enough to provide us the time evolution of our operators:

\subsection{Mean Squared Displacement calculations}\label{sec:Mean Squared Displacement-formulas}

As we showed in Sec.\ \ref{Section: quantum transport}, it proves useful for us to compute the mean squared displacement as a function of time while targeting the transport properties. For that, we can rewrite equation (\ref{eq:msd}) as 
\begin{equation}\label{eq:MSD LSQUANT}
    \rho (E) \Delta X ^2 = \mathrm{Tr}\left[  \left[U(t),X\right]^\dagger \delta (H-E) \left[U(t),X\right]\right],
\end{equation}
which provides us once again with the opportunity to use our recursive algorithm of Sec.\ \ref{Section: Spectral calculations} for $\bra{\phi_{msd}(t)}\delta(H-E)\ket{\phi_{msd}(t)}$ where we have defined $\ket{\phi_{msd}(t)}=\left[U(t),X\right]\ket{\phi_{msd}}$.

To compute this $\ket{\phi_{msd}(t)}$, we can apply the chain rule from its definition. For a finite time step $\Delta t$ we have
\begin{equation}\label{eq: chain rule}
    \left[\hat U(t+\Delta t),X\right]=\hat U(\Delta t )\left[\hat X,\hat U(t)\right]+\left[\hat X,\hat U(\Delta t)\right]\hat U(t).
\end{equation}
We divide the computation of the solution in two terms of the sum. To compute the first term, it suffices to remember that $\left[\hat X,\hat U(t)\right]\ket{\phi}=\ket{\phi_{msd}}$, so this term represents the action of the time-evolution operator on our $\ket{\phi_{msd}}$ and can be computed with the algorithm described in Sec.\ \ref{Sec:Time evolution} replacing $\ket{\phi}$ by $\ket{\phi_{msd}}$.

The right-hand term represents the time evolution of the commutator, to solve it we will apply \eqref{eq: Expansion of time evolution} to get
\begin{equation}
    \left[\hat X, U(\Delta t)\right]=\sum\Upsilon_m(\Delta t)\left[\hat X,\hat T_m(\tilde H)\right],
\end{equation}
where applying the recursion relation in (\ref{eq:chebyhev polinomia definition}) we get its recursive version.
\begin{equation}\label{eq: momenta conmuter}
    \left[\hat X,T_{m}(\tilde H)\right]= 2\tilde H\left[\hat X,T_{m-1}(\tilde H)\right]+2\left[\hat X,\tilde H\right]T_{m-1}(\tilde H)-\left[\hat X,T_{m-2}(\tilde H)\right].
\end{equation}
This defines a unique way of evolving our wave function under the effect of the commutator.

We recall that $\hat U(0)$ is the identity and $[\hat X, \hat H]=i\hbar V$ so the procedure to compute the right-hand side of equation (\ref{eq:MSD LSQUANT}) is:

\begin{enumerate}
    \item Generate the random phase $|\phi\rangle$ to approximate the trace.

    \item Choose $[\hat X,\hat U(t)]|\phi\rangle=0$, $|\phi(t)\rangle=|\phi\rangle$ for $t=0$.

    \item In each time step, compute $\hat U(\Delta t)|\phi\rangle$ and $\hat U(\Delta t)[\hat X,\hat U(t)]|\phi\rangle$. Applying the method of Sec.\ \ref{Sec:Time evolution}. We will store the current $|\phi(t+\Delta t)\rangle$ for the next iteration.

    \item We make use of equation (\ref{eq: momenta conmuter}) and compute $[\hat H,\hat U(\Delta t)]|\phi(t)\rangle$ following again the same procedure as in \ref{Sec:Time evolution}, but where now we make recursion in step 3 to be
    \begin{equation}
    \begin{array}{l}
        ([\hat X,\hat U (\Delta t)]|\phi\rangle)_{m}=\\\\\qquad2\hat H\left([\hat X,\hat U (\Delta t)]|\phi\rangle\right)_{m-1}+2V_x|\phi_{m-1}\rangle-\left([\hat X,\hat U (\Delta t)]|\phi\rangle\right)_{m-2}.\\\\
    \end{array}
    \end{equation}

    Note that this expansion will also depend on the expansion of the time-evolution operator $U(\Delta t)$, computed during step 4, so it will prove very convenient to do these two steps at the same time.

    \item We get $\left[\hat X,\hat U(\Delta t)\right]|\phi(t)\rangle$ using equation (\ref{eq: chain rule}).

    \item With these new wave functions, we do points 4-5 of Sec.\ \ref{Section: Spectral calculations} for $|\phi_A\rangle=|\phi_B\rangle=\left[\hat U(t),\hat X\right]|\phi\rangle$, and store the $\langle T_m (t)\rangle$ for each time step.

    \item We can now choose an energy and do steps 6-7 of Sec.\ \ \ref{Sec:Time evolution} for each time step, thus having the mean squared displacement at each $t$.

    \item We average this procedure for several choices of the random phase.

\end{enumerate}


\section{Self-consistent algorithms}

\subsection{Electrostatic potential calculations}\label{Sec: Poisson}

To achieve superperiodic potentials without the need of building moir\'e systems, an option is to mimic the superperiodicities making use of artificial external electric fields. 
Such fields will have an effect on the Hamiltonian of the material via the external field in \ref{eq: born oppenhaimer} with 
\begin{equation}
    V_{ext}=\sum_i V(\mathbf x_i)n_i,
\end{equation}
where $\mathbf x_i$ is the position of the $i$-th lattice site and $n_i$ is the occupancy operator and can be computed as 

\begin{equation}
    \langle n_i\rangle=\int_{-\infty}^{E_F} f(E')\rho_i(E')\mathrm d E',
\end{equation},
where $\rho_i$ is the density of states projected at each lattice site and defined as 
\begin{equation}
    \rho_i(E)=\mathrm{Tr}\left[P_i\delta(\hat H-E)P_i\right]
\end{equation}
and with $P_i$ the projection operator in the lattice site $i$. Note that one of the $P_i$ in the previous equation could be removed, and we keep them both for clarity and for connection with the actual calculation.

This $V(\mathbf x)$ can be obtained by the Poisson equation

\begin{equation}\label{eq: The Poisson equation}
    \nabla \left(\epsilon\nabla V(\mathbf{r})\right)=en(\mathbf{r}),
\end{equation}
where $e$ is the charge of the proton and $n(x)$ is the number of electrons at each point of the space, and in particular $n(\mathbf x_i)=\langle n_i\rangle$.

At this point, it is easy to see that the calculation of that potential function is non trivial. As we add some potential term $V_{ext}$ to our system, the occupancies at each lattice site will change. At the same time, this change in occupancies would lead to a different potential.

For that purpose, we need to find a self-consistent solution to the full Schr\"odinger-Poisson system of equations \cite{Ferry}. We will do as follows:

\begin{enumerate}
    \item discretize the space, so that we have a collection of points $x_{nlm}$ for the whole system.
    \item Generate a random distribution of the occupancies $\langle n_i\rangle$ in the position of our material.
    \item Use them to solve the Poisson equation $\nabla V=-en(x)$ with a finite-difference method.
    \item Use the new values of $V$ in the region of the material to compute the new set of occupancies $\langle \tilde n_{i}\rangle$ that solve \eqref{eq: The Poisson equation} (See appendix \ \ref{appendix poisson_solution}).
    \item repeat steps 3-4 until the difference in potentials for two consecutive steps is smaller than a tolerance.
\end{enumerate}

In addition, in Sec.\ \ref{sec:qdot_array}, we will make use of this procedure and give more details on how to perform steps 3 and 4.

\subsection{Electron-electron interaction}\label{sec: Hubbard model}

As mentioned previously, electron-electron interaction is fundamental for the correct understanding of the exotic properties of twisted moir\'e systems, especially in the case of MATBLG. The methods explained so far can  reproduce with high precision the transport properties of systems under the single electron approximation.

This interaction is mediated by the Coulomb interaction \cite{fetter-walleka}, which can be expressed as
\begin{equation}
    \sum_{i\neq j}\frac{1}{4\pi \varepsilon_0 (r_i-r_j)}n_in_j,
\end{equation}
where $n_i,n_j$ are the electron number operators $n_i=c_i^\dagger c_i$ we can see that this equation is beyond the one electron approach. Let us take the simplest form of these correlations. We can roughly approximate these correlations to assume that only on-site interactions matter. In doing so, we achieved the known Hubbard-term \cite{Marder}
\begin{equation}
    H_{Hubbard}=\sum_i U n_{i\uparrow}n_{i\downarrow}.
\end{equation}

To approach these many-body calculations, we will rely on the so-called mean-field calculations \cite{Claveau_2014}. For that, let us rewrite the value of those $n_{i\sigma}$ as $n_{i\sigma}=\langle n_i\rangle + \Delta n_i$. Now we can express the product $n_{i\sigma}n_{j\sigma}$ as
\begin{equation}
\begin{array}{c}
    n_{i\uparrow} n_{j\downarrow}=\langle n_{i\uparrow}\rangle \langle n_{j\downarrow}\rangle+\Delta n_{i\uparrow}\langle n_{j\downarrow}\rangle+\Delta n_{j\downarrow}\langle n_{i\uparrow}\rangle +\Delta n_{i\uparrow}\Delta n_{j\downarrow}=\\
    \Delta n_{i\uparrow}\Delta n_{j\downarrow}+n_{i\uparrow}\langle n_{j\downarrow}\rangle + n_{j\downarrow}\langle n_{i\uparrow}\rangle-\langle n_{i\uparrow}\rangle\langle n_{j\downarrow}\rangle.
\end{array}
\end{equation}

We call the mean-field approximation to the canceling of the quadratic term in the fluctuations so that we can express the Coulomb term as a single electron Hamiltonian.   

To compute the value of this Hamiltonian, we will need the average occupations at each site; this leaves us with a self-consistent problem similar to the one explained in the previous section. This self-consistent problem is frequently solved with a Hartree-Fock convergence of the expected occupations for each of the spin channels \cite{Claveau_2014}.

The procedure is as follows:

\begin{enumerate}
    \item We start with a random distribution of $\langle n_{i\uparrow} \rangle$ and $\langle n_{i\downarrow}\rangle$ that matches the uniformity criterion $\sum_{i\sigma}\langle n_{i\sigma}\rangle=N_{orb}$ with $N_{orb}$ the number of orbitals in our tight-binding model $N_{orb} = 2\times number\ of\ sites$.

    \item From there we generate our Hamiltonian 
    \begin{equation}
        H=H_{TB}+\sum_{i}\left[n_{i\uparrow}\langle n_{i\downarrow}\rangle+n_{i\downarrow}\langle n_{i\uparrow}\rangle -\langle n_{i\uparrow}\rangle\langle n_{i\downarrow}\rangle\right].
    \end{equation}

    \item We make a computation of the local density of states at each orbital using the methods described for equation\ \eqref{eq:LDOS}.

    \item When we have all the local densities of states computed, we may sum them to obtain the full DoS and compute the Fermi energy as the energy that matches the condition
    \begin{equation}
        N_{orb}/2=\int_{-\infty}^{E_F}\sum_iLDoS(E,i,\sigma)\;\mathrm d  E.
    \end{equation}

    \item We compute a new set of expected values for the occupations at each site as 
    \begin{equation}
        \langle \tilde n_{i\sigma}\rangle = \int_{-\infty}^{E_F} LDoS(E,i,\sigma) \;\mathrm d E.
    \end{equation}

    \item We compute our convergence criteria as the maximum difference between two consecutive computed occupations $\alpha=\text{Max}(|\langle \tilde n_{i\sigma}\rangle-\langle n_{i\sigma}\rangle|)$. 
    
    \item In case $\alpha$ is smaller than our tolerance, we take the set $\langle \tilde n_{i\sigma}\rangle$ as the expected value of our occupations. Otherwise, we make $\langle \tilde n_{i\sigma}\rangle=\langle n_{i\sigma}\rangle$ and we return to step 2.
\end{enumerate}

\begin{figure}
    \centering
    \includegraphics[width=\linewidth]{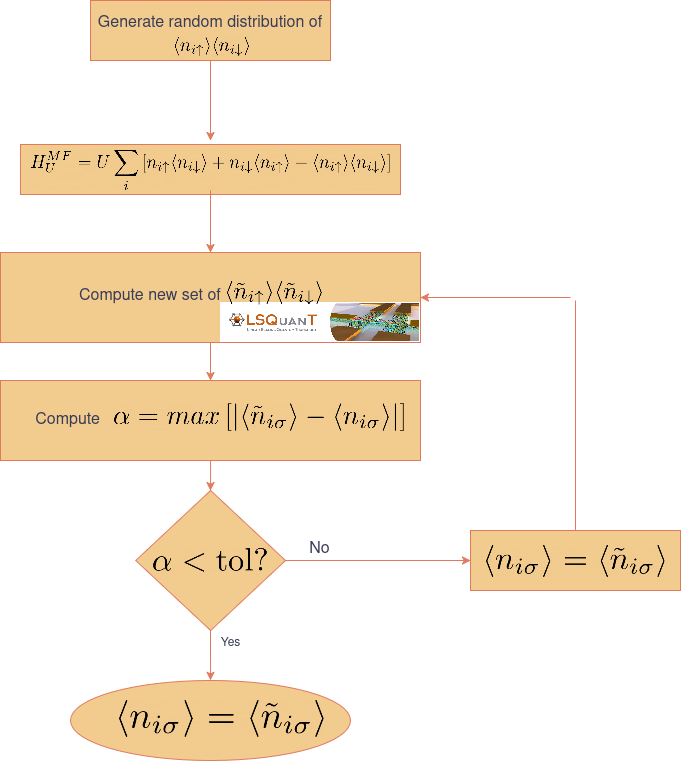}
    \caption{Scheme of the workflow for the mean-field Hartree-Fock calculation.}
    \label{fig:hubbard-workflow}
\end{figure}

In Fig.\ \ref{fig:hubbard-workflow} we can see a scheme of this workflow explicitly.

\subsection{The linear mixing}

Both schemes described in Sec.\ \ref{Sec: Poisson} and Sec.\ \ref{sec: Hubbard model} are self-consistent: starting from a trial set of initial occupancies, one computes the corresponding electronic response and feeds it back into the Hamiltonian, iterating until a fixed point is reached. In practice, a naive fixed-point iteration may oscillate or even diverge because the output of one iteration can over-correct the input of the next one. A standard stabilization strategy is to replace the complete update with a mixed update.

Here, instead of updating the occupancies in step 4 of Sec. \ref{Sec: Poisson} and step 5 of Sec. \ref{sec: Hubbard model} we introduce a mixing between the occupancies of the last two iterations \cite{ZERNER1979550}. We denote $\langle \tilde n_i\rangle$ as the theoretical new occupations after the self-consistent step. We can compute the updated occupancies as:

\begin{equation}\label{eq: linear mixing}
    \langle n_{i,new}\rangle=\beta_{mix}\langle\tilde n_i\rangle+\left(1-\beta_{mix}\right)\langle n_i\rangle ,
\end{equation}
with $\beta_{mix}$ the mixing parameter in the range $\beta_{mix}\in [0,1]$, chosen to guarantee sufficiently slow convergence.

In practice, we have observed that extremely small values of $\beta_{mix}$ (extremely slow convergences) are needed to avoid instabilities for the Schr\"odinger-Poisson convergence (up to $\beta_{mix}=0.005$) where the potential response to small differences in occupations may be amplified. Meanwhile, our results for the Hartree-Fock mean field are mostly independent of our choice of $\beta_{mix}$ where we have considered a wide variety of values in the range $\beta_{mix}\in[0.1,1]$.

\chapter{superperiodic conditions}\label{chapter:superperiodicities}

This chapter uses controlled model systems to isolate the transport effects that arise purely from superperiodic and quasiperiodic structure. We begin with the simplest case of a one-dimensional linear chain under externally imposed modulations, which serves to identify how an additional long-wavelength length scale introduces characteristic crossovers in wave-packet dynamics.

We then extend the analysis to monolayer graphene subject to electrostatic superlattice potentials of different symmetries, from quasi-one-dimensional to fully two-dimensional periodic patterns, and finally to a quasicrystalline modulation that lacks translational periodicity. In this way, the chapter builds a set of transport fingerprints that distinguish ordinary periodic propagation, superlattice-induced velocity renormalization, and genuinely anomalous quasiperiodic dynamics.

The chapter closes by considering a more realistic gate-defined platform based on Bernal bilayer graphene, where patterned electrostatics generate quantum-dot arrays and artificial moir\'e-like landscapes. Together, these examples provide a clean conceptual reference for the more realistic twisted multilayer systems studied in the following chapter.

\section{Linear chain under superperiodic conditions}\label{sec:linear-chain-superp}

We begin this section by analyzing the behavior of a linear chain under superperiodic conditions. We can understand this as the simplest case of a superperiodic system driven by differences in the tight-binding parameter \cite{Sun_2025} or the onsite energy \cite{CRUZ2023115779}.

We start by building a basic model for a linear chain
\begin{equation}
    H_{TB}=\sum_i \gamma_0 \left(c_i^\dagger c_{i+1}+c_{i-1}^\dagger c_i\right),
\end{equation}
with $c_i^\dagger,c_i$ the creation and annihilation operators at the sites $x=i\Delta x$ with $\Delta x$ the distance between points in the linear chain. Furthermore, we will take $\Delta x=1$ for simplicity. Here $\gamma_0$ is the tight-binding parameter.
This case is particularly easy to analyze, since the Hamiltonian for a defined $k$ point is 
\begin{equation}\label{eq:linear chain}
    H_k= 2\gamma_0\cos (k)c_k^\dagger c_k.
\end{equation}

Then, each wave in this Hamiltonian with a well-defined $k$ will have a well-defined energy. The inverse is not true, where within the first Brillouin zone, any $k$ will have a matching energy at $-k$. 

The choice of this model will simplify the solution of our problem; here, to grant real-space control, we will perform our simulations in real space. For that, we will choose a length $L$ of the chain that we want to model and run our simulations for that chain length, stopping before the wave function reaches the size of the simulation box. For the calculations presented in this thesis, we have chosen $L=2084$ to ensure an optimal balance between computational efficiency and precision. There, one can choose a version of equation (\ref{eq:msd}) for an initial state $|\psi\rangle$. Therefore,

\begin{equation}\label{eq:msd linear chain}
    \Delta X^2(t)=\langle\psi|\delta(H-E)\left(X(t)-X\right)^2|\psi\rangle.
\end{equation}
This equation is analytically solvable with a finite-difference method. We explain the solution for a given initial state in Appendix\ \ref{appendix poisson_solution}. 

This approach imposes restrictions on the initial state: as our simulation length is bounded in real space, so must the wave function. For that purpose, we will choose a Gaussian wave packet centered on a specific site of the lattice $j$. There, our chosen wave function is 
\begin{equation}\label{eq:gaussian-wave-packet}
    |\psi_j\rangle=\sum_{i}\left[\cos\left({k_0\Delta x_{ij}\lambda}\right)-i\sin\left(k_0\Delta x_{ij}\lambda\right)\right]e^{-\left(\frac{\Delta x_{ij}}{2\sigma}\right)^2}c_i^\dagger c_i
\end{equation}
with 
$$k_0=\frac{2\pi}{L}$$ the minimum choice of $k$ that grants the wave function will match the periodic boundary conditions of the simulation length. Here 
$\Delta x_{ij}$ is the distance from the lattice site $i$ to the center of the pulse $j$; and $\lambda,\sigma$ parameters that we freely choose corresponding to the wavenumber in units of $L^{-1}$ and the standard deviation of the wave packet, respectively.

This initial wave function has a trackable position in real space at the expense of not having  a well-defined energy, however, the average value of the energy is well known
\begin{equation}
    \langle E \rangle = \langle 2\gamma_0\cos(k)\rangle=2\gamma_0 \cos (\langle k\rangle)e^{-\frac{1}{8\sigma^2}}.
\end{equation}
where $\langle k \rangle=k_0$ represents the central wave number $k$ of the wave packet. In this way, we can achieve energy resolved properties similar to the ones explained in Chapter\ \ref{chapter:computational}.
In our case, as our $\sigma\approx10$ in units of the lattice constant, the exponential term becomes negligible, and we can make $\langle E\rangle\approx E(\langle k \rangle)$.

It is also noticeable for our calculations that the Fermi velocity, defined from equation (\ref{eq:fermi velocity}), if we follow a similar trick, can be expressed as 

\begin{equation}\label{eq: fermi-velocity-linear-chain}
\langle v(k)\rangle\approx\frac{2\gamma_0}{\hbar}\sin(\langle k\rangle).
\end{equation}
With that, we can treat this average value of $k$ for our pulse, similarly to the way we treat a plane wave evolving with the wave number $k$. For the remaining parts of the section, we will use $k$ to refer both to the specific value of $k$ and to the average $\langle k\rangle$ whenever no distinction is needed. We used this model against an Anderson potential in Sec.\ \ref{sec:Anderson-scaling-theory-of-localization} to illustrate the effects of superimposed disorder on a linear chain. In this section, we proceed to use it against super- and quasiperiodicities.

\subsection{Superperiodic case}\label{sec:superperiodic_cas}
In this section we will build a first toy model to approach superperiodicities in the linear chain. To do so, we will superimpose our linear chain to a superperiodic potential as
\begin{equation}
    V(x)=\left\lbrace
    \begin{array}{ccc}
         V/2 &&x\in[x_n,x_n+d]  \\
         -V/2 && \text{otherwise} 
    \end{array}
    \right.
\end{equation}
where $x_n$ is a collection of points in the space chosen at distances $x_n=2dn$. We can see that the linear chain defined in (\ref{eq:linear chain}) is periodic under transformations $x'=x+2d$, so we will call $2d$ the superperiodic length of our linear chain.
Note that in the equipotential sections, the group velocity can still be computed from equation (\ref{eq:fermi velocity}) and the energy can be represented by
\begin{equation}
    E(k)=2\gamma_0\cos(k)\pm \frac{V}{2}.
\end{equation}

\fullpagefigure[1]{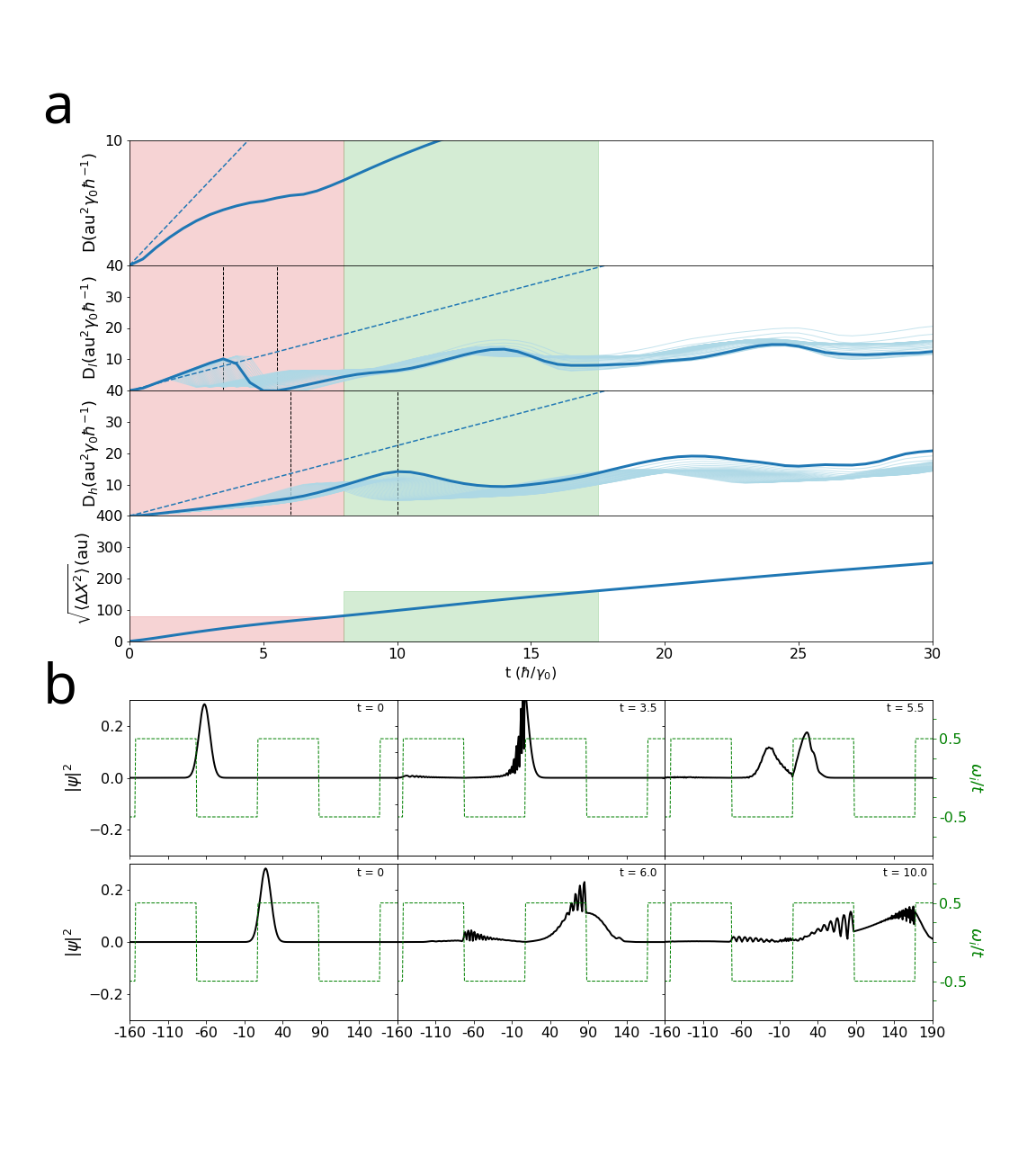}
{Transport in a superperiodic linear chain. a: Top panel represents the average diffusion coefficient. Second (third) panels represent the diffusion for initial wave packets in the low (high) potential regions, the darker blue curve represents a specific choice of starting position that will afterwards be represented in panel b. Bottom panel represents the electron spreading. The red and green areas in the figure represent the regions of time before the spreading reaches $d$ and $2d$ respectively. 
    b: Real-space probability density of the wave functions for initial conditions low (top row) and high (bottom row) energy regions for the dark colored lines in second and third panels of a. The left panels correspond to $t=0$ and the middle and right to the times marked as vertical dashed lines in second and third panels of figure a.}{fig:superperiodic_linear_chain}

    


There is yet another difficulty to cover. As real space is not homogeneous, we will have different results depending on our choice of the initial wave packet, in particular, we are varying the center of the wave packet $x_j$. To solve this issue, we will average several different values of $x_j$. In particular, we will average over different centers along the entire superperiodicity. For coherence between the different calculations, we will choose the energies to be the same for all the calculations. This also has the advantage of being comparable to the usual calculations we will perform making use of the methods described in Sec.\ \ref{sec:Mean Squared Displacement-formulas}. To do so, we will choose our initial value of $E$ , making 
\begin{equation}
    k=\arccos\left(\frac{E\mp\frac{V}{2}}{2\gamma_0}\right).
\end{equation}
In Fig.\ \ref{fig:superperiodic_linear_chain} we can see the short-time evolution of this Gaussian pulse traveling a linear chain for a certain value of $k$, this figure is the superperiodic version of \ref{fig:anderson_scaling} for the disordered linear chain. 

The top panel of Fig.\ \ref{fig:superperiodic_linear_chain}a represents the average diffusion coefficient generated for all possible choices of $j$ within a superperiodicity. There, the red and green shadowed regions are for the regimes below the average electronic spread (bottom panel of Fig.\ \ref{fig:superperiodic_linear_chain}a)  reaching $d$ and $2d$, respectively. We can see that the diffusion coefficient exhibits a renormalization of the average velocity at the times when the electronic spread has values between $d/2$ and $d$. Later in this text, we will show that this is shared by the quasi-1D case superperiodicity in graphene. After that point, the superperiodic system evolves linearly with a Fermi velocity different from that of the pristine system, as described in Sec. \ref{sec: ballistic regime}. We will call this new velocity the Fermi velocity of the superperiodic lattice.

In the second and third panels of Fig.\ \ref{fig:superperiodic_linear_chain} a, we can see the diffusion coefficient for a Gaussian wave packet with starting position in the low and high-energy regions, respectively. We can see that the wave packet with initial position in the low-energy region drastically reduces its diffusion coefficient at distances close to $d/2$. The reason for this can be found in the top panels of Fig.\ \ref{fig:superperiodic_linear_chain}b, where we can see the probability density of the wave packet at the starting time (left), the time where the renormalization starts (middle), and a time after the renormalization (right). Here we can clearly see that the source of this reduction of the diffusion coefficient is the scattering in the low-to-high-energy regions, the first time being the arrival of the packet to the barrier, and the second time, a time after the scattering has ended.

Similarly for the high-energy region, we can see in the third panel of Fig.\ \ref{fig:superperiodic_linear_chain}a that the velocity first increases before reaching $d$ and then reduces between $d$ and $2d$. In the bottom panels of Fig.\ \ref{fig:superperiodic_linear_chain}b we can see the state of this packet at the initial time (left), at the time of the increase (center) and at the time of the drop (right). We can clearly see that the reason for this speed increase is the scattering in the high-to-low energy region, where a drop in the potential energy is transformed into an increment of the kinetic energy. The drop in the diffusion coefficient occurs once again in the scattering from the low-to-high energy regions. 

To conclude the analysis of the superperiodic linear chain, we emphasize that the observed velocity renormalizations arise solely from the presence of a finite superlattice length and the associated scattering at the superperiodic interfaces. These effects establish clear time and length scales that govern the deviation from ideal ballistic propagation and constitute the simplest example of how externally imposed long-range structure modifies wave-packet dynamics.

With this minimal model as reference, we now extend the analysis to quasiperiodic modulations, where long-range order persists but no finite supercell exists. This allows us to contrast genuine periodic superlattices with deterministic aperiodic structures and to identify which dynamical features survive when periodicity is removed.

\subsection{Quasiperiodic case}\label{Sec: linear_chain_quasi}

Having introduced the superperiodic case, where the external modulation defines an enlarged but strictly periodic unit cell, we now turn to quasiperiodic structures. In this work, quasiperiodicity will arise from an electrostatic potential arranged according to a Fibonacci sequence. Unlike a superperiodic modulation, a Fibonacci pattern does not repeat at any finite length, yet it remains fully deterministic and exhibits well-defined long-range order. In the context of a linear chain, this quasiperiodic modulation will allow us to explore how self-similar order and the absence of periodicity affect wave-packet propagation.

The  potential for the Fibonacci chain is defined as 
\begin{equation}\label{eq:fibonacci potential}
    V(x_i)=V\cdot\mathcal F_\infty(i) ,
\end{equation}
being $\mathcal F_\infty(i)$ the $i$-th position of $\mathcal F_\infty$ defined by the inflation rule
\begin{equation}
    \begin{array}{ccc}
         \mathcal F_0&=&\{-1/2\}  \\
         \mathcal F_1&=&\{ 1/2\}  \\
         \mathcal F_n&=&\{F_{n-2},\mathcal F_{n-1}\}.
    \end{array}
\end{equation}
 
\begin{figure}
    \centering
    \includegraphics[width=0.9\linewidth]{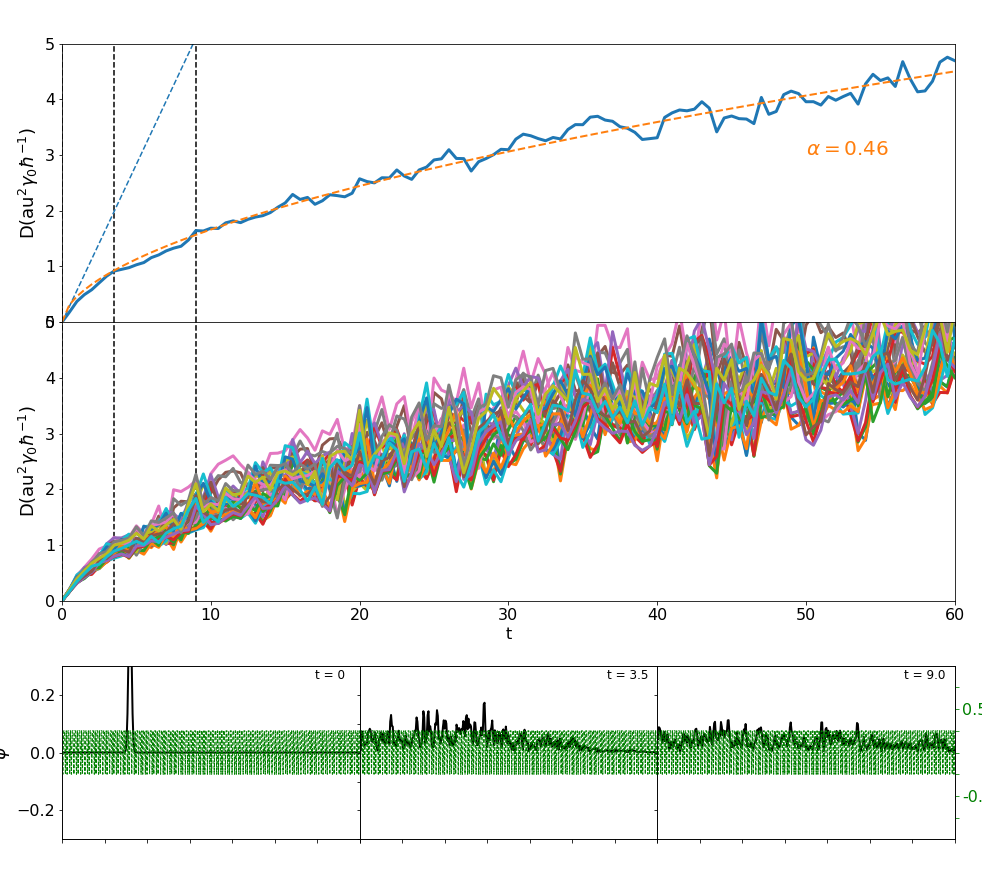}
    \caption{Evolution of a Gaussian wave packet as a function of time for a Fibonacci linear chain. Top panel represents the average diffusion coefficient (solid blue curve). The blue dashed lines in the top panel represent the ideal diffusion coefficient for the linear chain in the ballistic case, in the blue solid line we observe the transport when the Fibonacci potential is superimposed. Orange dashed line represents the fit to a function $D=at^\alpha$ (orange) (see text.). In the middle panel we observe each behavior of the diffusion coefficient for different wave packets at different starting centers in the Fibonacci chain. We can see all iterations behave subballistically. In the three bottom panels, we observe the short-time behavior of the wave packet as transport adapts to the quasiperiodicity of the chain.}
    \label{fig:qc_linear_chain}
\end{figure}

This way, following the inflations, we have $\mathcal F_2=\{-1/2,1/2\}$ and $\mathcal F_3=\{1/2,-1/2,1/2\}$. This chain is a well studied case of quasicrystal \cite{StephanQC}. In Fig. \ref{fig:qc_linear_chain} we can explicitly see that the transport through this linear chain exhibits a sub-ballistic behavior, where the scattering at different scales induced by the quasicrystalline potential decreases the velocity. Here, the diffusion coefficient evolves as a power law as $D\propto t^\alpha$ with $-1<\alpha <1$. Anomalous wave-packet spreading of this type has been thoroughly documented in quasiperiodic systems, including the one-dimensional Fibonacci chain \cite{Roche1997ConductivityOQ} and quasicrystalline alloys \cite{MAYOU2008209}.
 This behavior can be understood as a direct consequence of the hierarchical structure of the quasiperiodic potential. Each inflation level of the Fibonacci sequence introduces partial reflections, so the wave packet encounters scattering interfaces at all spatial scales. This suppresses both ballistic propagation and exponential localization and gives rise instead to critical wave functions. As a result, the velocity correlations decay algebraically and the diffusion coefficient follows the anomalous power-law behavior. In the top panel of Fig.~\ref{fig:qc_linear_chain} we see explicit confirmation for this anomalous transport behavior where we compare our diffusion coefficient (solid blue line), with a power law $D\propto t^{\alpha}$ (orange line) obtaining an $\alpha$ value of $\alpha=0.46$.  It is important to note here that, by changing the strength of the superimposed potential, this model can span values of $-1<\alpha<1$.

 As we mentioned, this mechanism has been rigorously established in one-dimensional Fibonacci systems \cite{PhysRevLett.76.4372}, but is a general effect of fractality in quasicrystals. During this text, we will review how this effect is present in other quasicrystals with different inflation patterns, and especially in $30^\circ$ twisted bilayer graphene.

\subsection{The sub-ballistic exponent in quasicrystals}

In the literature, the exponent $\alpha$ extracted directly from $D(t)$ is seldom reported; instead, anomalous transport is most commonly characterized by a spreading exponent $\beta_{QC}$, defined  as $\langle\Delta X(t)\rangle\propto t^{2\beta_{QC}}$, so $2\beta=\alpha+1$. Consequently, comparisons with experiments are usually made via how $\beta_{QC}$ (or, equivalently, $\alpha$) manifests indirectly in macroscopic observables such as $\sigma_{\mathrm{dc}}(T)$, $\sigma(\omega)$, or quantities set by a finite time cutoff (inelastic decoherence, finite size, etc.), rather than through $D(t)$ itself.

In practice, it has been found to have an impact on optical conductivity as it decreases the strength of the Drude peak with respect to the standard crystals, achieving plateaus for small values of $\alpha$ and even Drude dips for $\alpha<0$ \cite{PhysRevLett.85.1290}. In addition, the impact of $\beta$ on temperature behavior is often given by the scaling laws $\sigma(T)\propto T^{-p(2\beta-1)}$ \cite{MaciaQC}.

In other 1 and 2 dimensional quasicrystals values for $\beta$ have been predicted to fall in this subballistic and superdiffusive regime $0<\alpha<1$, as is the case of the Penrose tiling ranging from diffusive to ballistic \cite{Actaphystrambly}. Others fall in the sub-diffusive regime $-1<\alpha<0$ \cite{PhysRevB.107.054206}. Some experiments have connected this value of $\alpha$ for inter-metallic alloy quasicrystals with the absence of the Drude peak characteristic of sub-diffusive transport \cite{MAYOU2008209}.

\section{Graphene under superperiodic conditions}
The superperiodic potentials introduced in the previous section for a linear chain can be generalized to a two-dimensional honeycomb lattice. For a monolayer graphene sheet, a periodic (or quasiperiodic) modulation can be imposed by combining several plane waves in the electrostatic potential. The resulting on-site modulation takes the form
\begin{equation}\label{eq:freeform potential}
    V(x,y)=\sum_{n=1}^{N_{per}}\sum_i^N A_n cos (g_n[x_i\ cos\theta_n+y_i\ \sin\theta_n])c_i^\dagger c_i\quad ,
\end{equation}
where $A_n$ is the amplitude of each wave, $g_n$ the wavelength, and $\theta_n$ the angular direction. This kind of potential can be experimentally induced in graphene in proximity to patterned hBN. Nolan Lassaline et al. generated this kind of sinusoidal pattern in graphene using a combination of thermal scanning-probe lithography and reactive-ion-etch pattern transfer \cite{Lassaline2021,Lassaline2025}. 

Depending on the choice of parameters, we can generate different patterns with distinct periodicities (See Fig. 4 in \cite{Lassaline2021} for some examples of these superimposed patterns). We will focus here on three choices of parameters that respect the symmetries of graphene.

\begin{itemize}
    \item With $N_{per}=1$ we can generate a controllable superperiodicity in one direction, this is an example of a quasi-1D superperiodicity, as even if the system is bidimensional, the superperiodicity is only imposed along one of the directions.
    \item With $N_{per}=3$ and $\theta_N=n\frac{\pi}{3}$ we generate a superperiodicity with respect to the symmetries of graphene, this case stands for a full 2D superperiodicity, and we will see that the effects on the transport from the one-electron perspective are similar to those of the bilayer graphene systems.
    \item With $N_{per}=6$ and $\theta_n=\frac{\pi}{6}$ we generate a quasicrystalline potential. This potential generates a 12-fold rotational symmetry with an inflation pattern similar to the structure of the graphene quasicrystal introduced in Sec.\ \ref{sec:quasicrystal intro}.
\end{itemize}

\begin{figure}
    \centering
    \includegraphics[width=\linewidth]{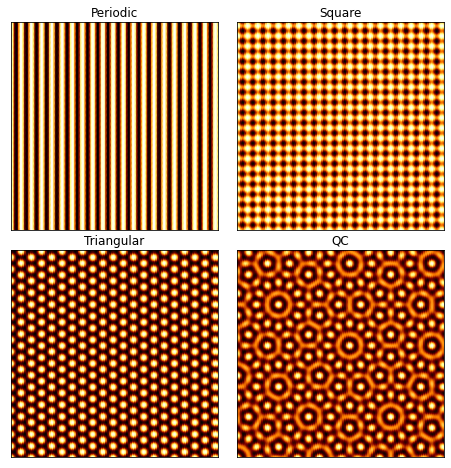}
    \caption{Shape of possible potentials emerging from equation (\ref{eq:freeform potential}) on a graphene lattice. }
    \label{fig:freeform potentials}
\end{figure}

There is also the possibility to study other superperiodicities that respect those of graphene, such as some rectangular lattices that match the unit cell square of graphene. In this text, our emphasis will be on the three described above. However, a similar version for a square potential will be used later to model quantum dots in bilayer graphene.

\subsection{Computational details}\label{sec: computational-details-freeform}

All the results for the 2D systems in this section are performed making use of the KPM techniques described in Sec.\ \ref{Sec: KPM-intro}, where in particular the calculation of spectral operators described in Sec.\ \ref{Section: Spectral calculations} with the choice of $A=B=\mathbb 1$ was used to compute the densities of states. Transport was measured computing the mean-squared-displacement $\left(\langle\Delta X^2\rangle\right)$ using the methods explained in Sec.\ \ref{sec:Mean Squared Displacement-formulas}. From the mean squared displacement, we computed both the electronic spreading $\left(\sqrt{\langle\Delta X ^2\rangle}\right)$ and the diffusion coefficient $\left(D=\partial_t \langle \Delta X^2\rangle/2\right)$ from \eqref{eq:diffusion coefficient definition}. From there, the Fermi velocity is extracted from the $t= [300,600]\ \mathrm{fs}$ region as the square root of the slope of the diffusion coefficient \eqref{eq: diffusivity ballistic} in order to establish the Fermi velocity after superperiodic scattering.

All calculations were performed on system sizes of $\sim 10^7$ lattice sites ($\sim 460\times 500\ \mathrm{nm}$) to allow a proper time evolution of the states. To give sufficient precision in energy $3500$ KPM moments were used in an energy width of $30\ \mathrm{eV}$ granting a broadening of $\sim 27\ \mathrm{meV}$.
 Time-evolution calculations were performed in intervals of $600\ \mathrm{fs}$ where discrete time steps were performed every $1\ \mathrm{fs}$.

\subsection{Superperiodicity in one dimension: the quasi-1D case}\label{Sec:quasi-1D}

We will start studying graphene for $N_{per}=1$. This is the result of superimposing a plane-wave-shaped potential in the x direction onto a graphene layer (top left panel of Fig. \ \ref{fig:freeform potentials}). This case has previously been studied in the literature, presenting anisotropy in transport along the $X$ and $Y$ directions \cite{Li2021,Lassaline2025}. 

\begin{figure}
    \centering
    \includegraphics[width=\linewidth]{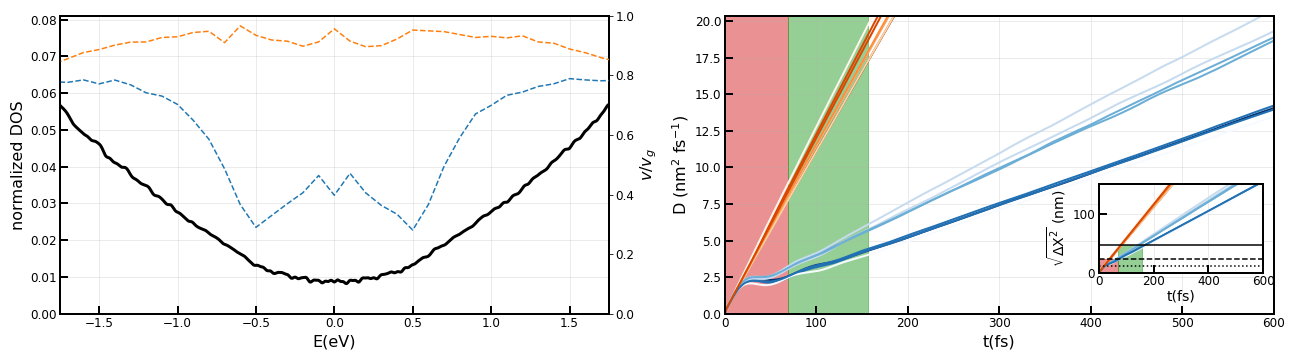}
    \caption{Density of states (left), diffusion coefficient (right) and electronic spreading (inset) in graphene for the quasi-1D superperiodicity obtained for $N=1$. The blue and orange curves represent transport along X and Y directions respectively, where the intensity of the colors represents the proximity to charge neutrality point $|E|$ in the range $E\in [-0.2,0.2] $. The shadowed regions represent the range before reaching half and full moir\'e lengths. The dashed lines in the left plot correspond to the Fermi velocity along X (blue) and Y (orange). }
    \label{fig:superperiodicities-freeform}
\end{figure}

In the left panel of Fig.\ \ref{fig:superperiodicities-freeform} we can see the low-energy density of states, superimposed with the Fermi velocity along the $X$ and $Y$ directions. Although propagation along the $y$-axis is essentially governed by the kinematics of the Dirac cone, as seen in the almost unrenormalized orange curve, motion parallel to the modulation direction is strongly modified by the imposed superperiodicity (blue curve). This effect has been attributed to the appearance of mini-gaps in the electronic structure in \cite{PhysRevB.97.195410}, that appear due to the interactions between different $k$-points coupled by the superperiodicity, and we have previously observed a similar behavior in Sec.\ \ref{Sec: super-quasiperiodicity in tblg}. The presence of such mini-gaps can be intuited by the presence of a spiky density of states close to the charge neutrality point; however, we do not have enough precision to resolve them in our current computational approach, being only able to observe their impact on the transport properties.

In the right panel of Fig.\ \ref{fig:superperiodicities-freeform}, we can see the diffusion coefficient along the X and Y directions. There we can see how the presence of this one-dimensional modulation alters the electronic transport in a direction-dependent way. The transport along $X$ (blue lines) suffers a renormalization similar to the one seen in the linear chain of Fig.\ \ref{fig:superperiodic_linear_chain}. In contrast, the orange lines show no renormalization, following a more conventional ballistic transport, with a Fermi velocity characteristic of graphene.

The intensity of the color at each curve represents the proximity in the energy of the calculation to the charge neutrality point. 
Our calculations confirm this phenomenon is due to superimposed periodicity, and once again we can find in the diffusion coefficient, along $X$, fluctuations in $D_x(t)$ at characteristic times when the spatial extent of the wave packet reaches half of the moir\'e length.

We find a positive comparison in the literature for this anisotropic transport in similar electrostatically modulated potentials achieved by proximity to lithographycally patterned hBN \cite{Li2021}. These results also present similarities to other similar superperiodic systems, such as nanoporous graphene \cite{Alc_n_2024}, where the relation between geometrical lengths and characteristic length scales for transport was reported. Some theoretical works have also shown that one-dimensional superlattices can generate additional Dirac points and strong anisotropic effects \cite{PhysRevLett.101.126804} supporting the idea of the generality of this effect for a quasi-1D superperiodic lattice superimposed on graphene. 

From a semi-classical point of view, this renormalization can be understood as the 2D case of the previously studied superperiodicity in the linear chain (Fig.\ \ref{sec:linear-chain-superp}). In the $x$ direction, these reductions originate from the partial reflection of the wave packet at the periodic interfaces. However, in contrast to that case, transport along $y$ remains almost ballistic with no significant modifications, proving the anisotropy of the system and preventing the appearance of electron pockets.

\subsection{ Superperiodicity along both dimensions}\label{sec:super2D}

\begin{figure}[h]
    \centering
    \includegraphics[width=\linewidth]{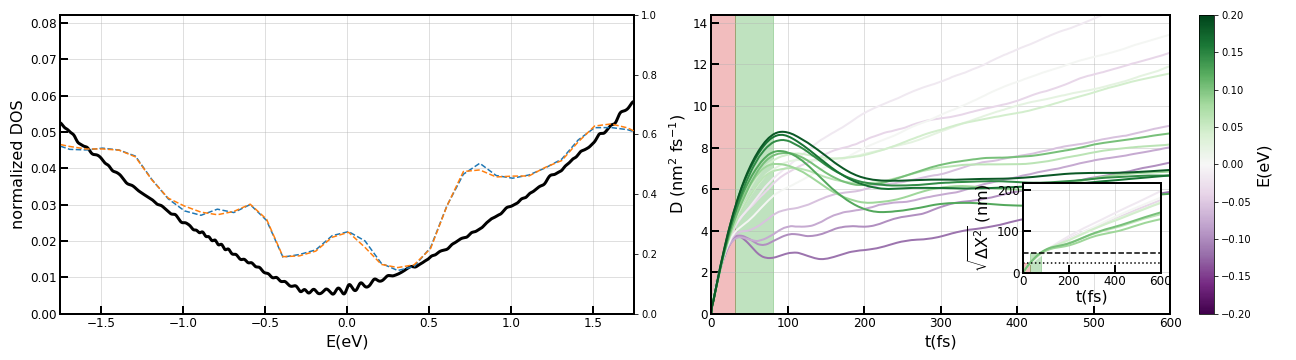}
    \caption{Density of states (left), diffusion coefficient in X (right) and electronic spreading in X (inset) in graphene for the triangular superperiodicity obtained for $N=3$. The color scale in the right plot and inset corresponds to the Energy distance to CNP where green corresponds to conduction band and purple to valence band. The shadowed regions represent the range before reaching half and full moir\'e lengths. The left panel only shows transport along X for clarity, as the system is isotropic and transport behaves the same in X and Y. The dashed lines in the left panel correspond to the Fermi velocity along X (blue) and Y (orange) exposing this isotropic behavior.}
    \label{fig:superperiodicities_freeform_super}
\end{figure}

To examine the effect of a two-dimensional superperiodic modulation, we set $N_{\mathrm{per}} = 3$ in equation \eqref{eq:freeform potential} and orient the three plane waves at angles $\theta_n = n\pi/3$. This choice yields a commensurate triangular superlattice that conserves the threefold rotational symmetry of the graphene lattice. Such modulations have previously been studied in the literature \cite{ead3d867-d652-325d-be20-3775e3412c44}. Because the potential preserves inversion symmetry, the original Dirac cones remain gapless, but similarly to the quasi-1D case, the folding of the Brillouin zone into a mini-Brillouin zone introduces additional Dirac points at high-symmetry points and opens minigaps whose magnitude depends on the modulation amplitude.

The black curve in the left panel of Fig.\ \ref{fig:superperiodicities_freeform_super} shows the density of states for this triangular superlattice, where, similarly to the previous case, these minigaps can be intuited in the shape of peaks. On the other hand, the Fermi velocity (orange and blue curves) presents an $X$-axis to $Y$-axis exchange symmetry, granted now by the conservation of the underlying symmetries in the graphene superperiodic lattice which reduces the in-plane transport anisotropy.

\begin{figure}
    \centering
    \includegraphics[width=\linewidth]{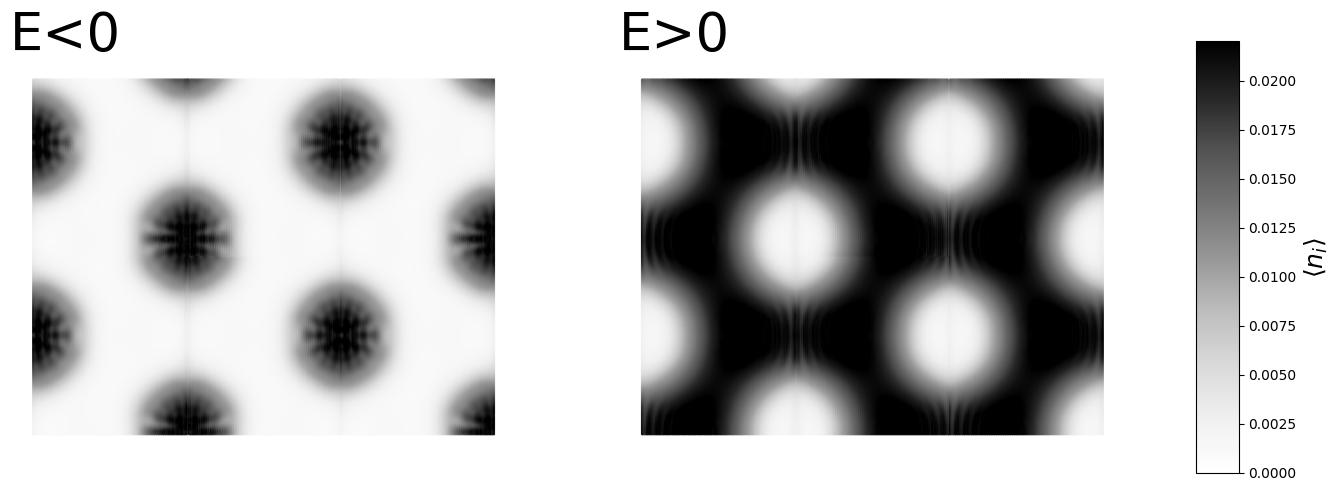}
    \caption{Local densities of states in the valence and conduction band regions where darker regions indicate the presence of electrons. We can see that, in the conduction band, electrons can freely move while in the valence bands electronic pockets are formed in real space. }
    \label{fig:graphenelike_LDOS}
\end{figure}

In the right panel of Fig.\ \ref{fig:superperiodicities_freeform_super} we can see that the transport simulations at different energies, where the intensity marks the energy distance of each curve from the CNP and green (purple) colors mark the conduction (valence) energy regions. We can clearly distinguish two different behaviors. One set of energies suffers a renormalization at half the moir\'e lengths and the other at a full moir\'e length. The origin of this behavior can be found in Fig.\ \ref{fig:graphenelike_LDOS}, where we observe the local density of states in the valence and conduction bands. The disposition of these states follows the two possible patterns that occur in this kind of superperiodicity. On one side in the conduction energy region, electrons follow a homogeneous distribution with low density zones present. These zones can be seen as individual scatterers, and thus these states suffer renormalizations at consecutive multiples of half the moir\'e length. There, renormalization occurs smoothly further from a full moir\'e length. 
We can see this region as the right-hand plot of Fig.\ \ref{fig:graphenelike_LDOS}. On the other hand, in the valence bands, we see that electrons are organized in the shape of pockets in real space with a higher density of states. This can be seen as the full 2D case of results viewed in Sec.\ \ref{Sec:quasi-1D} with similar renormalizations around half a moir\'e length, where again most of the transport derives from tunneling between the electronic pockets through the regions with low conductive channels. This behavior can be seen in the left-hand side panel of Fig.\ \ref{fig:graphenelike_LDOS}.

The transport properties reflect the threefold symmetry of the modulation. The diffusion coefficients $D_x(t)$ and $D_y(t)$ shown in Fig.\ \ref{fig:superperiodicities_freeform_super} are nearly identical, indicating that the triangular potential renormalizes the group velocity uniformly for all in-plane directions. As a result, the in-plane transport anisotropy is reduced with respect to the quasi-1D case, and we can spot no differences in the transport at low energies in our systems. Our results are in good agreement with previous results in the literature, such as \cite{ead3d867-d652-325d-be20-3775e3412c44}.

\subsection{Quasiperiodicity}\label{sec:Quasiperiodic_potential}

\begin{figure}
    \centering
    \includegraphics[width=\linewidth]{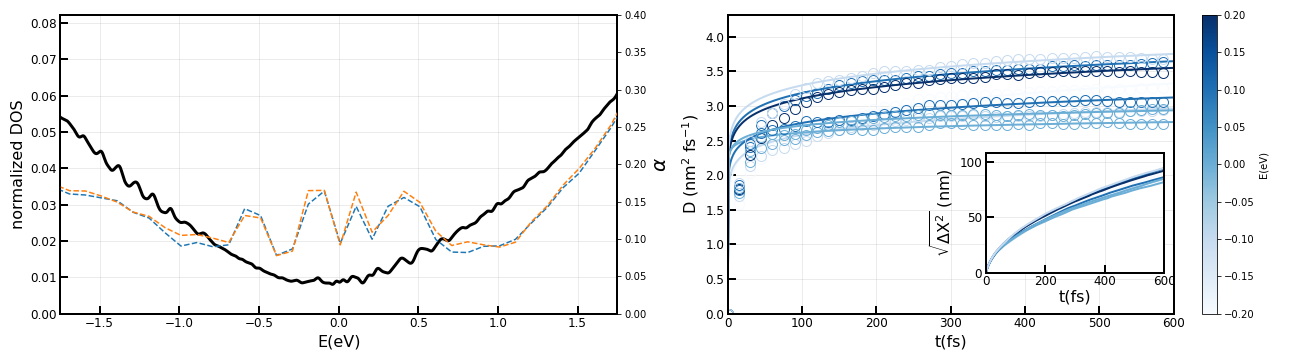}
    \caption{Figure similar to Fig. \ref{fig:superperiodicities-freeform} for the quasicrystalline case. In the right plot we can see the subballistic behavior generated by the superperiodicity, where circles mark our simulated results and the solid lines show the fits to a power law $D\propto t^{\alpha}$. The intensities in the right plot mark proximity in energy to CNP. Due to the symmetries, for clarity, we have only plotted the transport along $X$ in the right panel. The dashed lines in the left plot now represent the $\alpha$ fitted from the diffusion coefficient.}
    \label{fig:freeform_potential_quasi}
\end{figure}

We finally turn to the most intriguing limit of our toy model: a modulation that is not strictly periodic, but instead combines several plane-wave potentials with angles that are mutually incommensurate. This modulation of $N_{per}=6$, in a similar fashion as in the case of $30^{\circ}$ twisted bilayer graphene, generates a twelvefold quasicrystalline pattern without translational invariance where the absence of a superperiodic length
implies that there is no well-defined mini-Brillouin zone, and consequently the band structure becomes a dense set of narrow bands and pseudogaps typical of quasicrystals \cite{10.1063/1.531914}.

This kind of system is not periodic, and thus is complicated to study transport in it. Unlike calculations done in Sec.\ \ref{Sec: linear_chain_quasi}, the states considered for our full KPM calculation are extended states to properly apply the stochastic trace approximation (See Sec.\ \ref{Sec: KPM-intro}). In previous sections, we have seen that a superperiodic potential does not interfere with transport until at least $1/4$ of the superperiodic distance is covered. In order to avoid finite-size effects in such non-periodic systems, we will take these values and only consider valid measurements those for which the electronic spreading has not reached $1/4$ of the system dimensions. We emphasize that, in practice, we have only observed qualitative effects of the boundaries in calculations where the spreading reaches $\sim 1/2$ of the edge length of our systems where we see a drastic drop in the diffusion coefficient due to the scattering at the edge boundaries. In the inset of the right panel of Fig.\ \ref{fig:freeform_potential_quasi} we see that our maximum distance in our calculation is $\sim 100\  \mathrm{nm}$, below our safety threshold.

Figure\ \ref{fig:freeform_potential_quasi} summarizes our numerical results for this quasicrystalline superlattice.  The right panel shows the time-dependent diffusion coefficient
$D(t)$.  In contrast to the periodic cases studied in
the previous two sections, where $D(t)$ fluctuates sharply at short times and then recovers a ballistic behavior, here $D(t)$ follows a slow (sub-linear) power-law growth.   This sub-ballistic behavior is a hallmark of quasiperiodic systems as we already showed in Sec.\ \ \ref{Sec: linear_chain_quasi} and reflects the presence of critical eigenstates whose group velocities are drastically renormalized at each time step.

In our quasicrystalline linear chain, we computed this sub-ballistic behavior with an exponent of $\sim 0.46$. In our two-dimensional quasicrystalline lattice, we achieve critical exponent values in the range $\alpha\in [0.05,0.2]$, generating a highly tunable resistive material whose resistivity will vary by differences in the superimposed potential.

These findings highlight once again the qualitative difference between periodic and
quasiperiodic superlattices and confirm a certain generality of the results previously obtained for the linear chain in Sections \ref{sec:superperiodic_cas} and \ref{Sec: linear_chain_quasi}.  

In this section, we have studied the electrostatic modulation of the superperiodic and quasiperiodic graphene lattice with the objective of proving the generality of the results previously obtained for the linear chain and the results for the twisted graphene, which we will review in the next chapter. In addition, they establish a criterion for what is expected in a superperiodic system. However, the
results shown in this section also demonstrate that effects on transport similar to those of moir\'e patterns can be replicated by an electrostatic modulation of graphene. This may open a pathway to engineer
critical states in graphene, and therefore, we will use the next section to revisit the effects of such superperiodicities in a more detailed and realistic case of superperiodic potential, the quantum-dot arrays.

\section{In-depth study of a superperiodicity: the quantum-dot array}
\label{sec:qdot_array}

In the previous sections, we have shown that superperiodic electrostatic modulations can strongly renormalize transport in both one-dimensional chains and monolayer graphene. We saw how quasiperiodic modulations generically lead to anomalous reductions in velocity. In addition, in Sec.\ \ref{Sec: Multilayer AA, AB, ABC graphene} we highlighted that AB-stacked bilayer graphene can be used to model a behavior similar to magic-angle twisted bilayer graphene by an inhomogeneous electric field. In this section, we combine both ideas and consider a more realistic and experimentally relevant situation in which a superperiodic potential is engineered in Bernal-stacked bilayer
graphene using electrostatic gates. The aim is to emulate, as closely
as possible, the long-wavelength moir\'e modulation of twisted systems without
actually twisting the layers.

From a theoretical point of view, this gate-defined moir\'e strategy is appealing for at least two reasons. First, the superlattice potential is set by lithographically defined gate geometries and dielectric environments, which can be modeled at the electrostatic level with high precision \cite{Huber2020}. This contrasts with twisted multilayers, where the actual moir\'e pattern is affected by lattice relaxation, twist-angle inhomogeneity, and strain. Second, the underlying AB bilayer is structurally simple, and its low-energy bands under a perpendicular displacement field are well understood, providing a clean starting point to disentangle the roles of band topology, bandwidth, and interactions in moir\'e-like systems. This procedure has already been followed for the monolayer case in \cite{BarconsRuiz2022}.

In this context, Bernal-stacked bilayer graphene is a particularly versatile platform. A perpendicular electric field opens a tunable band gap and redistributes charge between layers, so that a spatially varying displacement field naturally realizes a superlattice of regions with larger and smaller gap. By patterning top and/or bottom gates on length scales comparable to those of realistic moir\'e periods, one can create an artificial landscape of quantum-dot-like puddles and interconnecting channels that closely mimics the long-wavelength modulation present in magic-angle twisted structures. Recent theoretical and experimental work has begun to explore precisely this route, proposing gate-defined superlattices in AB bilayer graphene as controllable emulators of correlated moir\'e physics \cite{PhysRevB.107.165158}.

The central objective of this section is to analyze a specific case of the AB-graphene quantum simulator. For that purpose, instead of prescribing an ad hoc superlattice potential as we previously did in this chapter, we will explicitly solve the Poisson-Schr\"odinger equations and predict the potential generated in the bilayer depending on the chosen gates and environmental conditions, with the objective of guiding future experiments in that area.

\subsubsection{A model for the Bernal graphene}

In order to solve the electrostatic conditions in our device, we are going to use a similar Schr\"odinger-Poisson approach to the one described in Sec.\ \ref{Sec: Poisson}. This procedure could be done by a full tight-binding description of the superperiodic unit cell and using \eqref{eq:LDOS}, yet it might prove convenient to reduce the complexity of this problem by obtaining a minimal model for the AB-graphene Hamiltonian.

Using a nearest-neighbor tight-binding description, we can build the simplest case of Hamiltonian for the graphene bilayer under a perpendicular electric field to be \cite{navarrorodriguez2023}:

\begin{equation}\label{eq:AB_graphene_hamiltonian}
    H=\left(\begin{array}{cc}
         H_G+V_{avg}+{\Delta V}&\begin{array}{cc}
              0&t_\sigma  \\
              0&0 
         \end{array}\\
         \begin{array}{cc}
              0&0  \\
              t_\sigma&0 
         \end{array}&H_G+V_{avg}-{\Delta V}
    \end{array}\right),
\end{equation}
where $H_G$ is the tight-binding Hamiltonian of monolayer graphene ($H_{TB}$ in \eqref{nearestneighborsgraphene}), $V_{avg}$ corresponds to the average potential between layers, and $\Delta V$, also called the interlayer asymmetry, is half the difference in potential between the top and bottom layer.

\begin{figure}[h]
    \centering
    \includegraphics[width=0.9\linewidth]{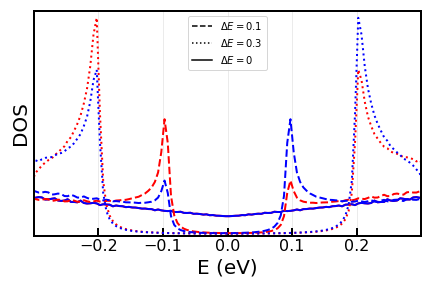}
    \caption{Projected density of states computed from \eqref{eq:AB_graphene_hamiltonian} for top (blue) and bottom (red) layers at different values of the interlayer asymmetry $\Delta V$. We can explicitly see the opening of a tunable gap with the applied electric field.}
    \label{fig:AB_DOS}
\end{figure}

The term $\Delta V$ models the effect of an electric field applied in the $z$ direction that crosses from the bottom to the top layer. In Fig.\ \ref{fig:AB_DOS} we can see how the density of states produced by this electric field generates an asymmetric gap opening with more electrons located on the bottom layer (red part).

If we assume that the electric field varies sufficiently slowly (much slower than the unit cell of our AB-graphene), we can locally assume that in each region of the space the Hamiltonian behaves as \eqref{fig:AB_DOS} and we vary the values of $\Delta V(\mathbf{r})$ and $V_{avg}(\mathbf{r})$.

\subsubsection{Device Geometry}

\begin{figure}[h]
    \centering
    \includegraphics[width=\linewidth]{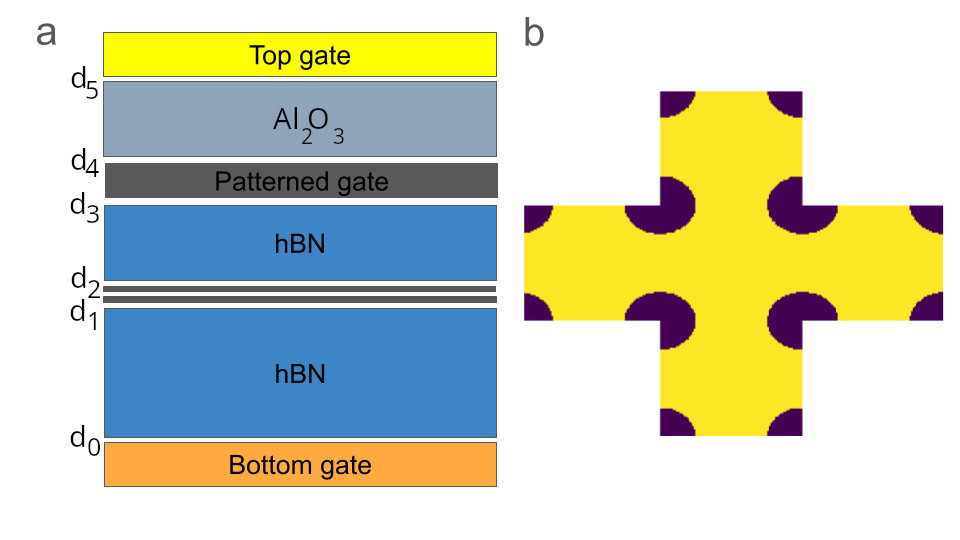}
    \caption{Description of a device to model the electrostatic potential modulations. Panel a shows a schematic picture of the device, the characteristic distances are marked (see text). Panel b shows a top view description of the middle gate where the yellow zones represent the graphene and the holes are marked in purple.}
    \label{fig:device-geometry-qdots}
\end{figure}

In order to generate a square lattice of potential in our Bernal graphene, we are going to model a heterostructure that uses a lithography-made superperiodic graphite gate as a way to modulate the spatial distribution of potential. In practice, our device will consist of a top Au gate and a bottom graphite gate, a thin middle graphite gate layer patterned by lithography, which will model the shape of our potential; and our bilayer graphene, all stacked with dielectric environments made of aluminum oxide ($Al_2O_3$) and hexagonal boron nitride (hBN) disposed as in Fig.\ \ref{fig:device-geometry-qdots}.

This device is general and can produce different physics. In order to simplify the analysis, for this thesis, we will keep the top and bottom layers fixed and vary the strength of the applied potential to the middle layer.

\subsubsection{Solution to the Poisson equation}

The coupling between the electrostatic potential and the electronic degrees of
freedom is implemented following the procedure described in Sec.\ \ref{Sec: Poisson}. 

From \eqref{eq:AB_graphene_hamiltonian} we can compute our occupations by integrating the DoS. For simplicity here, instead of using the KPM calculation for the density of states, we will use a simpler approach following the procedure shown in Appendix \ref{Appendix: green's function} where the DoS and LDoS are computed using a Green's function method \cite{economou}.

The only remaining ingredient is the explicit solution for the Poisson equation \eqref{eq: The Poisson equation} in this device. The problem of solving the Poisson-Schr\"odinger equation in our 2D materials heterostack is reduced to
\begin{equation}\label{eq:1-d-version-poisson}
   \frac{\partial}{\partial z}\left(\epsilon_{zz}(z)\frac{\partial}{\partial z}\hat V_{k_xk_y}(z)\right)=4\pi^2\left( \epsilon_{xx}(z)\frac{k_x^2}{L_x^2}+\epsilon_{yy}(z)\frac{k_y^2}{L_y^2}\right) \hat V_{k_xk_y}(z)-e\hat n_{k_xk_y}(z) 
\end{equation}
with $\epsilon_{ii}$ the diagonal elements of the dielectric tensor, $L_x,L_y$ the dimensions considered on the $x$ and $y$ axes of the stack, and $\hat V_{k_xk_y},\hat n_{k_x,k_y}$ the Fourier transform of the potential and the electronic density, respectively. In Appendix\ \ref{appendix poisson_solution}, we give a detailed description of the derivation of this equation for a 2D material heterostack which can be done for an arbitrary choice of constant diagonal permittivities $\epsilon_{xx} (z),\epsilon_{yy}(z),\epsilon_{zz}(z)$. In our case, we will choose the permittivity profile in $z$ to be:

\begin{equation}
    \epsilon_{ii}(z)=\left\lbrace
    \begin{array}{ccc}
         \epsilon_{hBN,ii}& &z<d_1  \\
         1& &z>d_1\quad\&\quad z<d_2  \\
         \epsilon_{hBN,ii}&  &z>d_2\quad\&\quad z<d_3\\
         1&  &z>d_3\quad\&\quad z<d_4\\
         \epsilon_{\mathrm{Al_2O}_3,ii}&  &z>d_4
         
    \end{array}
    \right. 
\end{equation}
where $\epsilon_{hBN,xx}=\epsilon_{hBN,yy}=\epsilon_{hBN,X}$, $\epsilon_{hBN,zz}=\epsilon_{hBN,Z}$ are the average dielectric constants in and out-of the plane of hBN, respectively, and $\epsilon_{\mathrm{Al_2O_3},ii}=\epsilon_{\mathrm{Al_2O_3}}$ the average isotropic dielectric constant of $\mathrm{Al_2O_3}$. This is equivalent to assuming homogeneous media in Fig.\ \ref{fig:device-geometry-qdots} with sharp interfaces between different slabs.

The top and bottom gates are introduced as boundary conditions for $V(d_0)=V_{bot}$ and $V(d_5)=V_{top}$. For the patterned gate, on the other hand, a boundary condition treatment is not enough, given the reciprocal space nature of \eqref{eq:1-d-version-poisson}, we cannot easily discriminate between fixing the potential in some regions of space (graphitic regions) and allowing variations inside the holes. For that reason, our middle layer will be modeled as a capacitor where we will induce two surface charges in $d_1$ and $d_4$ in the interface between the graphitic gate and the dielectric. The surface charges potential are going to be

\begin{figure}[t]
    \centering
    \includegraphics[width=\linewidth]{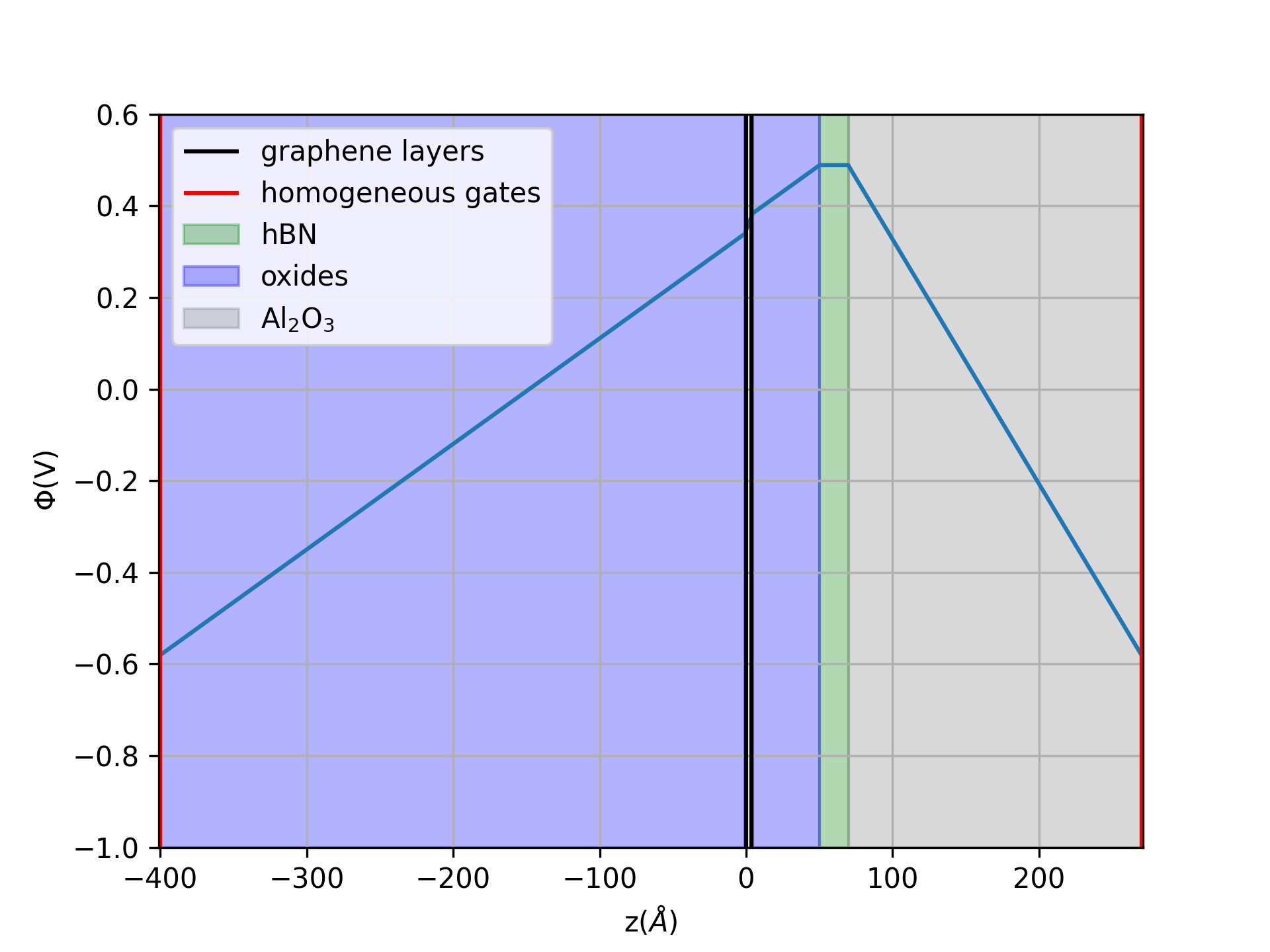}
    \caption{$z$ profile for a Poisson in a gap with parameters extracted from Table \ref{tab:choice of parameters} before starting the convergence for $k=0$.}
    \label{fig:z_profile_poisson}
\end{figure}

\begin{table}[]
    \centering
    \begin{tabular}{c c c c c}
        \hline
        parameter && Description && Value\\
        \hline
        \hline
         $V_{top}$&& top gate voltage &&  -580.5 meV\\
         $V_{bot}$&& bottom gate voltage &&  -580.5 meV\\
         $V_{mid}$&& patterned gate voltage &&  487.5 meV\\
         $\epsilon_{hBN,X}$&& in-plane dielectric constant of hBN && 6.7 \\
         $\epsilon_{hBN,Z}$&& out-of-plane dielectric constant of hBN &&  3.56\\
         $\epsilon_{\mathrm{Al_2O_3},Z}$&&  dielectric constant of $\mathrm{Al_2O_3}$ &&  9\\
         \hline
         $a_{SL}$&& Lattice parameter of the middle gate&&20 nm\\
         $d_{hole}$&&Diameter of the holes in Fig.\ \ref{fig:device-geometry-qdots}&&$a_{SL}/2$\\
         $d_0$&&See Fig.\ \ref{fig:device-geometry-qdots} &&-40 nm\\
         $d_1$&&See Fig.\ \ref{fig:device-geometry-qdots} &&0 nm\\
         $d_2$&&See Fig.\ \ref{fig:device-geometry-qdots} && 0.35 nm\\
         $d_3$&&See Fig.\ \ref{fig:device-geometry-qdots} &&5 nm\\
         $d_4$&&See Fig.\ \ref{fig:device-geometry-qdots} &&7 nm\\
         $d_5$&&See Fig.\ \ref{fig:device-geometry-qdots} &&27 nm\\
         & 
    \end{tabular}
    \caption{Choice of parameters for our twisted bilayer graphene simulator}
    \label{tab:choice of parameters}
\end{table}

\begin{equation}
    \begin{array}{ccc}
         \rho_{G,bot}&=&-\frac{V_{top}-V_{mid}}{d_5-d_3}\epsilon_0\epsilon_{\mathrm{Al_2O_3}}  \\\\

         \rho_{G,top}&=&-\frac{\epsilon_0\epsilon_{hBN,zz}(V_{bot}-V_{mid})+\rho_{bot}(d_1-d_0)+\rho_{top}(d_1-d_0+(d_2-d_1)\epsilon_{hBN,zz})}{((d_3-d_0-(d_2-d_1))+(d_2-d_1)\epsilon_{hBN,zz})}
    \end{array}
\end{equation}
with these absorbent surface densities, we guarantee that, for $k=0$, the potential is homogeneous in the middle gate region. 
And, at the same time, we can expand its structure into the Fourier space, adding them as virtual charges in the equation \eqref{eq:1-d-version-poisson}. In Fig.\ \ref{fig:z_profile_poisson} we can see the $z$-profile of the potential in the stack and the effect of these charges in stabilizing a constant potential in the middle gate at $k=0$ before starting the convergence, so $\rho_{bot}=\rho_{top}=0$.

\begin{figure}
    \centering
    \includegraphics[width=\linewidth]{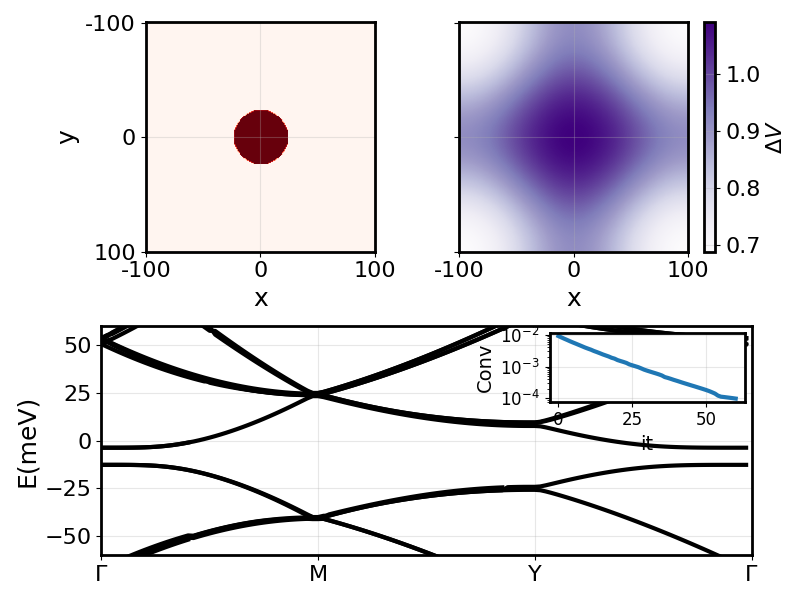}
    \caption{Top left panel shows the region of the unit cell where bands cross the Fermi level. Top right panel shows the distribution of the gap $\Delta V$ across the unit cell. Bottom panel shows the band dispersion of the whole supercell along the high-symmetry path $\Gamma\to M\to Y\to \Gamma$. Inset displays the evolution of the convergence parameter with iterations.}
    \label{fig:quantum_dots}
\end{figure}

We have found a convergence up to $0.1\ \mathrm{meV}$ for this system, in the inset of the bottom panel of Fig.\  \ref{fig:quantum_dots} we can see the evolution of the convergence parameter. In this specific choice  of parameters of Table\ \ref{tab:choice of parameters}, we can find well-defined quantum dots. That is, regions in space where the electrons are maximally confined. 

In the dark region of the top left panel of Fig.\ \ref{fig:quantum_dots}  we can see the location of the quantum dots; this is the point where the local bands generated by \eqref{eq:AB_graphene_hamiltonian} cross the Fermi level, while in the clear regions the system has the Fermi level inside the gap. This produces a characteristic localization of the quantum dots.

In the top right panel of Fig.\ \ref{fig:quantum_dots} we can see the map of the potential difference in one superperiodic unit cell. Following the discussion of equation \ref{eq:AB_graphene_hamiltonian}, the gap depends on $\Delta V$, according to \cite{navarrorodriguez2023}, quantum dots are achievable with a difference in the gap so that the density of states is suppressed outside of the dot. On the other hand, in our case, the suppression generated by the non-homogeneous distribution of $V_{avg}$ causes, even if the gap is maximal inside the dot, the energy shift generated by the distributions in the layer is enough to set the system inside (outside) the gap when outside (inside) the quantum dot.

In the bottom panel of Fig.\ \ref{fig:quantum_dots} we can see the band-structure of the full superperiodic cell, and we observe that the dots' bands present curvature. This is a characteristic of tunneling between different quantum dots \cite{navarrorodriguez2023}. This interaction can be modulated depending on the strength of the potentials chosen or the distance between the dots.

With that, we have generated a superperiodic potential in a realistic device. These kinds of devices also present interactions, similar to those generated by the moir\'e systems, and can be used to engineer these interactions or study the physics generated by moir\'e in different interacting ranges \cite{BarconsRuiz2022}.

\chapter{Transport in twisted Multilayer Graphene}\label{Chapter: Transport in twisted multilayer graphene}

Twisted multilayer graphene provides a natural route to engineering long-wavelength superlattices (moir\'e patterns) and, consequently, electronic states with strongly renormalized kinetic energy. In the magic-angle regime, the moir\'e bands become extremely narrow and the low-energy wave functions concentrate in the AA regions, creating an ideal setting where even weak perturbations can have an outsized impact on transport and localization. At the same time, real samples are inevitably affected by disorder (electrostatic fluctuations, local strain, and atomic-scale defects), which makes it essential to establish how robust the key single-particle fingerprints of twisted systems are against realistic perturbations before addressing interaction-driven phases.

This chapter applies the real-space transport framework developed earlier to realistic twisted graphene multilayers, with the aim of understanding how structural complexity and disorder reshape electronic propagation. The main focus is magic-angle twisted bilayer graphene, where the narrowness of the flat bands and the strong real-space modulation of the wave functions make transport especially sensitive to even weak perturbations. Within a single-particle, zero-temperature description, we quantify how Anderson disorder modifies the density of states, diffusion, and mean free path, and show that in a finite regime where flat-band features remain identifiable, moderate disorder can counterintuitively enhance transport through a disorder-induced delocalization mechanism.

We then connect this behavior to the evolution of the quantum metric extracted from optical conductivity, providing a geometric interpretation of the real-space extent of the electronic states. Beyond the magic-angle case, the chapter extends the analysis to graphene quasicrystal approximants and to hybrid trilayer stacks that combine moir\'e and quasicrystalline interfaces, allowing us to test the robustness of high-energy resonant and anomalous transport signatures against both disorder and proximity effects.



\section{Computational details}

Methodologically, we combine atomistic modeling of the relaxed geometries of twisted structures, a realistic $p_z$ tight-binding description of the full multilayer Hamiltonian, and linear-scaling KPM/real-space quantum-transport techniques to access large samples beyond direct diagonalization. Throughout the chapter, we characterize each system using the density of states, velocity renormalization, diffusion coefficients, and mean free paths, and we complement the transport analysis with geometric indicators (quantum metric) that connect naturally to the real-space extent of the electronic ground state and to the modern theory of localization.

\subsection{Structural relaxation}

To build realistic twisted graphene superlattices, we use molecular dynamics simulations with classical potentials to relax the structures \cite{PhysRevLett.134.126301}. The importance of such relaxation procedures for the correct modeling of the electronic properties of twisted bilayer graphene for small angles has been reported in the literature \cite{Hung2021,Nguyen_2022}. However, for large angles, we have found an almost negligible difference between the relaxed and unrelaxed cases in our calculations \cite{22s2-vpm5}.

We start with uniform Bernal-stacked bilayer graphene, twisted to an angle of ${\sim}1.1^\circ$ for the magic-angle and ${\sim 29.8^\circ,31^\circ}$ for the quasicrystalline approximants (see Sec. \ref{sec:Quasicrystalline-approximants}), and optimize the structure until all force components are smaller than 0.5 meV/atom. Intralayer forces are computed using optimized Tersoff and Brenner potentials \cite{Lindsay2010}, while interlayer forces are modeled using Kolmogorov-Crespi potentials \cite{PhysRevB.71.235415,Leven2016}. In reality, these materials are often deposited on substrates such as silicon oxide or encapsulated with other materials such as hexagonal boron nitride or stacked on top of graphite gates \cite{Nguyen_2022}. To model the effect of an underlying substrate, we force the flatness of the bottom layer of each stack not allowing relaxation in the vertical direction.


\subsection{Tight-binding models}

The electronic properties of the twisted multilayer systems are then computed using the $p_z$ tight-binding (TB) Hamiltonian,

\begin{equation}
\mathcal{\hat{H}} = \sum_{n} \varepsilon_n \ket{\phi_n} \bra{\phi_n} + \sum_{n,m} t (\vec{r}_{nm}) \ket{\phi_n} \bra{\phi_m},
\end{equation}
where $\ket{\phi_n}$ describes the orbital $p_z$ at the carbon site $n$ with position $\vec{r}_{n}$, $\varepsilon_n$ is the electrostatic potential at the carbon site $n$, and $\vec{r}_{nm} = \vec{r}_{m} - \vec{r}_{n}$. The hopping energies $t (\vec{r}_{nm})$ between carbon sites are given by the standard Slater-Koster expression \cite{Trambly2010,Hung2021}.

\begin{align}\label{eq:charlier_model}
t ({\vec r}_{nm}) &= \cos^2 (\phi_{nm}) V_{pp\sigma} (r_{nm}) \nonumber \\
&+ \sin^2 (\phi_{nm}) V_{pp\pi} (r_{nm}),
\end{align}
where the direction cosine of $\vec{r}_{nm}$ along the $z$-axis is $\cos(\phi_{nm}) = z_{nm}/r_{nm}$. The distance-dependent Slater-Koster parameters are \cite{Trambly2012}
\begin{align}\label{eq:decay-Magic-angle-hoppings}
V_{pp\pi} (r_{nm}) &= V_{pp\pi}^0 \exp \left[ q_\pi \left( 1 - \frac{r_{nm}}{a_0}  \right) \right] F_c (r_{nm}), \nonumber \\
V_{pp\sigma} (r_{nm}) &= V_{pp\sigma}^0 \exp \left[ q_\sigma \left( 1 - \frac{r_{nm}}{d_0}  \right) \right] F_c (r_{nm}),
\end{align}
with a smooth cutoff function $F_c (r_{nm}) = \left[ 1 + \exp \left(( r_{nm}-r_c)/\lambda_c\right) \right]^{-1}$.
To model the flat electronic bands of relaxed TBLG at the magic angle $\sim$$1.1^\circ$, the TB parameters are adjusted to $V_{pp\pi}^0 = - \gamma_{0}= -2.7 \,~\mathrm{eV}$, $V_{pp\sigma}^0 = 367.5 \,~\mathrm{meV}$, $q_\pi/a_0 = q_\sigma/d_0 = 22.18 \,~\mathrm{nm}^{-1}$, $a_0 = 0.1439 \,~\mathrm{nm}$, $d_0 = 0.33 \,~\mathrm{nm}$, $r_c = 0.614 \,~\mathrm{nm}$, and $\lambda_c = 0.0265 \,~\mathrm{nm}$ \cite{Nguyen_2022}.

\subsection{KPM calculations}

After relaxation of the lattices, we used a $16\times 16$ tiling magic-angle unit cell in our simulations, corresponding to $\sim 2.8\cdot 10^6$ atoms, to ensure a proper evolution of the electronic states and proper convergence of our methods. Similarly, for quasicrystalline approximants, a $4\times 4$ tiling of each unit cell, giving $\sim 3.4\cdot 10^6$ and $\sim 5.5\cdot10^6$ for $31^\circ$ and $29.8^\circ$ approximants, respectively. For the trilayer, the same $4\times 4$ tiling was chosen, giving $\sim 5\cdot 10^6$and $8.3\cdot10^6$ for the $31^\circ$ and $29.8^\circ$ approximants.

After that, the KPM methods described in Sec. \ref{Sec: KPM-intro} were used to compute (local) densities of states, and the time evolution of the mean-squared displacement. Then, we followed a similar approach to the one described in Sec.\ \ref{sec: computational-details-freeform} to extract the diffusion coefficient. 
For the time-evolving properties, a time step of $1\ \mathrm{fs}$ was considered, evolving to $600\ \mathrm{fs}$ in all our simulations. The only exceptions are the long-time calculations for clean systems, where we reached $6000\ \mathrm{fs}$ in order to find a precise description of the Fermi velocity and confirm the ballistic behavior of the system.

\section{Magic angle twisted bilayer graphene}\label{Sec:MATBLG_paper}
In this section, we present the results obtained for magic-angle twisted bilayer graphene, including the disorder-induced delocalization mechanism revealed in \cite{PhysRevLett.134.126301}.

\subsection{Structural relaxation}\label{sec:structural-relaxation-magic-angle}

From a structural point of view, we start with relaxed structures for the MATBLG. The effect of interatomic forces on the electronic properties of small-angle twisted bilayer graphene has been well established in \cite{PhysRevB.101.195425} and \cite{Nguyen_2022}.

In these works, three different effects are clearly established: 
\begin{itemize}
    \item In-plane carbon atom displacements tend to reduce the size of the AA regions; this can be understood from the fact that, as we introduced in Sec.\ \ref{Sec: Multilayer AA, AB, ABC graphene}, AA graphene is less stable than AB graphene.
    \item This relaxation results in large out-of-plane displacements of the AA zones, increasing the interlayer distance in those regions. These can be viewed as frustrated regions.
    \item This relaxation effect is critical for the flat-band formation, helping to reduce the dispersion and is necessary to achieve a proper real-space modeling of the flat-band states.
\end{itemize}

We have relaxed our structure, achieving the described features. More specifically, we have replicated the results in \cite{Hung2021}, taking then the relaxed systems as a starting point for our electronic calculations.

\subsection{Electronic structure}
\begin{figure}[h!]
\includegraphics[width=\columnwidth]{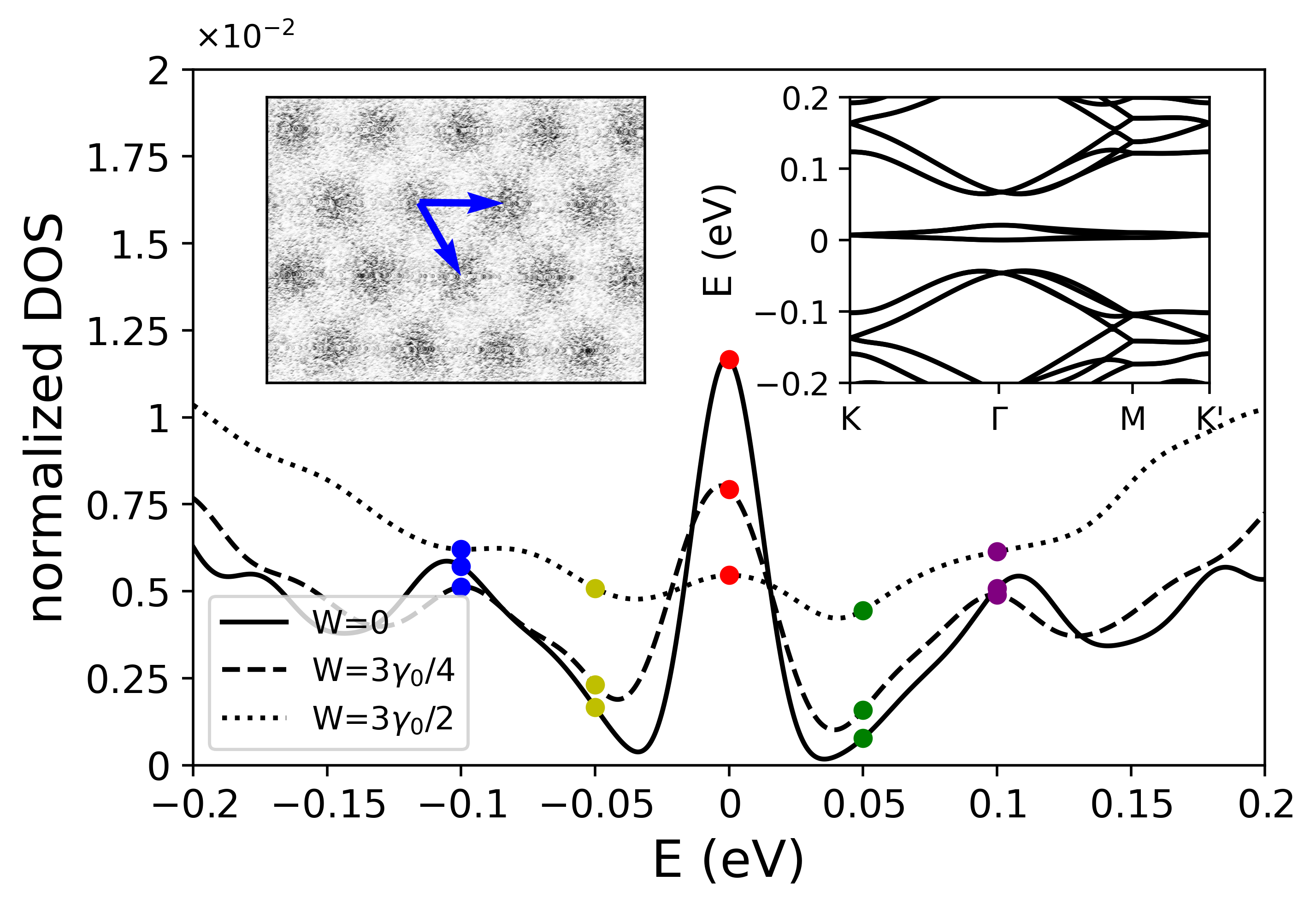}
\caption{Total density of states of MATBLG for disorder strengths of $W={3\gamma_0/2}$ (dotted line), $W=3\gamma_0/4$ (dashed line), and no disorder (solid line). Left inset: local density of states (LDoS) of the clean system at charge neutrality where higher LDoS (darker zones) is concentrated in the AA moir\'e regions. The blue arrows indicate the moir\'e unit cell lattice vectors. Right inset: band structure of MATBLG, with the flat bands clearly present around $E=0$. }
\label{fig:electronic-struct-matblg}
\end{figure}

 We start by analyzing the impact of disorder on the electronic structure of MATBLG through its impact on the total (DoS) and local density of states (LDoS). In Fig.\ \ref{fig:electronic-struct-matblg} we plot the DoS for Anderson disorder strengths of $W = 0$, $3\gamma_0/4$, and ${3\gamma_0/2}$, where $\gamma_0$ is the tight-binding parameter of graphene. For reference, we show the band structure of the clean case in the right inset and the LDoS at charge neutrality ($E = 0$) in the left inset. The flat bands and the corresponding localization of the states in the AA regions are well reproduced in the absence of disorder. In the main panel, the presence of a strong peak in the DoS at $E=0$ highlights the presence of the moir\'e-induced flat bands. The role of disorder is to broaden and reduce this peak, which remains clearly visible at $W=3\gamma_0/4$ before finally being washed out for $W\geq{3\gamma_0/2}$, coinciding with the disappearance of AA spatial localization (see the right inset of Fig.\ \ref{fig:msd-deloc-matblg} for $W=2\gamma_{0}$).

\subsection{Transport in the clean case}

\begin{figure}[tbh]
\includegraphics[width=\columnwidth]{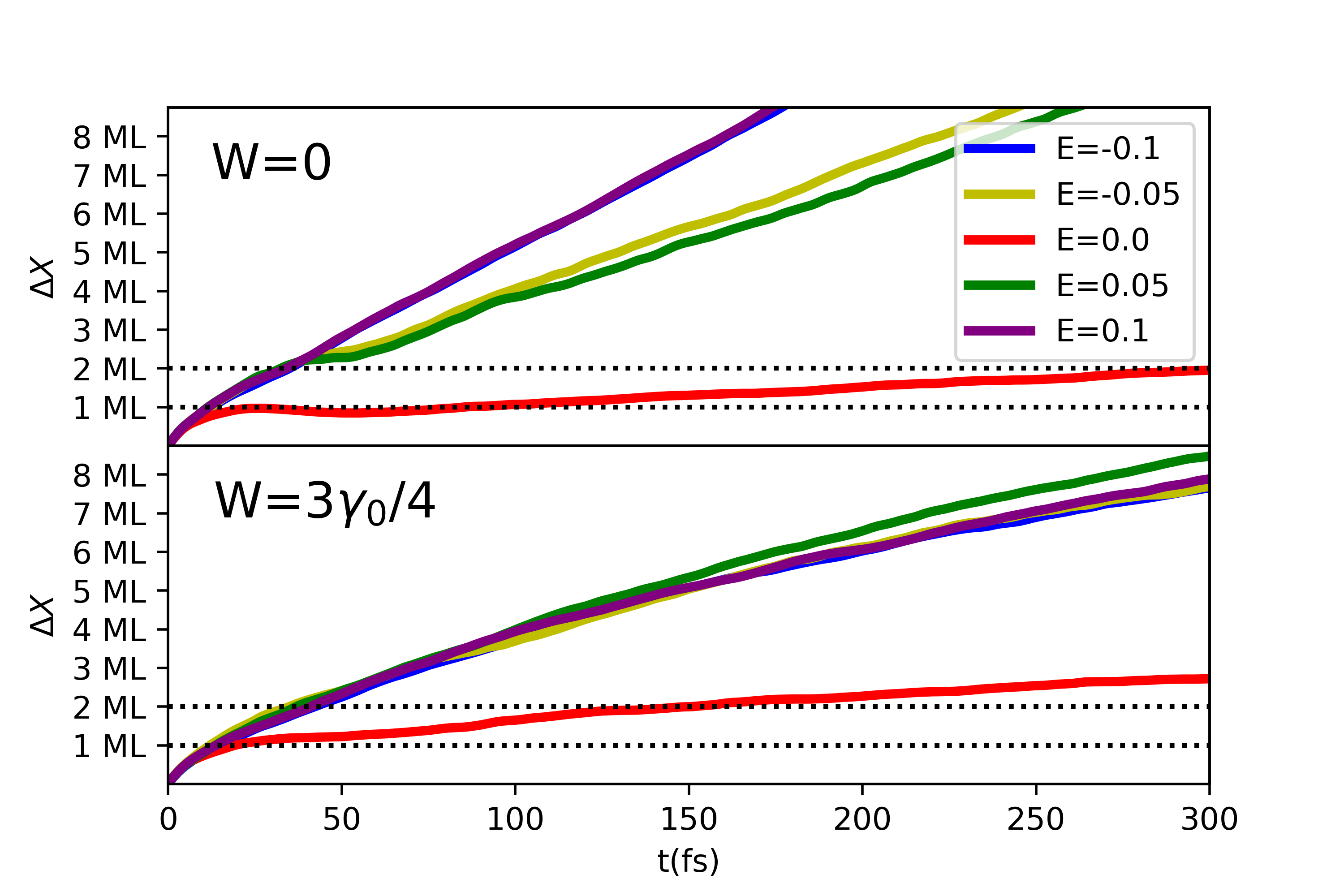}
\caption{Electronic spreading as a function of time for the clean system (upper panel) in units of the moir\'e length (ML) and the disordered system with Anderson disorder W=$3\gamma_0/4$ (lower panel), black dotted lines are intended just as a guide to the eye.}
\label{fig:superperiodic-eff-matblg}
\end{figure}

 In Fig.\ \ref{fig:superperiodic-eff-matblg} we show the spreading of the wave packet at short times, given by $\Delta X = \sqrt{\Delta X^2}$, where $\Delta X^2$ is the mean-squared displacement as defined in equation\ (\ref{eq:msd}). We will follow a similar analysis as in Sec.\ \ref{sec:super2D} to bound the MSD with the characteristic length scales of the MATBLG. The top panel shows the clean case, and the bottom panel corresponds to a disorder strength of $W = 3\gamma_0/4$. On the $y$-axis, the spreading of the propagation of the wave packet is shown in units of the moir\'e length of the MATBLG system, which is $\sim$13 nm.

In the clean case (top panel), at very short times, before reaching the first moir\'e length (ML), states propagate with the same velocity at all energies. However, upon reaching the first ML, the low-energy states in the flat band (red curve) undergo a dramatic slowing of their propagation velocity. In these states electrons are localized in the AA regions similarly to the left panel of Fig.\ \ref{fig:graphenelike_LDOS} (see left inset of Fig.\ \ref{fig:electronic-struct-matblg}). The states adjacent to the flat band (yellow and green curves) experience a slowing at a spreading of 2ML. These states have been found to present antilocalization in the AA zone \cite{PhysRevLett.128.126401}, similar to the right panel in Fig. \ref{fig:graphenelike_LDOS}. Meanwhile, the high-energy states (blue and purple curves) experience little velocity renormalization. In general, for the clean case, we observe a very similar transport behavior as the one observed for the superperiodic potentials of chapter\ \ref{chapter:superperiodicities}, where the real-space pattern of LDoS behaves as a predictor for characteristic transport length scales.

In the disordered case (bottom panel), similar behavior is seen for the flat band (red curve), whereas states at all other energies converge to the same behavior owing to the Anderson disorder.

These results present a direct visualization of the accommodation of the wave function with the moir\'e lattice and the complex flat-band physics of MATBLG. They also provide lower spatial and temporal bounds for the appearance of such flat-band physics, which do not appear until the wave function has evolved long enough to see the superperiodicity of the moir\'e lattice. This also remarks that the inherent physics of superperiodic transport studied in the previous chapter only happens at the energies of the flat band, confirming the difference against transport between inter- and intra-layer states presented in Sec.\ \ref{Sec: super-quasiperiodicity in tblg}.

\begin{figure}[tbh]
\includegraphics[width=\columnwidth]{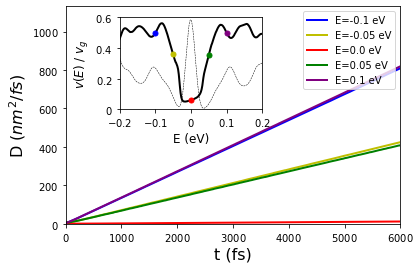}
\caption{Diffusion coefficient as a function of time for clean MATBLG at energies of $E = -100$ meV (blue), $-50$ meV (yellow), charge neutrality (red), $50$ meV (green), and $100$ meV (purple). Inset: the corresponding energy-dependent Fermi velocity (solid line), superimposed with the DoS of the clean case (dotted line, rescaled for clarity). }
\label{fig:fermi-vel-matblg}
\end{figure}

In Fig.\ \ref{fig:fermi-vel-matblg} we plot the long-time evolution of the diffusion coefficient of clean MATBLG. In the absence of disorder, the transport is ballistic and the diffusion coefficient increases linearly with time $D(E,t)=v^2(E) t$. This behavior is seen at all energies (different colored curves) in Fig.\ \ref{fig:fermi-vel-matblg}. From the slope of these curves we then extract the Fermi velocity of the MATBLG system, which we plot in the inset, relative to the velocity of single-layer graphene $v_\text{g}$.  {Note that while the Fermi velocity may not be uniform around the Fermi surface, here we plot its average over the Fermi surface at the indicated energies $E$.} Here we see that around charge neutrality the Fermi velocity is very low $v < 0.1 v_\text{g}$, characteristic of the flat nature of the moir\'e bands.

\subsection{Disorder-induced delocalization}\label{sec:Disorder-induced-delocalization}

\begin{figure}[tbh]
\includegraphics[width=\columnwidth]{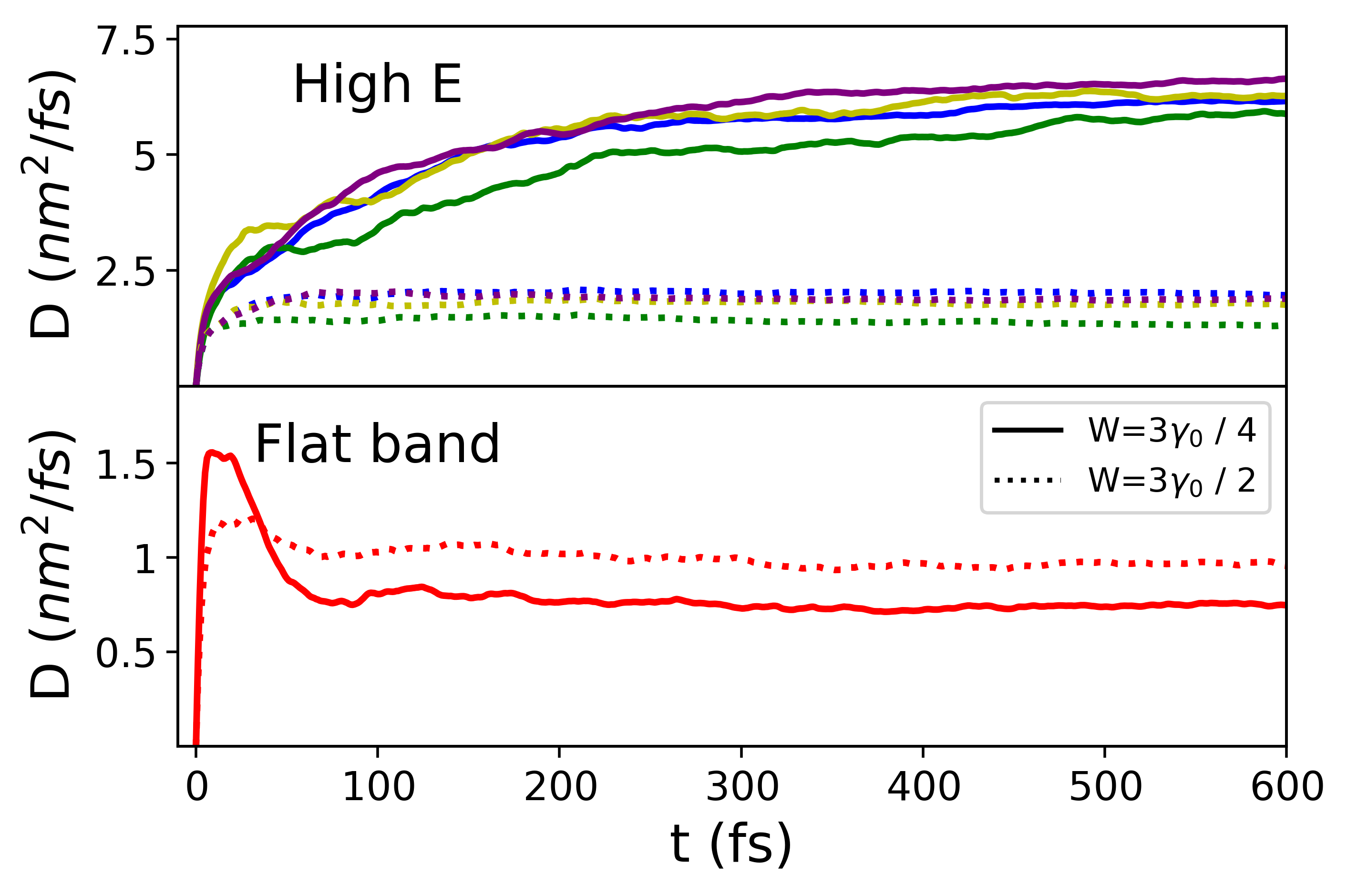}
\caption{{Time dependent diffusion coefficient at energies of $E = -100$ meV (blue), $-50$ meV (yellow), charge neutrality (red), $50$ meV (green), and $100$ meV (purple) for MATBLG under Anderson disorder strengths of $W=3\gamma_0/4$ (dashed) and $W={3\gamma_0/2}$ (dotted).} }
\label{fig:disorder-deloc-diff}
\end{figure}

  We use this section to study the weakness of the MATBLG localized states against disorder. In Fig.\ \ref{fig:disorder-deloc-diff} we examine electronic transport in disordered MATBLG. In the presence of Anderson disorder, the diffusion coefficients now saturate at long times for all energies. However, we observe a qualitative difference between transport within the flat bands compared to that at higher energies. When increasing the strength of the disorder from  $W = 3\gamma_0/4 \rightarrow {3\gamma_0/2}$, $D$ decreases by a factor of $\sim$$4$ for all energies except at charge neutrality (red curve), where $D$ actually increases, opposite to typical behavior. This is illustrated further in Fig.\ \ref{fig:msd-deloc-matblg}, where we plot the mean free path $\ell(E)$ for three different strengths of the disorder. In the energy range corresponding to the moir\'e flat bands, for weaker disorders we see a clear increase of the mean free path with increasing disorder strength, opposite to the scaling behavior at higher energies. This increase in $\ell(E)$ actually coincides with a delocalization of the LDoS around charge neutrality in the presence of disorder, as highlighted in the left and right insets of Fig.\ \ref{fig:msd-deloc-matblg}. \footnote{Here we note a slight electron-hole asymmetry in the mean free path, arising from an asymmetry in the band structure (inset of Fig.\ \ref{fig:electronic-struct-matblg}) and correspondingly in the Fermi velocity (inset of Fig.\ \ref{fig:fermi-vel-matblg}).} A similar behavior of the electron-phonon coupling in MATBLG has been reported in Refs. \cite{Choi2018,Gadelha2022}. 
  This disorder-induced delocalization was later found in a qubit array quantum simulator \cite{rosen2025flat}, and in other theoretical works \cite{hou2025fate}.

We note that a stronger disorder will eventually suppress any remnant of the flat bands and thus reduce the mean free path following the scaling behavior $\ell\sim (\gamma_{0}/W)^{2}a_{cc}$, where $a_{cc}$ is the carbon-carbon spacing. This is seen when the disorder increases from $W={3\gamma_0/2}$ to $W=2\gamma_{0}$ in Fig.\ \ref{fig:msd-deloc-matblg}.

Therefore, the observed anomalous ``disorder-induced delocalization'' exists over a finite range of disorder strengths and is maintained when disorder is low enough to preserve the moir\'e-induced flat-band states. This effect is driven by the disorder-induced broadening of the flat bands and the corresponding  delocalization of states in real space. Using a simple argument based on the Fermi golden rule, the increase in the mean free path is driven by the reduction of the DoS and the corresponding scattering rate. For weaker disorder, following the scaling theory of localization, one expects that near flat bands the localization length (related to the mean free path through the Thouless relationship) will reach values on the order of 100 nanometers for a disorder strength corresponding to the effect of electron-hole puddles generated by a silicon oxide substrate \cite{VanTuan2016}. Finally, we note that in the strong Anderson disorder limit, the mean free paths in disordered MATBLG are similar to those found in disordered monolayer graphene \cite{Lherbier2008}.

\subsection{Impact on quantum metric and correlations}

 To compute the quantum metric, we make use of the SWM rule \cite{SWMrule}, which relates it to the optical conductivity (via the fluctuation-dissipation theorem) as
\begin{equation}\label{eq:SWMrule}
    \int_0^{\infty}\dfrac{d\omega}{\omega}\Re{\sigma_{xx}(\omega)}=\dfrac{\pi e^2}{h}{\cal G}_{xx},
\end{equation}
where ${\cal G}_{xx}$ is the dimensionless quantum metric, $e$ is the electric charge, $h$ is the Planck constant, $\sigma(\omega)$ is the optical conductivity, and $\omega$ is the frequency. This QM is related  to the invariant part of the spread of the Wannier functions \cite{MaximallylocalizedWannierfunctions},
$\Omega_{\mathrm{I}} = \mathrm{Tr}\{{\cal G}_{xx}\} \cdot V/(2\pi)^2$,
where the trace runs over the Cartesian indices. Here, a small QM relates to a small Wannier spreading, and thus to a strongly localized ground state. On the other hand, when the QM continuously increases with system size, it indicates delocalization of the Wannier functions and the ground state. Here we compute $\sigma(\omega)$ from the Kubo formula \cite{KUBO}, making use of a Chebyshev polynomial expansion \cite{opticalconductivity} with a broadening of 66 meV, and obtain convergence of ${\cal G}_{xx}$ with system size and number of polynomials.

 Fig.\ \ref{fig:msd-deloc-matblg} (bottom panel) shows ${\cal G}_{xx}$ for different strengths of the disorder, and we compare its disorder-dependent evolution with the mean free path. Interestingly, the evolution between $W=0.75\gamma_{0}$ and $W=2\gamma_{0}$ is qualitatively similar for ${\cal G}_{xx}$ and $\ell$, indicating a disorder-induced delocalization mechanism. The increase in QM at lower disorder is expected for weakly disordered cases, since the cleaner the system, the longer the corresponding mean free path and localization length. Note that the precise value of the mean free path is not shown for the lowest disorders due to computational difficulty accessing the diffusive regime in the simulations.
However, we can make a rough estimate using the Fermi golden rule when disorder only weakly affects the band structure, for which $\ell \propto 1/W^2$. Taking the smallest value of the disorder with a converged mfp ($W = 3/4$), for which $\ell = 33.8$ nm, we extrapolate $\ell \approx 76$ nm for $W = 1/2$ and $\ell \approx 306$ nm for $W = 1/4$.

\fullpagefigure[1]{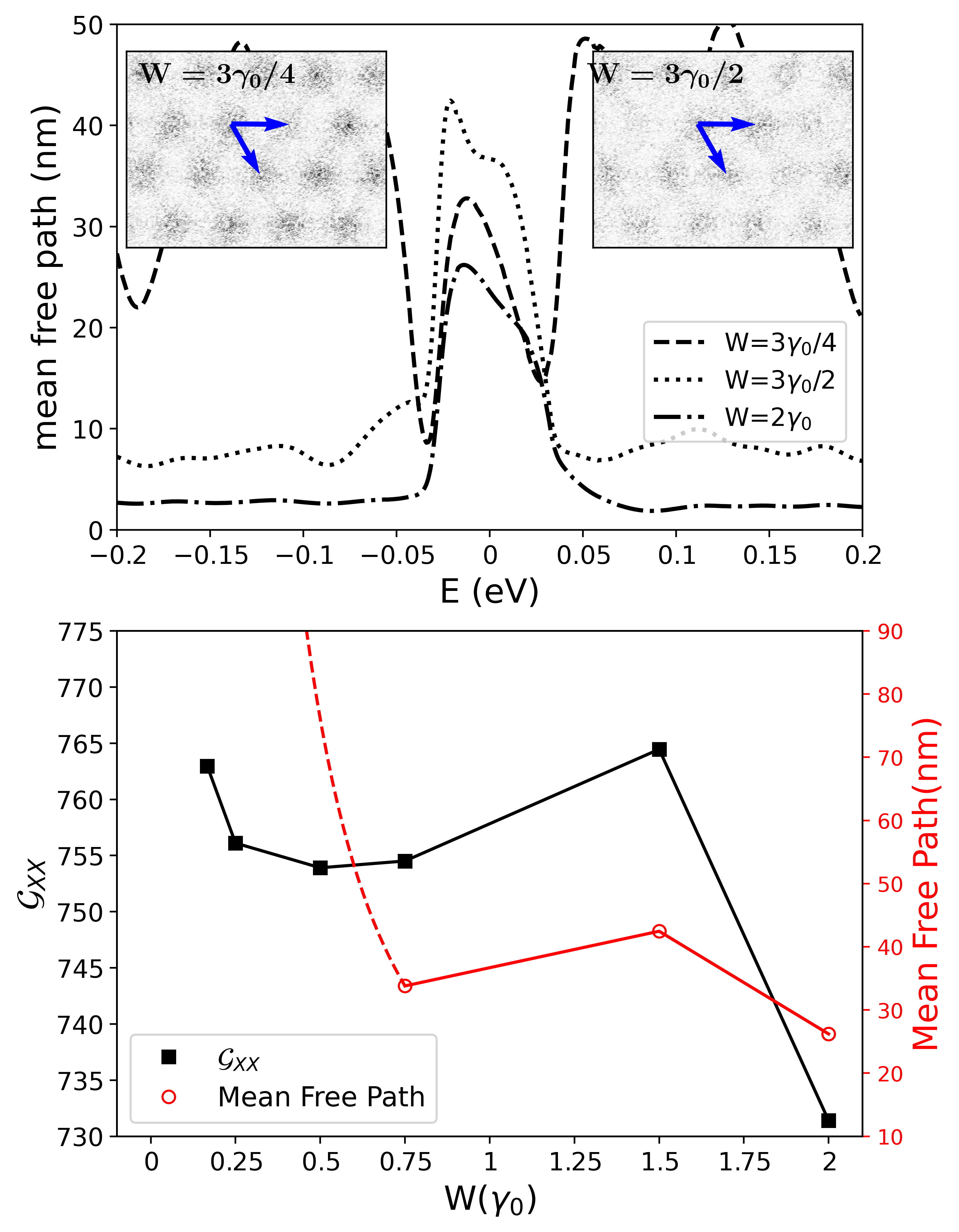}
{Top panel: mean free path for Anderson strengths of  $W=3\gamma_0/4$ (dashed line),{$W=3\gamma_0/2$} (dotted line) and $W=2\gamma_0$ (dash-dotted line). The insets show the LDoS at charge neutrality with disorders of  $W=3\gamma_0/4$ (left) and {$W=3\gamma_0/2$} (right). Bottom panel: quantum metric ${\cal G}_{xx}$ and mean free path at the flat band for different disorder strengths. {The red dashed line represents the expected mean free path computed using the Fermi golden rule for smaller disorders.}}{fig:msd-deloc-matblg}
{Here it is important to note that such disorder-induced changes in the integrated quantum metric could also have an impact on MATBLG superconductivity and the superfluid weight of the states following some recent works relating the contribution of the integrated quantum metric to the superfluid weight and the critical temperature \cite{Julku2020Feb, Torma2022}}.

\section{The graphene quasicrystal}\label{Sec: Graphene Quasicrystal}
\subsection{The quasicrystal approximants}\label{sec:Quasicrystalline-approximants}
As we saw in Sec.\ \ref{Sec: super-quasiperiodicity in tblg}, a quasicrystal has, by definition, no real-space periodicity. In Sec. \ref{sec:Quasiperiodic_potential} we faced the difficulties of studying transport with open boundary conditions. Due to computational limitations, the system sizes studied for twisted bilayer graphene do not achieve the real-space extension of the superimposed potential artificial moir\'es displayed in chapter\ \ref{chapter:superperiodicities}. There are several other strategies one could follow to mitigate these effects, such as the use of absorbing boundary conditions \cite{10.1063/1.470477,de_Castro_2023,CRPHYS_2024__25_S2_A11_0}.

In our case, with the purpose of avoiding contamination of our results from the boundary conditions, we will rely on the quasicrystal approximants. This is a common approach \cite{RevModPhys.65.213,PhysRevB.99.165430} in the quasicrystal community where the aperiodic structures are approximated for increasingly similar, longer periodicity systems. More specifically, we will approximate our graphene quasicrystal with two levels of approximation, the $31^\circ$ and $29.8^\circ$ angles that have unit cells of sizes $\sim 56\ \mathrm{nm}$ and $\sim 73\ \mathrm{nm}$ respectively.

\subsection{Structural relaxation results}

We report that the results obtained for structural relaxation in the two systems under study are negligible both from the in-plane and out-of-plane displacements. The impact of such relaxation is also negligible from an electronic point of view. This lack of relaxation was previously predicted for large angle twisted graphene bilayers \cite{Nguyen_2022} as a result of the homogeneity of its stacking where there are no long-range AA or AB zones.

As a result, we have carried all our calculations in relaxed structures for the quasicrystal approximants, but confirm that these results are independent of the structural relaxation.

\subsection{Transport in the clean case}

\begin{figure}[h!]
    \centering
    \includegraphics[width=\linewidth]{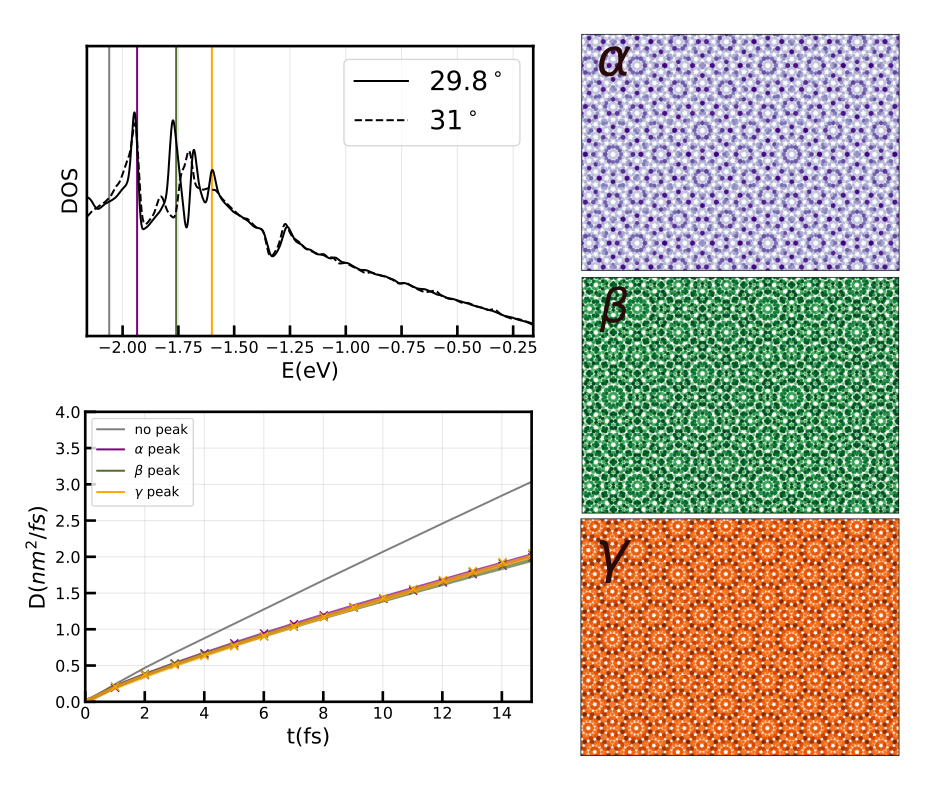}
    \caption{Top left: Density of states of $29.8^\circ$ and $31^\circ$ approximants, we have marked the quasicrystalline peaks $\alpha$ (purple), $\beta$ (green) and $\gamma$ (orange). Bottom left: Diffusion coefficient of the $29.8^\circ$ approximant as a function of time, we observe a sub-ballistic behavior at the three resonant peaks (colored lines) while outside the peaks (gray line) we recover usual ballistic transport. The right panels show the quasicrystalline real-space localization patterns for the same region in space at the three quasicrystalline points $\alpha$, $\beta$ and $\gamma$ of the $29.8^\circ$ approximant.}
    \label{fig:QCproperties}
\end{figure}

In the top left panel of Fig.\ \ref{fig:QCproperties}, we can see the density of states of the $31^\circ$ and $29.8^\circ$ approximants, where we see $29.8^\circ$ has a much closer DoS to the one expected for the quasicrystal (See the resonant states of Fig.\ \ref{fig:KM}). In $29.8^\circ$ we can clearly distinguish the resonant states associated with the $\alpha$ (purple), $\beta$ (green), and $\gamma$ (orange) marked  as the three vertical lines.

In the bottom left panel, we can see the diffusion coefficient at very short times, here we have considered the times before the electronic spreading reached the superperiodicity of the approximant, so we have only considered times $t<13\ \mathrm{fs}$. The transport within this constraint shows a clear difference between the states in and outside the resonant peaks. In the purple, green, and orange curves, we see how the transport at the resonant peaks $\alpha,\beta, \gamma$ presents sub-ballistic features, confirming the quasicrystalline nature of those states. For all three, we have predicted similar values for the exponent $\alpha$ of $\sim 0.84$ thus providing, to our knowledge, the first estimate in the literature for the sub-ballistic exponent of the dodecagonal graphene.

In the three panels on the right of Fig.\ \ref{fig:QCproperties}, we see confirmation of the quasicrystalline nature of these states, achieving for $\alpha, \beta$ and $\gamma$ quasicrystalline states that follow inflations of the Stampfli pattern, as we already introduced in Sec.\ \ref{sec:quasicrystal intro} and was shown in other results in the literature\cite{10.3389/frcrb.2024.1496179,doi:10.1126/science.aar8412}.

\subsection{Weakness against disorder}\label{Sec:QC weakness against disorder}

Disorder-induced delocalization, in a way similar to how it was introduced for magic-angle twisted bilayer graphene in Sec.\ \ref{sec:Disorder-induced-delocalization}, is a longstanding conundrum for studies based on intermetallic-alloy quasicrystals. Based on the results presented in that section, we can infer that such delocalization could be caused by the breaking of the subtle fractal ordering offered by such quasicrystals due to the addition of these new disordered states.

\begin{figure}
    \centering
    \includegraphics[width=\linewidth]{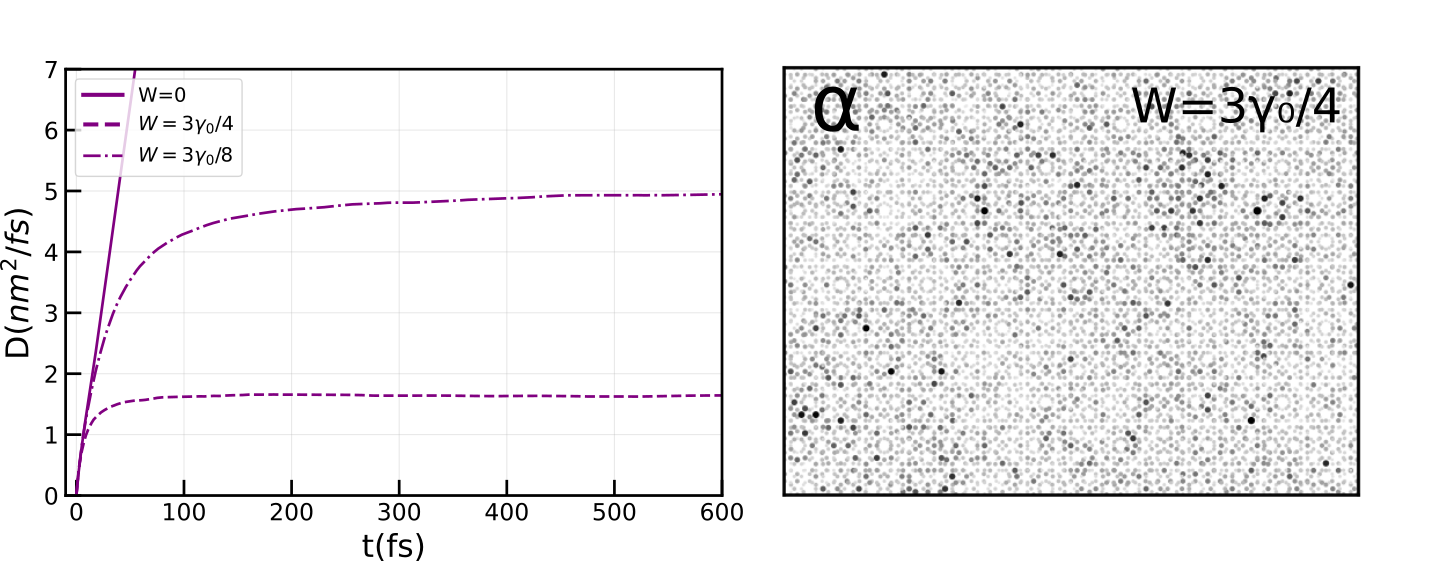}
    \caption{Left: Time evolution of the diffusion coefficient of the $\alpha$-peak states of the $29.8^\circ$ quasicrystal twisted bilayer graphene approximant against three different disorder strengths. Right: LDoS of the approximant under disorder $W=3\gamma_0/4$. We can explicitly see the lack of quasicrystalline patterns of localization present in Fig.\ \ref{fig:QCproperties}.}
    \label{fig:Disorderqc}
\end{figure}

We try here to identify this disorder-induced delocalization within our limitations for the quasicrystalline approximants. In the left panel of Fig.\ \ref{fig:Disorderqc} we can see the diffusion coefficient for disorders of $3\gamma_0/4$ and $3\gamma_0/8$ of the states present in $\alpha$ peak of dodecagonal graphene. At a specific choice of disorder, for all times, transport is always decreasing for increasing values of disorder, not showing this exotic feature, and instead following the usual Fermi golden rule. Note that we extended the calculations to smaller disorders than those studied for the magic-angle case of Sec.\ \ref{sec:Disorder-induced-delocalization}, where, as we stated, $3\gamma_0/4$ for monolayer graphene produces mobilities similar to those induced in graphene by electron-hole puddles of an hBN substrate. 

We can see the reason in the bottom panel of Fig.\ \ref{fig:Disorderqc} where the local density of states of dodecagonal graphene is displayed. We observe that the $\alpha$-peak localization pattern present in Fig.\ \ref{fig:QCproperties} has vanished for $W=3\gamma_0/4$. This is a sign of the weakness of such quasicrystalline states against disorder and predicts that the resistive features of quasicrystals can only be observed in extremely small disorders.

However, it is also important to make two considerations here regarding the limiting factors for our calculations: 

\begin{itemize}
    \item The amount of disorder that we can achieve with precision in our calculations is lower bounded due to the finite size of our systems and the presence of an electronic broadening in the electronic states. This implies the possibility that this system can present disorder-induced delocalization mechanisms at extremely weak disorders $W<25\ \mathrm{meV}$.
    \item As we saw in Fig.\ \ref{fig:QCproperties}, as a result of the use of approximants, we observe traces of the resistivity of quasicrystals in transport only at relatively short distances (below the superperiodicity). Loss of the localized fractal picture is not expected to be limited by the approximant unit cell, but to evolve continuously towards the formation of the fully $12$-fold symmetric states. Yet we remark that more studies with larger approximants are needed in order to completely discard the delocalization induced by disorder.
\end{itemize}

\section{A trilayer with quasicrystal and magic-angle}\label{Sec: trilayer}

So far, we know that: (i) in magic-angle twisted bilayer graphene (MATBLG), interlayer hybridization at small twist produces low-energy flat bands and associated transport anomalies (Section\ \ref{Sec:MATBLG_paper}), and (ii) in $\sim 30^\circ$ twisted bilayer graphene the strongest hybridization is shifted to high energies and gives rise to resonant quasicrystalline states with characteristic dodecagonal real-space patterns (Section \ \ref{Sec: Graphene Quasicrystal}). A natural next step is to combine both ingredients in a single stack and ask whether the two energy sectors remain effectively decoupled, or instead imprint measurable proximity effects on one another.

In this section, we therefore consider a trilayer hybrid heterostructure in which a MATBLG bilayer is interfaced with a graphene quasicrystal approximant (near $30^\circ$), following the large-scale tight-binding and real-time Kubo simulations.
Because the relevant phenomena live at well-separated energies (flat-band physics in the $\sim 10$-$30\ \mathrm{meV}$ window around charge neutrality versus quasicrystalline resonances in the $\sim 1.5$--$2\ \mathrm{eV}$ range) the analysis is most transparent when organized into a low-energy and a high-energy regime. We will first examine how the additional layer at $\sim 30^\circ$ affects the spectral fingerprints and transport of the flat bands, and then turn to the high-energy quasicrystalline peaks and their anomalous dynamics. 

For that, in the rest of this section, we set a $30^\circ$ angle rotation in the \textbf{bottom} and \textbf{middle} layer interface; and a $1.1^\circ$ rotation in the \textbf{middle} to \textbf{top} interface.

\subsection{Structural relaxation}

\begin{figure}
    \centering
    \includegraphics[width=\linewidth]{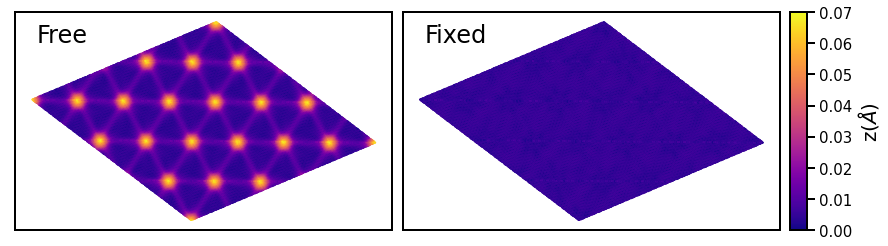}
    \caption{Middle layer of the QC+MATBLG bilayers for the $31^\circ$ approximant. In the left panel we see the effects of relaxation where the bottom layer is allowed to freely relax. In the right panel we see the same middle layer where we fix the bottom layer out-of-plane displacements to be zero.  }
    \label{fig:relaxationQC}
\end{figure}

Similarly to the previous sections, for our electronic calculation, the displacements on the $z$ axis of the bottom layer are fixed to mimic the effect of a substrate. However, we want to make a further comment here on the relaxation in the trilayer. We observed a dominance in the structural relaxation of the QC+MATBLG trilayer of the $AA$-zones in the $1.1^\circ$ twisted interface. 

In Fig.\ \ref{fig:relaxationQC} we observe the difference in the out-of-plane displacements between the relaxation of the full $29.8^\circ$ bilayer when the bottom layer is set free (left panel) and when it is fixed in the out-of-plane coordinates. In the free case, we observe the impact of the MA interface (middle and top layers), where the out-of-plane displacements in the $AA$ regions of this interface tend to increase the interlayer distance between the middle and top layers, while the interlayer distance between the two bottom layers remains almost constant. This is maintained in the bottom layer, where these regions present a displacement that keeps the interlayer constant approximately close.

In contrast, when we fix the out-of-plane displacements of the bottom layer to zero, the $30^\circ$ interface also forces the flatness of the middle layer, as we can see from the right-hand panel of Fig.\ \ref{fig:relaxationQC} where the vertical displacements can be considered negligible.

However, we found no significant differences in the electronic structure or transport in the initial calculations and decided to choose the Fixed version because of experimental relevance.

\begin{figure}[h]
    \centering
    \includegraphics[width=\linewidth]{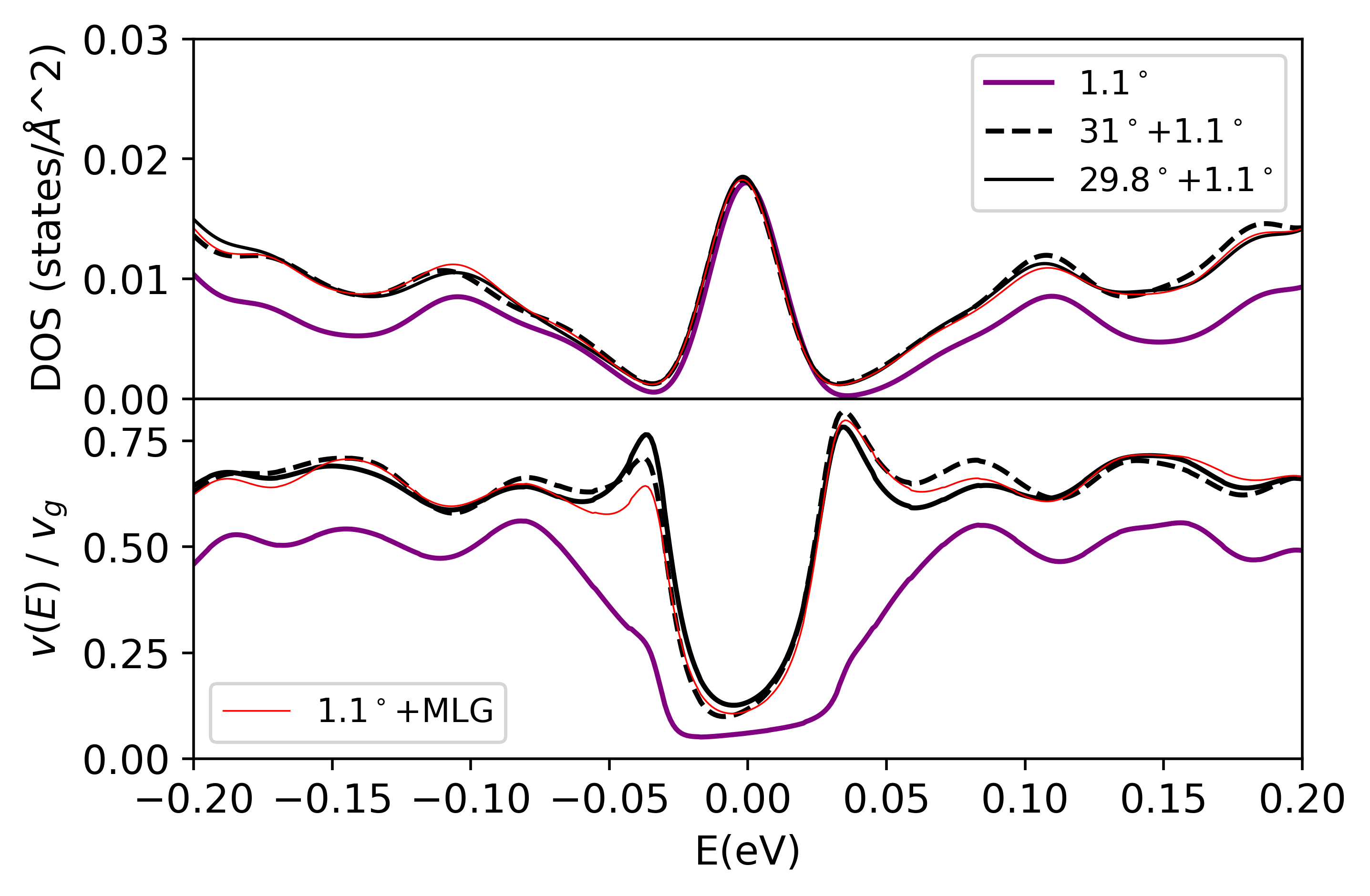}
    \caption{Density of states (top panel) and Fermi velocities (bottom panel) of MATBLG (purple solid line, replicated from Fig.\ \ref{fig:electronic-struct-matblg}) and the two trilayer systems (black solid and dashed lines). The averaged Fermi velocity of the trilayer system is depicted by the red line, considering a trilayer system where the MATBLG and the quasicrystalline layer are decoupled (see main text).}
    \label{fig:fig2QC}
\end{figure}

\subsection{The low-energy regime}\label{Sec:QC_MATBLG_low_energy}

We now turn to the low-energy window, where the electronic structure of magic-angle twisted bilayer graphene (and in our trilayer) is dominated by the moir\'e flat bands discussed in Sec.~\ref{Sec:MATBLG_paper}.

\subsubsection{DoS and averaged Fermi velocity.}
Fig.\ \ref{fig:fig2QC} compares the DoS of MATBLG (reproduced from Fig.\ \ref{fig:electronic-struct-matblg}) in purple; with that of the two trilayers built from the $31^\circ$ and $29.8^\circ$ quasicrystal approximants in dashed and solid black lines, respectively.
Here, within the flat-band window $|E| \lesssim 25\,\mathrm{meV}$, the DoS remains essentially unchanged when the quasicrystalline layer is added: the low-energy peak associated with the moir\'e flat bands remains nearly identical in the bilayer and trilayer stacks.
At somewhat larger energies, a modest increase in DoS appears in the trilayers, consistent with the emergence of additional dispersive channels provided by the extra graphene layer \cite{22s2-vpm5}.

The bottom panel of Fig.\ \ref{fig:fig2QC} shows the corresponding energy-dependent Fermi velocity, extracted from the ballistic scaling of the diffusion coefficient.
Although the DoS suggests that the flat bands themselves remain intact in the clean limit, the averaged Fermi velocity can increase substantially in the trilayer, reaching factors of $5\times$ in the most severe cases.
Importantly, this increase does not necessarily imply a genuine steepening of the MATBLG flat bands: since the velocity is averaged over the full Fermi surface, it can reflect parallel transport through a largely decoupled quasicrystalline layer with a graphene-like velocity scale.
To disentangle these effects we introduce a ``decoupled'' control calculation in which the interlayer hoppings between the MATBLG bilayer and the third layer are removed, mimicking an infinite separation between the layers. The resulting curve (red in Fig.~\ref{fig:fig2QC}) shows that, at low energies, both the DoS and the averaged velocity remain essentially unchanged relative to the coupled trilayer.
This supports the interpretation that the low-energy flat-band manifold is only weakly hybridized with the $\sim 30^\circ$ layer in the pristine case, and that the apparent velocity enhancement is dominated by the additional conducting channel rather than a strong modification of the MATBLG states.

\subsubsection{Layer-projected spectral weight and LDoS at charge neutrality.}

\begin{figure}
    \centering
    \includegraphics[width=\linewidth]{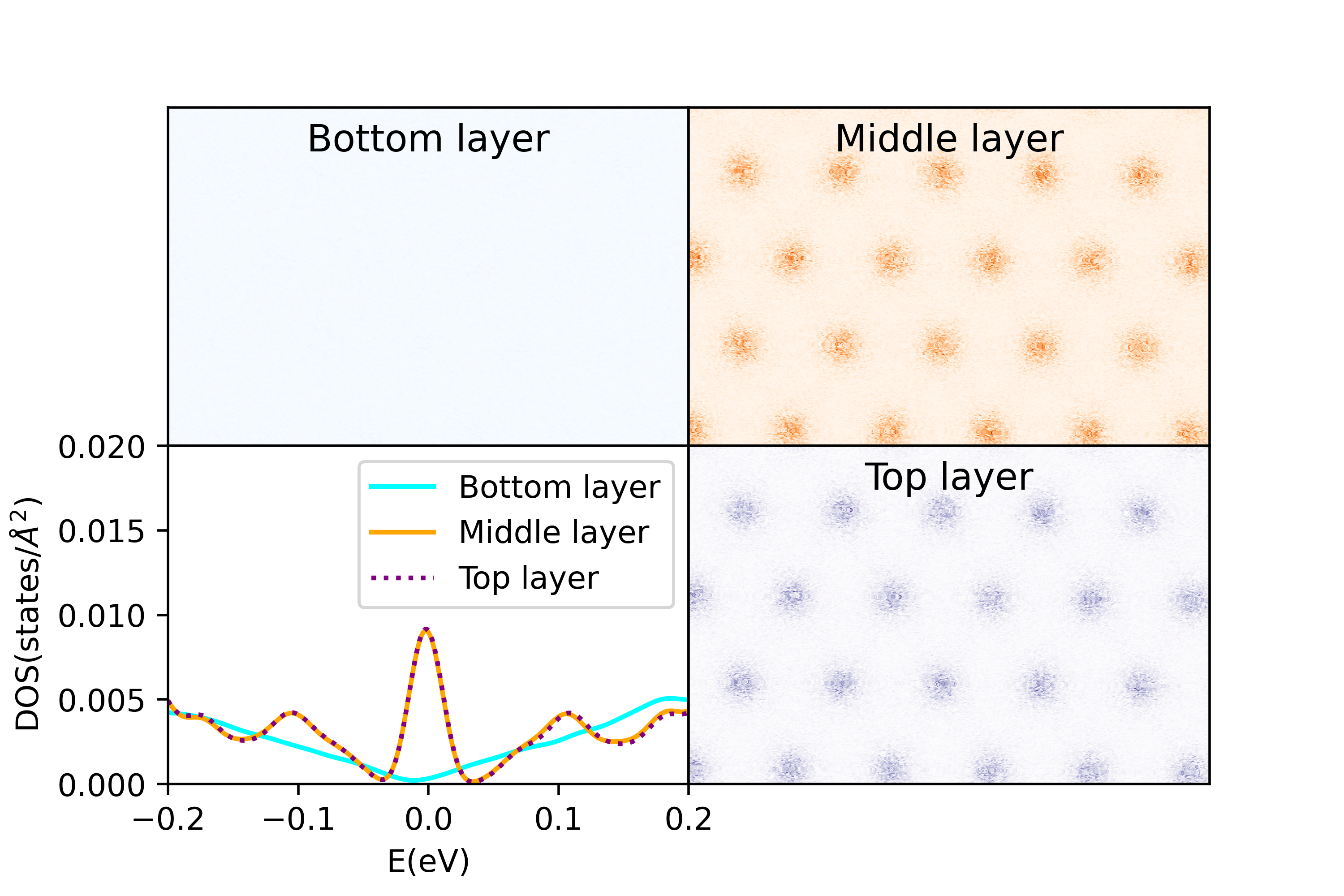}
    \caption{Layer-projected DoS (bottom left panel) and LDoS of each of the layers for the $31^\circ$+$1.1^\circ$ trilayer hybrid system at charge neutrality ($E=0$). The moir\'e pattern is visible in the two $1.1^\circ$-rotated layers, while the extra quasicrystalline layer does not display any signal of the flat band.}
    \label{fig:fig3QC}
\end{figure}

The weak hybridization picture is further corroborated by the layer-resolved quantities in Fig.~\ref{fig:fig3QC}.
The layer-projected DoS demonstrates that the flat-band peak is carried almost entirely by the two $1.1^\circ$-twisted layers (the MATBLG bilayer), whereas the $\sim 30^\circ$ layer contributes a spectrum that resembles that of (nearly) monolayer graphene. Consistently, the LDoS maps reveal the characteristic moir\'e periodic modulation in the MATBLG layers, with increased weight in the AA overlap regions as discussed in Sec.\ \ref{Sec:MATBLG_intro} and \ \ref{Sec:MATBLG_paper}, while the quasicrystalline layer shows no imprint of the moir\'e superlattice at low energy.
Altogether, Fig.~\ref{fig:fig3QC} indicates that the third layer behaves largely as a spectator in the clean low-energy regime: it provides additional (graphene-like) states, but does not absorb spectral weight from the MATBLG flat bands.

\subsubsection{Disorder and the suppression of disorder-induced delocalization.}

\begin{figure}
    \centering
    \includegraphics[width=\linewidth]{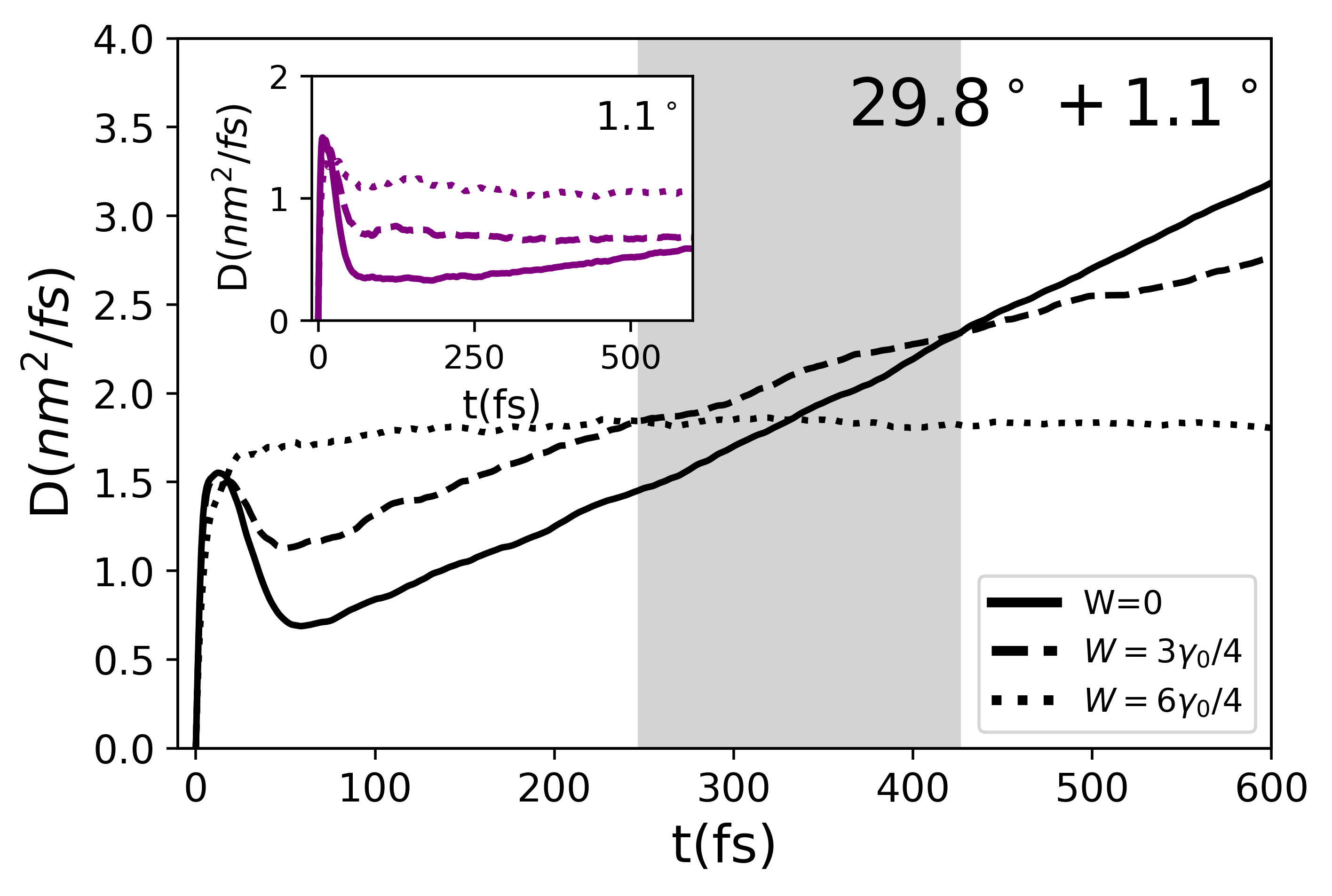}
    \caption{Time-dependent diffusion coefficient in the flat band for different disorder strengths, for MATBLG (inset) and the trilayer with the $29.8^\circ$ approximant (main panel).
}
    \label{fig:fig4QC}
\end{figure}
We now test whether this apparent resilience extends beyond spectral properties by introducing Anderson disorder, as in Sec.~\ref{sec:Disorder-induced-delocalization}. We consider the same disorder strengths as in that section $W=3\gamma_0/4$ and $W=6\gamma_0/4$ and compute the time-dependent diffusion coefficient at the center of the flat band ($E=0$).
For MATBLG alone, the inset of Fig.~\ref{fig:fig4QC} (adapted from Fig.\ \ref{fig:disorder-deloc-diff}) presents the anomalous trend previously reported and discussed in Sec.~\ref{sec:Disorder-induced-delocalization}: at long times, increasing disorder can lead to an increase of the diffusion coefficient.

Interestingly, this behavior is reversed in the trilayer stack.
As shown in the main panel of Fig.~\ref{fig:fig4QC}, once the quasicrystalline graphene layer is added, the diffusion coefficient follows a conventional trend: a larger disorder systematically reduces $D(E=0,t)$ at long times, consistent with standard disorder-limited transport. This occurs despite the fact that, in the clean limit, the low-energy states remain strongly projected onto the MATBLG layers. We attribute this difference to the new electronic channels enabled by the presence of the extra layer twisted $30^\circ$ that enables the scattering of states from the MA interface in the presence of disorder.

We want to remark that at very short times this disorder-induced delocalization (see left white region of Fig. \ref{fig:fig4QC}) can be recovered and even increased for shorter times. For other times, the scaling from Sec.\ \ref{sec:Disorder-induced-delocalization}, where for increasing disorder we find delocalization $\to$ localization, can be recovered (gray region of Fig.\ \ref{fig:fig4QC}). This predicts that for a sufficiently short sample these effects could be observed even in our trilayer.

 \subsection{High-energy states}\label{Sec:QC_MATBLG_high_energy}
\begin{figure}[h]
    \centering
    \includegraphics[width=\linewidth]{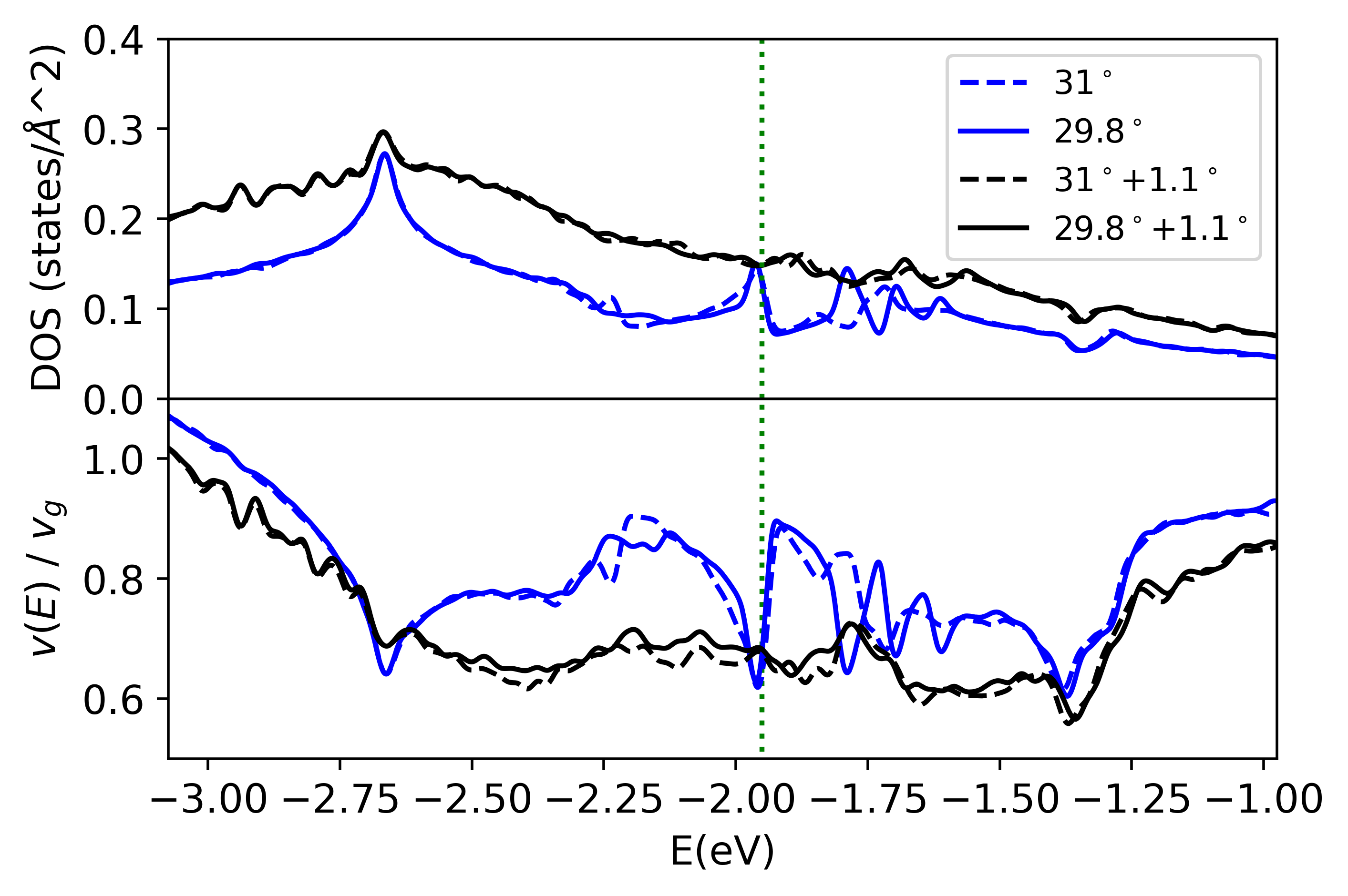}
    \caption{DoS (top panel) and Fermi velocity (bottom panel) in the high-energy quasicrystalline-state regime. The vertical green dotted line marks the energy of the $\alpha$ quasicrystalline peak ($E=-1.95$ eV) \cite{PhysRevB.99.165430}.}
    \label{fig:fig5QC}
\end{figure}

We now turn to the high-energy window where $30^\circ$-twisted bilayer graphene (tBLG) quasicrystals display their most distinctive electronic fingerprints in the shape of a set of resonant peaks in the $[-1.5,2]\ \mathrm{eV}$ energy range. Fig.\ \ref{fig:fig5QC} summarizes the main impact of the proximity effect with the $1.1^\circ$ layer in this energy range. For the quasicrystalline bilayer approximants, the DoS shows prominent resonant structures, and we single out the $\alpha$ peak at $E=-1.95~\mathrm{eV}$ (vertical line). Remarkably, when the magic-angle bilayer is brought into proximity, forming the $29.8^\circ + 1.1^\circ$ trilayer, these resonant fingerprints are strongly suppressed: the DoS becomes significantly smoother and the corresponding Fermi velocity profile turns comparatively featureless.
Instead of resonant peaks, the combination between the quasicrystal and the $1.1^\circ$-rotated layer induces a systematic reduction in velocity throughout the high-energy range of interest, consistent with an enhanced hybridization background but a strong degradation of the quasicrystalline resonant response.

 \begin{figure}[h]
     \centering
     \includegraphics[width=\linewidth]{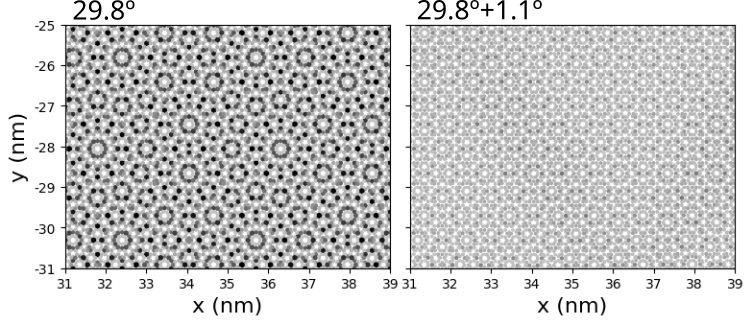}
     \caption{Local density of states of $29.8^\circ$-twisted bilayer graphene (left) and the two bottom layers of the $29.8^\circ + 1.1^\circ$ trilayer at the $\alpha$ peak (vertical dotted line in Fig.\ \ref{fig:fig5QC}). The fractal localization pattern is destroyed by the addition of the extra layer and replaced with a homogeneously distributed local density of states.}
     \label{fig:figldosQC}
 \end{figure}

The same conclusion emerges from the probes in real space. In the pristine quasicrystalline bilayer, the LDoS evaluated at the $\alpha$ peak displays a highly inhomogeneous, self-similar localization landscape (left panel of Fig.~\ref{fig:figldosQC}), reflecting the emergence of a dodecagonal order that is incompatible with translational invariance, as we previously saw in Fig.\ \ref{fig:QCproperties}. In sharp contrast, evaluating the LDoS on the same atomic coordinates for the bottom and middle layers (quasicrystalline interface) within the trilayer, we observe that this fractal localization pattern is destroyed and replaced by a much more homogeneous distribution.
This highlights a key qualitative asymmetry of the proximity effects explored in this section: while low-energy flat-band physics can remain comparatively robust to adding a quasicrystalline layer in the clean limit, the high-energy quasicrystalline states are instead fragile to the presence of a magic-angle layer.

 \begin{figure}
     \centering
     \includegraphics[width=\linewidth]{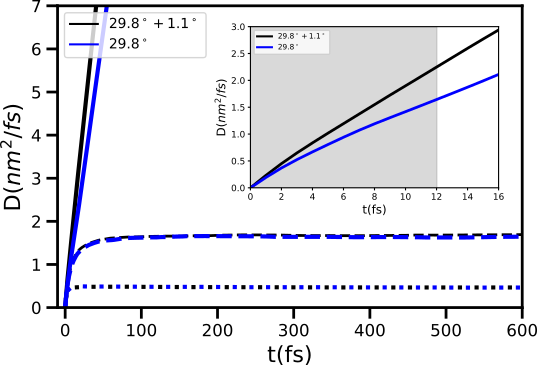}
     \caption{Time-dependent diffusion coefficient of the quasicrystalline approximant and the trilayer system at the energy of the $\alpha$-peak (vertical dotted line in Fig.\ \ref{fig:fig5QC}) for disorder strengths of $W=0$ (solid), $W=3\gamma_0/4$ (dashed) and $W=6\gamma_0/4$ (dotted). Inset: zoom of the short-time diffusion coefficient, indicating the sub-ballistic behavior of the quasicrystal (blue line). The shadowed region corresponds to times where the mean squared displacement has not reached the periodicity of the approximant.}
     \label{fig:Fig7QC}
 \end{figure}

Transport simulations reinforce this picture. Fig.\ \ref{fig:Fig7QC} compares the time-dependent diffusion coefficient $D(t)$ at the $\alpha$-peak energy for the quasicrystalline approximant and the trilayer. In the inset, we can see a comparison between the short-time behavior of the quasicrystalline bilayer (blue curve, extracted from Fig.\ \ref{fig:QCproperties}) and trilayer (black). Surprisingly, the quasicrystalline resistivity was washed out even in this pristine case, where the states are now conventional ballistic states. 

Introducing Anderson disorder drives both systems into standard diffusive behavior in this high-energy regime: increasing the disorder strength monotonically reduces $D(t)$ and shortens the effective mean free path. Here, with the addition of disorder, the transport in the bilayer and the trilayer cannot be distinguished from each other.

The combination of these results supports a central idea: The broadening of the states induced by the sole proximity of the $1.1^\circ$ layer is strong enough to mix the resonant states, whose fragility to disorder we spotted in Sec.\ \ \ref{Sec:QC weakness against disorder}. Creating a region of the space with a reduced Fermi velocity as a result of the impact of the resonant states of the pristine bilayer but with no signature of quasicrystalline features itself. With the addition of disorder, the broadening induced by the extra layer becomes negligible compared to the broadening induced by Anderson disorder, and the transport becomes indistinguishable between the tri- and bilayer.

\chapter{Magnetism in Twisted bilayer Graphene}\label{chapter:magnetism}

In Sec.\ \ref{Sec:MATBLG_intro} we saw that one of the most important aspects of MATBLG comes from the discovery of highly correlated phases. There, the emergence of narrow moir\'e bands in MATBLG places the system in a regime where Coulomb interactions compete on the same footing with kinetic energy, allowing for a complex phase diagram in the presence of exotic phenomena such as the correlated insulator or the superconductor phases. \cite{Cao2018,Cao2018CI,Balents2020}.

These observations motivate a microscopic characterization of magnetic ordering tendencies in moir\'e flat-band systems, and provide a concrete arena where spin, valley, and sublattice degrees of freedom intertwine with topology. As we discussed previously in Sec.\ \ref{sec: Hubbard model} a simple study of these magnetic properties can be understood as a first approach to the electron-electron correlations.

This chapter provides an exploratory extension of the thesis from transport and localization toward interaction-driven physics in magic-angle twisted bilayer graphene. In particular, it focuses on magnetic and topological phases reported experimentally in this platform and on the microscopic modeling strategies commonly used to describe them. The purpose of Chapter\ \ref{chapter:magnetism} is not to present a complete theory of correlated magnetism, but rather to establish the first steps toward incorporating electron-electron interactions into the real-space framework developed in the previous chapters.
The motivation is to eventually jointly address the impact of disorder and Coulomb interaction (limited to a treatment in the mean-field approximation) on the delocalization-localization effects we have found in Chapter\ \ref{Chapter: Transport in twisted multilayer graphene}. Indeed, if the measured cleanest samples of twisted bilayer graphene at magic angles clearly exhibit highly exotic physics that clearly demand to go beyond the mean field, the physics of lower-quality samples (presence of twist angle disorder, electron-hole puddles, proximity effect,..) might be partly captured by a mean-field approach due to the disorder-induced broadening of flat bands, and thus reduction of local electron-electron interaction strength. Additionally, the presence of fragile magnetic states in part of the phase diagram has also been the focus of intense studies, whose origin remains debated.

In that sense, this chapter serves as a bridge between the single-particle transport perspective of the previous one and the broader problem of how disorder, flat-band localization, and many-body instabilities may interplay in moir\'e systems. Because this part of the project remains computationally challenging, the discussion is necessarily more programmatic and methodological, but it clarifies both the physical motivation and the limitations that define the next stage of the research.


\section{A landscape of exotic phases}
\label{Sec:Correlations bibliographic intro}
This section reviews the advances in the study of correlations for magic-angle twisted bilayer graphene achieved to date, as well as their main results concerning the onset of magnetic ordering.

\subsection{Experimental phenomenology: from anomalous Hall response to quantized Chern insulators}
\label{Sec:MATBG_expt_pheno}

In the first experimental reports on MATBLG, Cao et al. already revealed a close proximity between the insulating and superconducting ground states as a function of carrier density \cite{Cao2018CI,Cao2018}. Shortly thereafter, transport experiments uncovered clear signatures of magnetic ordering in parts of the moir\'e phase diagram. A particularly direct indicator is the appearance of a large anomalous Hall response accompanied by hysteresis, consistent with a ferromagnetic state that breaks time-reversal symmetry. In 2019 Sharpe et al. reported emergent ferromagnetism near the three-quarter filling of a moir\'e miniband, together with a giant anomalous Hall effect and current-driven switching of the magnetic state \cite{doi:10.1126/science.aaw3780}. Although the Hall response in that work was not quantized, it suggested that MATBLG can realize an incipient Chern-insulating regime when interactions polarize the relevant internal degrees of freedom.

A decisive step towards topological magnetic order came with the observation of a robust quantized anomalous Hall (QAH) effect at (effectively) zero magnetic field. In devices where MATBLG is crystallographically aligned with hexagonal boron nitride (hBN), Serlin et al. observed a Hall resistance quantized consistent with a Chern insulator around 3/4 filling of the flat band and accompanied by ferromagnetic order \cite{doi:10.1126/science.aay5533}. Importantly, the QAH response was shown to be electrically switchable with extremely small currents, emphasizing the unusually soft magnetic energy scales typical of moir\'e flat-band systems.

Subsequent experiments have broadened this picture, revealing that correlated topological phases in MATBLG are not limited to a single filling factor or a single device architecture. Gate-tunable Chern insulating states with different Chern numbers have been reported near several integer fillings, often in close competition with superconducting domes or other correlated insulators \cite{0256b11272db4e4fb10d4e4bb162df16}. Thermodynamic and transport probes further indicate that flavor symmetry breaking is a recurring motif at integer fillings, and that a ``Hund-like'' energetic hierarchy between spin/valley flavors can influence the gap formation and the anomalous Hall response \cite{Park2021}. Scanning probe and compressibility measurements have also revealed an unexpectedly rich sequence of incompressible states, including phases interpreted as correlated Chern insulators and, at finite fields, fractional Chern insulators emerging from moir\'e Chern bands \cite{Pierce2021,Xie2021}.

Taken together, these observations suggest that magnetic order in MATBLG is often entangled with band topology and with the multi-component (spin/valley) structure of the moir\'e bands. The experimental phase diagram is therefore best viewed not as a single ordered state but as a landscape of competing instabilities whose energetic separation can be comparable to disorder, strain, screening, and small symmetry-breaking fields. The question of how disorder influences and impacts this landscape, where subtle changes in the occupations can substantially reshape the observed ordering tendencies remains an open question, and even if current techniques don't allow us to address this problem in all its complexity, it motivates first studies on the effects of disorder on the description of such magnetic states.

\subsection{A mean-field approach to correlations}\label{Sec:MAgnetism in TBLG_theory}

From a theoretical point of view, the natural starting point for most works is the continuum description of moir\'e bands using a Bistritzer-MacDonald (BM) model that can be considered a generalization of the one presented in \eqref{eq:Bistritzer} \cite{Wu_2018,doi:10.1073/pnas.1108174108}. While the BM framework is fundamentally a single-particle theory, its ability to capture the formation of extremely narrow bands makes it an ideal platform to investigate correlation effects either by adding interactions at the mean-field level or by constructing effective low-energy lattice Hamiltonians where many-body methods can be applied.

One of the most direct strategies is to incorporate Coulomb interactions in the continuum model through self-consistent Hartree or Hartree-Fock (HF) approximations. These approaches exploit the fact that the moir\'e wave functions have strong real-space modulation, which enhances exchange and can naturally drive flavor symmetry breaking (spin/valley polarization, intervalley coherence, etc.). In this spirit, continuum HF calculations have been used to account for correlated insulating behavior at commensurate fillings and to map out competing broken-symmetry states as a function of filling and interaction strength. First, Liu et al. \cite{Liu_2021} found an antiferromagnetic phase with a bandgap of $\sim 1-5\  \mathrm{meV}$, qualitatively in agreement with the measured gap of $\sim 0.86\ \mathrm{meV}$. In \cite{PhysRevB.102.035136,Bultinck_2020} this gap is predicted to be on the order of meV. The key message emerging from these studies is that in MATBLG, the energetic hierarchy between different ordered states can be extremely small, so that weak perturbations (substrate alignment, strain, screening environment, displacement fields) may play a decisive role in selecting the experimentally realized phase.

A real-space perspective becomes particularly valuable when discussing magnetic textures and spatial inhomogeneity of the order parameter within the moir\'e unit cell. The methods for these self-consistent algorithms are similar to the ones described in Sec.\ \ref{sec: Hubbard model} which differ in the computation of the local density of states. In this direction, Vahedi et al. made use of Green's function recursion to simulate the real-space HF (as the one presented in Sec.\ \ref{sec: Hubbard model}) both with mean-field theory (MFT) and with dynamical mean-field theory (DMFT) \cite{RevModPhys.68.13}. With that, they could investigate magnetic instabilities close to charge neutrality \cite{10.21468/SciPostPhys.11.4.083}. These studies show how different approximations (HF, DMFT) support the formation of magnetic trends. The simplest approximations (as the one generated by a HF) proved to result in the same antiferromagnetic phases at zero filling of the band-gap, while differing on quantitative parametric values in the interactions for achieving similar magnetizations.

In that line, Vidarte et al. \cite{Vidarte_2025} used a similar approach with a combination of Lanczos-like \cite{RHaydock_1972,RHaydock_1975,Lanczosbook} real-space spectral calculations and HF, finding both the presence of anti-ferromagnetism at zero filling of the flat band and a ferromagnetic phase at 3/4 filling of the flat band, potentially inducing Chern phases similar to the ones predicted by Serlin et al. 

At zero filling Vidarte et al found antiferromagnetism with local values for the gap in the AA zones in the order of magnitude of $0.1-1\ \mathrm{meV}$ for a Hubbard parameter ratio $U/V_{pp\pi}^0=1$ and at $3/4$ filling, they found ferromagnetic gaps up to $36\ \mathrm{meV}$.

\section{Mean-field simulations}\label{sec:mean-field-simulations}
In the following sections, we follow the steps of Vahedi and Vidarte \cite{Vidarte_2025} with our KPM approach and target a mean-field convergence by making use of the algorithm described in Sec.\ \ref{sec: Hubbard model}, with the objective of studying such magnetic phases under disorder. To do so, we will make use of our KPM approach to compute local densities of states as shown in \eqref{eq:LDOS}.

\subsection{The structural relaxation}

The original study by Vidarte et al. showed results for a relaxed and an unrelaxed structure. In a similar fashion, for relaxed cases, we have considered both the fixed and the free bottom layer with the computational details given in Sec.\ \ref{sec:structural-relaxation-magic-angle}.

\subsection{Tight-binding model}

In order to obtain results as close as possible, we use the tight-binding model proposed in \cite{Vidarte_2025, 10.3389/frcrb.2024.1496179}, defined as 

\begin{align}\label{eq:lewenkopf_model}
t ({\vec r}_{nm}) &= \cos^2 (\phi_{nm}) V_{pp\sigma} (r_{nm}) \nonumber \\
&+ \sin^2 (\phi_{nm}) V_{pp\pi} (r_{nm}),
\end{align}
with
\begin{align}
V_{pp\pi} (r_{nm}) &= V_{pp\pi}^0 \exp \left[-\frac{R-a_0}{r_0} \right] \nonumber \\
V_{pp\sigma} (r_{nm}) &= V_{pp\sigma}^0 \exp \left[ -\frac{R-d_0}{r_0}  \right] 
\end{align}

This can be understood as a version of \eqref{eq:charlier_model} with slightly modified exponential decay. The parameterization of the model is reported in Table \ref{tab:params_lewen}. In Fig.\ \ref{fig:overlapdecay} (See section \ref{Sec:Tight-Binding-graphene}), we showed that this model and the one used in \eqref{eq:charlier_model} give very similar results. 

\begin{table}[]
    \centering
    \begin{tabular}{c c c c c}
        \hline
        parameter && Description && Value\\
        \hline
        \hline
         $V^0_{pp\pi}$&&  $\pi$-bond $p_z$ transfer integral at equilibrium &&  $-2.7\ \mathrm{eV}$\\
         $V^0_{pp\sigma}$&&  $\sigma$-bond $p_z$ transfer integral at equilibrium  &&  $0.48\ \mathrm{eV}$\\
         $a_0$&& in-plane nearest-neighbor average distance && $1.42\ $\AA\\
         $d_0$&& out-of-plane nearest-neighbor average distance &&  $3.35\ $\AA \\
         $d_0$&& decay length && 0.319$a_0$ \\
         & 
         
    \end{tabular}
    \caption{Choice of parameters for\ \eqref{eq:lewenkopf_model}}
    \label{tab:params_lewen}
\end{table}

This model precisely reproduces the flat band and the electronic localization in the AA zones of the MATBLG shown in Fig.\ \ref{fig:electronic-struct-matblg}. 

\subsection{Computational details}

To date, our KPM approach has not been able to reproduce the results shown by other real-space studies that report magnetizations in the range of 0.1-10 meV for a wide variety of values of the Hubbard parameter $U$ \cite{10.21468/SciPostPhys.11.4.083,Vidarte_2025}. For this reason, we report the computational details of the calculations carried out so far and provide a careful discussion of the possible origins of these discrepancies. Following the steps of the two articles cited, we have considered two different twist angles $\theta$ for our calculations, $1.47^\circ$ and $1.08^\circ$ and, for each of them, two different fillings of the flat band $\nu=0,\nu=3/4$, with expected antiferromagnetic and ferromagnetic states, respectively. 
In order to achieve sufficient precision, we have chosen $9\times 9$ tilings of the unit cell of $1.47^\circ$, formally equivalent to the $9\times 9 $ $k$-point mesh grid considered in \cite{10.21468/SciPostPhys.11.4.083}. For the twist angle $\theta\approx 1.08^\circ$ we were unable to achieve the precision required for  the LDoS calculation of such a large lattice in a reasonable amount of computational time, so the tilings are as large as $4\times 4$ times the unit cell. For the energy broadenings we have taken a constant value of $\eta=20\ \mathrm{meV}$, narrower than the $25\ \mathrm{meV}$ regularization parameter considered in \cite{Vidarte_2025} for the Lanczos recursion. Meanwhile, the number of moments at each iteration changes in order to keep this broadening constant according to \eqref{eq:Jackson-kernel-broadening}.
The energy step considered for the integral is $4\cdot 10 ^{-5}\ \mathrm{eV}$, orders of magnitude below the minimal gap considered for the two previous works.

\subsection{Initial conditions}
In order to conduct the self-consistent procedure exposed in Sec.\ \ref{sec: Hubbard model}, we need to make a choice of the initial conditions for step 2. Here, we have considered three possible initial conditions for magnetization.
\begin{itemize}
    \item \textbf{Random} initial conditions: we initialize all magnetizations with a uniform distribution of random values in the range $[-1,1]$.
    
    \item \textbf{Antiferromagnetic} initial conditions: We induce an antiferromagnetic phase localized in the AA zones, where the magnetization is set to $m_i=\sigma\exp\left(-|\mathbf{r}_i-\mathbf{r}_{AA}|^2/l_{dec}^2\right)$ where $l_{dec}=4a_{cc}$ is the decay length and $\sigma=\pm 1$ with positive (negative) sign assigned to the A (B) sublattice. 

    \item \textbf{Ferromagnetic} initial conditions: We initialize all lattice sites to a finite magnetization $m_i\in[-1,1]$.
\end{itemize}

After setting the magnetizations, the initial occupations are computed as
\begin{equation}
\begin{array}{ccc}
    \langle n_{i\uparrow}\rangle&=&(1+m_i)/2+\nu/N\\
    
    \langle n_{i\downarrow}\rangle&=&(1-m_i)/2+\nu/N ,
\end{array}
\end{equation}
with $N$ the number of lattice sites in the unit cell.

\subsection{Results}\label{Sec: results mf+hf}

\begin{figure}[h]
    \centering
    \includegraphics[width=\linewidth]{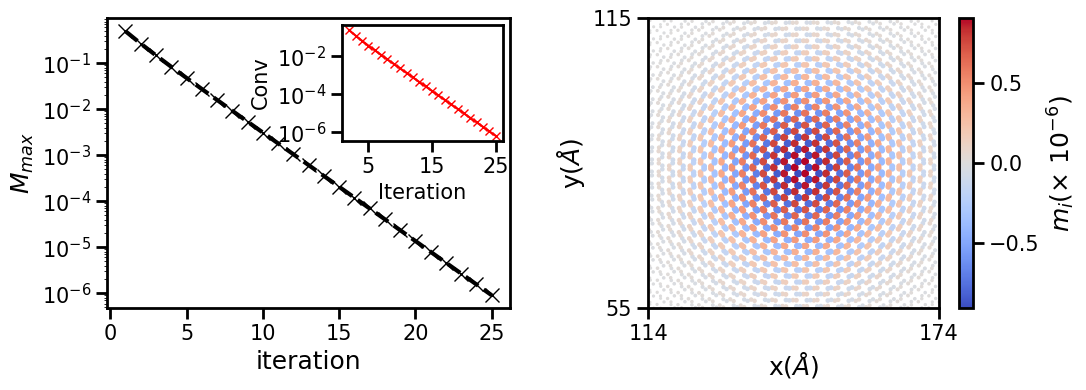}
    \caption{Convergence behavior for $\theta\approx 1.47^\circ$ and $U/V_{pp\pi}^0=2$. Left: Evolution of the maximum magnetization with the number of iterations. We see the magnetization reach values comparable with the convergence threshold. Inset of the left panel shows the evolution of the convergence parameter. Right: Real-space distribution of the magnetization from \eqref{eq:mag_deffinition}. The AA zone is located in the center of the plot. We observe the qualitative distribution is similar to that of \cite{10.21468/SciPostPhys.11.4.083,Vidarte_2025} but with a reduced absolute value for the magnetization.}
    \label{fig:not_convergence}
\end{figure}

Several $U$ values have been reported in both \cite{10.21468/SciPostPhys.11.4.083} and \cite{Vidarte_2025}. The configurations are similar, both in the potential and in geometry, having the second case also considered relaxed lattices. Both articles first discuss the $1.47^\circ$ angle as a simple case and then focus on the magic-angle. We have tried to run convergences in both $1.47^\circ$ and $1.08^\circ$ where, at the time, we have not achieved the proper convergence in either of the cases. In general, we see both the convergence parameter and magnetization at all lattice sites decrease as the convergence steps increase. In Fig.\ \ref{fig:not_convergence} we can see an example of convergence behavior where the magnetizations at each site reach values smaller than the tolerance in our convergence criterion. We report here only the results related to $1.47^\circ$ because of the simplicity in their treatment, together with the dimensionless magnetizations \begin{equation}\label{eq:mag_deffinition}
    m_i=\langle n_\uparrow\rangle-\langle n_\downarrow\rangle
\end{equation}

Overall, both articles show good qualitative agreement in their results. However, we can spot some differences between them. In \cite{10.21468/SciPostPhys.11.4.083}, they report a critical value for $U/V_{pp\sigma}$ of $1$, while in \cite{Vidarte_2025} the system clearly shows magnetization on the order of $M_{max}\approx 10^{-4}$, showing a band-gap in the AA dimer sites of $~31\ \mathrm{meV}$. A smaller gap between spin channels is reported in the AA dimer site in \cite{10.21468/SciPostPhys.11.4.083} in the case of $1.08^\circ$ with a reported larger magnetization $M_{max}\sim 2.5\cdot 10^{-2}$. 

The combination of these results suggests that the critical energy splitting between states could be highly dependent on the simulation method. We consider here some plausible reasons for the failure to achieve satisfactory convergence and future routes to achieve precise results:

\begin{itemize}
    \item \textbf{Effect of the broadening}. The precision of KPM is known to be highly dependent on its broadening \cite{FAN20211}. The broadening within our computational capabilities, as small as $20\ \mathrm{meV}$, could not have been precise enough to achieve satisfactory resolution of such small differences in spin channel occupations. In addition, the Jackson kernel broadening we use does not generate a uniform broadening along the whole unit cell, and it is possible that the effective broadenings in the flat band are smaller than what we report here given the electron-hole asymmetric nature of the problem. To mitigate that, we have also considered the Lorentz kernel \cite{FAN20211} with similar results for both kernels.

    \item \textbf{Numerical error}. We acknowledge that the techniques involved in the proposed algorithm are highly sensitive to numerical precision and the coordination of two different self-consistent methods. Any uncontrolled numerical instability during the calculation could be a source of randomness large enough to remove the spin imbalances between channels. 

    \item \textbf{Finite-size effects}. The feasibility of this work depends on finding a good compromise between computational time and accuracy. For that, we have considered grids up to $9\times 9$ replications in real space. This number is chosen to consider a similar number of bands as \cite{10.21468/SciPostPhys.11.4.083} while \cite{Vidarte_2025} does not report the exact real-space tiling of their calculations. We have compared our densities of states in the unperturbed case and considered that it converged sufficiently compared to $16\times 16$ tilings, but the cumulative error during the convergence could be too large to achieve correct results.
    
    \item \textbf{The energy grid}. To maintain reasonable amounts of computational time, we have chosen energy grids spaced $\sim 2\cdot 10^{-5}$. Small differences in the exact point of the Fermi level could lead to a lack of detection. We have tried to refine this energy grid at the expense of some precision in terms of broadening, but the results were not satisfactory.

\end{itemize}

\begin{figure}
    \centering
    \includegraphics[width=\linewidth]{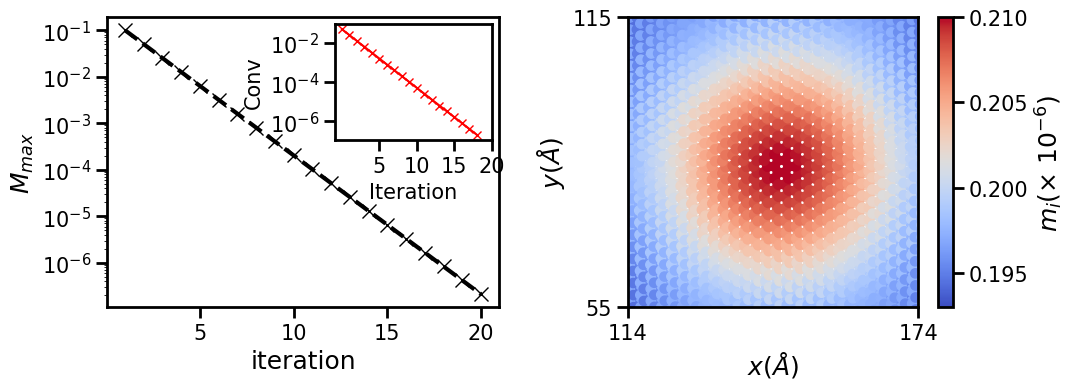}
    \caption{Figure analogous to that of Fig.\ \ref{fig:not_convergence} for the ferromagnetic case.}
    \label{fig:not_convergence_ferro}
\end{figure}

In Figs.\ \ref{fig:not_convergence} and \ref{fig:not_convergence_ferro} we can see a typical convergence for the mean-field problem for the 0 filling and 3/4 filling of the flat band. Here we observe the convergence of a twisted bilayer graphene with a twist-angle of $\sim 1.47^\circ$. 

It is clear that the magnetization (left panels in Figs.\ \ref{fig:not_convergence} and \ref{fig:not_convergence_ferro}) reaches values compatible with the precision (we recall here that our convergence parameter is the difference between two consecutive steps) with no clear indicators of convergence. In contrast, in the upper left panel of Fig.\ \ref{fig:nanoribbon convergence}, we observe a typical convergence behavior of the problem. Despite the lack of convergence, in the right panels of Figs.\ \ref{fig:not_convergence} and \ref{fig:not_convergence_ferro} we  observe a qualitative distribution of the local magnetizations similar to the one expected from \cite{10.21468/SciPostPhys.11.4.083,Vidarte_2025}.

\subsection{Validity of the method}
\begin{figure}[h]
    \centering
    \includegraphics[width=\linewidth]{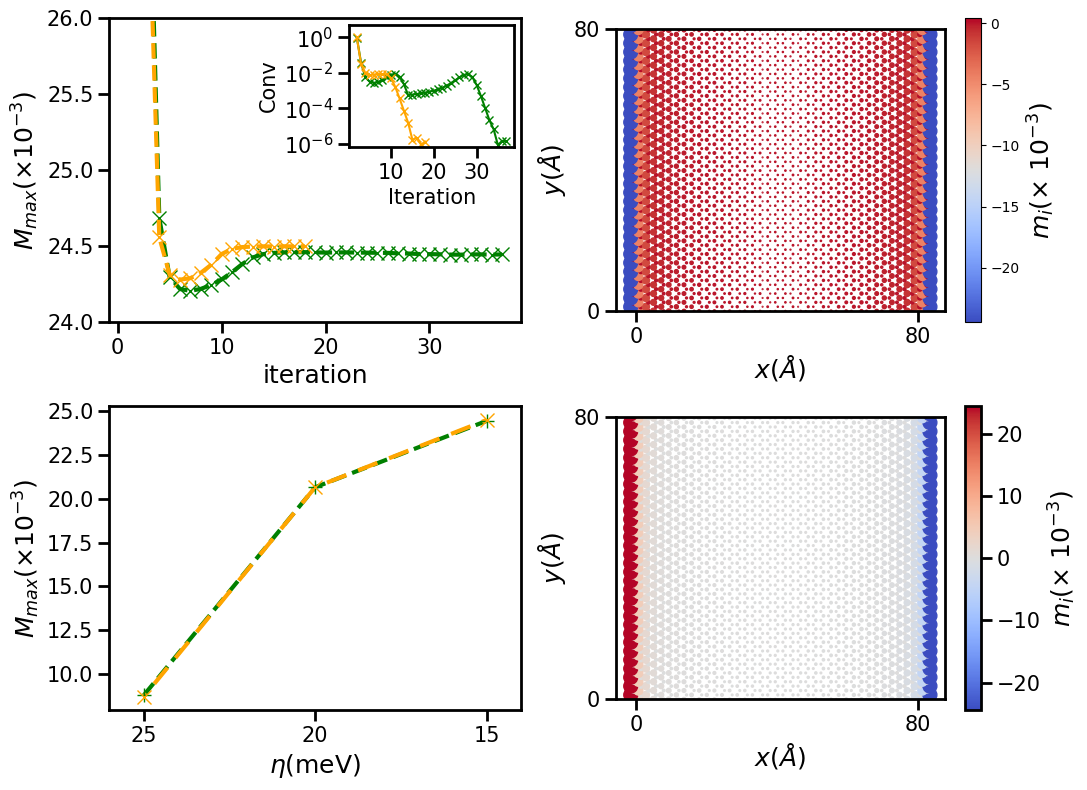}
    \caption{Top left: Evolution of magnetizations for the nanoribbon  KPM+Hartree-Fock Hubbard mean field with $\eta=15$ meV for the ferromagnetic (orange) and antiferromagnetic (green) solutions. Inset shows the evolution of the convergence parameter. Bottom left: Magnetization dependence on the broadening. Right top (bottom) shows the magnetization distribution for ferromagnetic (antiferromagnetic) solutions from \eqref{eq:mag_deffinition}. }
    \label{fig:nanoribbon convergence}
\end{figure}

To better understand the effects of the broadening, one can present an easier problem concerning zig-zag graphene nanoribbons (zGNR). In Fig.\ \ref{fig:nanoribbon convergence} we observe a convergence of the KPM-mean-field calculation for a zGNR, computed with the same method that was used for section \ref{Sec: results mf+hf}. In the top and bottom right panels of Fig.\ \ref{fig:nanoribbon convergence} we observe the converged results for the magnetization of the zig-zag graphene nanoribbon in its ferro- and antiferromagnetic phases respectively, computed for a broadening of $\eta\approx 0.15$ meV. These results are qualitatively correct when compared to previous results in the literature \cite{6tny-vt8q}, while quantitatively underestimating the magnetizations. In the upper left panel, we can see that the self-consistent algorithm is properly converged for these results.

In the bottom left panel, we observe the predicted maximum local magnetization for antiferromagnetic (green) and ferromagnetic (orange) solutions. We note that the quantitative values for these local magnetizations show a strong dependence on the broadening. This suggests that the effect of the broadening may be substantial in KPM calculations and that further optimization is needed, which is currently beyond our computational capabilities. Yet, we note that this subtle dependence on the broadening of the states can be achieved by disorder-induced broadening and suggests a large variation of magnetization for these flat-band systems against small amounts of disorder.

\chapter{Spin transport in corrugated monolayer graphene}\label{chapter:spin}

This chapter turns to spin transport in suspended monolayer graphene, using it as a complementary setting in which structural complexity affects dynamics in a way that differs markedly from charge transport. After reviewing the basic spin-orbit coupling terms relevant to graphene and the main semiclassical mechanisms of spin relaxation, the chapter introduces an atomistic description of thermally corrugated graphene in which local curvature generates position-dependent hopping, doping, and spin-orbit fields. Realistic corrugation profiles are obtained from molecular-dynamics simulations, and these are combined with a curvature-dependent tight-binding Hamiltonian to study both charge diffusion and spin dynamics.

A central result is that atomic-scale corrugations leave charge transport comparatively close to the ballistic regime over broad energy windows, while simultaneously producing strong short-range fluctuating Rashba fields that drive efficient spin relaxation and limit spin lifetimes to the nanosecond scale. In this way, the chapter shows that spin transport is an especially sensitive probe of microscopic structural disorder and provides a concrete mechanism to reconcile the high mobility of suspended graphene with the much shorter spin lifetimes observed experimentally.

\section{Spin-orbit coupling in graphene}
\label{subsec:SOC_graphene}

This chapter mainly follows the work developed by Fabian et al. in \cite{Fabian_2007}. We start our work by introducing a simple low-energy model for spin-orbit coupling (SOC) in graphene. We will consider intrinsic and extrinsic contributions to the spin-orbit coupling. In pristine free-standing graphene the intrinsic SOC is very weak due to the small atomic number of carbon and the planar bonding environment $sp^2$ \cite{PhysRevB.74.155426}. However, there are many other contributions to SOC that can arise in proximity to other materials \cite{Song2018}.

\begin{figure}
    \centering
    \includegraphics[width=\linewidth]{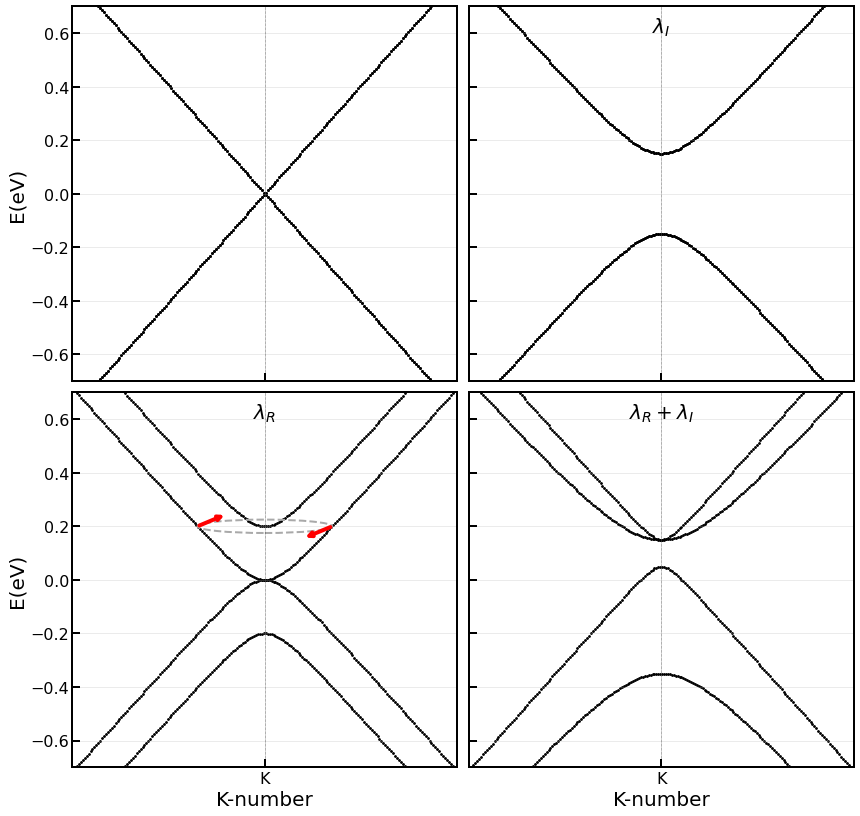}
    \caption{Effects of different spin-orbit couplings in the $K$ cone of graphene. Top left: Monolayer graphene without spin-orbit coupling. Top Right: Monolayer graphene with Kane-Mele spin-orbit coupling, we can explicitly see the topological gap opening. Bottom left: Rashba spin-orbit coupling produces an in-plane spin-texture (red arrows). Bottom right: Combination of Rashba+Kane-Mele produces a topological gap at Fermi level when $\lambda_I>\lambda_R$.}
    \label{fig:SOC graphene}
\end{figure}

Starting with the intrinsic SOC, one can consider the usual interaction between spin and orbit, the Kane-Mele spin-orbit coupling \cite{PhysRevLett.95.226801}. This is an intrinsic SOC generated by the symmetries of the honeycomb lattice: it is the only spin-dependent term allowed by time-reversal symmetry, inversion symmetry between the two sublattices, and the threefold rotational symmetry of graphene, and it manifests itself as a next-nearest-neighbor spin-dependent mass term that preserves all lattice symmetries.
\begin{equation}
    H_I = \lambda_I \, \tau_z\otimes\sigma_z \otimes s_z.
    \label{eq:HI_graphene}
\end{equation}
Microscopic tight-binding calculations show that in monolayer graphene $\lambda_I$ \cite{PhysRevB.82.245412} is extremely weak. Under typical conditions, $\lambda_I \sim 10$ $\mu$eV is found in clean graphene \cite{PhysRevB.80.235431}. We can see the effect of this term on the band structure of the top right panel of Fig.\ \ref{fig:SOC graphene}. There we can see that it produces a topological gap that can then be experimentally measured \cite{PhysRevLett.122.046403}. Experimental works have found that it has a value of $\lambda_I\sim 42\ \mu$eV.

However, when inversion symmetry is broken, as is the case of proximity with a substrate or an applied transverse electric field, a Bychkov-Rashba type SOC can be induced. This term is frequently considered to be the dominant mechanism in spin manipulation and, in many experimentally relevant situations, for spin relaxation \cite{RevModPhys.76.323}.

In order to model this Rashba term, we must take an extended version of the $K\cdot p$ model of graphene displayed in equation (\ref{KdotP}) in spin state. Then in the presence of a substrate-induced asymmetry or a transverse electric field $E_z$, the inversion symmetry is broken and an extrinsic SOC of Bychkov-Rashba \cite{PismaZhETF.39.66} type is allowed by the symmetry. At the level of the Dirac Hamiltonian, it takes the form  \cite{Song2018}
\begin{equation}
    H_R = \lambda_R 
    \left(
        \tau_z \otimes \sigma_x \otimes s_y - \sigma_y \otimes s_x
    \right),
    \label{eq:HR_graphene}
\end{equation}
where $\lambda_R$ is the Rashba SOC strength. This term is the natural generalization of the standard two-dimensional electron gas Rashba Hamiltonian \cite{PhysRevB.82.245412} to the case of Dirac fermions in graphene. The effect of this term can be seen in the bottom left panel of Fig.\ \ref{fig:SOC graphene} producing a ``massive behavior'' close to the Dirac point and generating an in-plane spin texture for the Dirac cone.


Thus, a full Hamiltonian for graphene with broken out-of-plane symmetry may be written as
\begin{equation}
    H = H_0 + H_I + H_R
    \label{eq:H_eff_Rashba} .
\end{equation}
It is interesting to note that even if we are not going to discuss in detail the topology of this Hamiltonian, the combination exposed in \eqref{eq:H_eff_Rashba} of the two kinds of spin-orbit coupling in graphene yet opens a topological gap at the Fermi-level when $\lambda_I>\lambda_R$. This can be seen from the bottom right panel of Fig. \ref{fig:SOC graphene} and was originally shown by Kane and Mele in 2005 \cite{PhysRevLett.95.226801}.

From the point of view of spin dynamics, the Rashba term can be regarded as a momentum-dependent effective magnetic field acting on the electron spins. In a semiclassical picture, each carrier with momentum $\boldsymbol{k}$ experiences an effective field $\boldsymbol{\Omega}(\boldsymbol{k})$ proportional to $\lambda_R$, around which its spin precesses between successive scattering events. In diffusive transport, this leads to different mechanisms of spin relaxation, in which the spin lifetime is controlled by the strength of the Rashba field and the momentum scattering time \cite{RevModPhys.87.1213}. This Rashba-induced dephasing mechanism will compete with the Elliott-Yafet process, where spin relaxation arises instead from spin-flip components of the impurity scattering potential in the presence of SOC.

\section{Spin relaxation in graphene}

We will dedicate this section to the main mechanisms that are frequently used to describe spin relaxation in monolayer graphene. As mentioned in the previous sections, the Rashba spin-orbit coupling, in combination with disorder, produces spin relaxation. 
Spin evolution can be complex, due to the spin precession generated by $\boldsymbol{\Omega}$ and spin-flips, but for spin relaxation we will see that the average value of spin follows an exponential decay as

\begin{equation}
    \langle S(t)\rangle\propto e^{-\frac{t}{\tau_s}} ,
\end{equation}
where $\tau_s$ is the spin relaxation time. In the next section, we will cover the main mechanisms for spin relaxation that have been studied to predict and explain the experimental values of this $\tau_s$ for graphene in the semiclassical approximation.

\subsection{Main relaxation mechanisms}\label{Section:Main-relaxation mechanisms}

Many studies have been conducted to understand and predict the value $\tau_s$, and as a result some mechanisms have been proposed, mediated, for example, by the hyperfine interaction or the Bir-Aronov-Picus process \cite{RevModPhys.76.323}. For graphene, the main processes studied and correlated to spin relaxation are the Dyakonov-Perel \cite{Dyakonov-Perel} and Elliott-Yafet \cite{PhysRev.96.266,YAFET19631} mechanisms. Both of them have a strong dependence on the Rashba spin-orbit coupling and originate in the electron-impurity scattering, which generates decoherence in the spin wave packets. The first is more focused on the differences induced by scattering in the spin precession and the second on the possibility of spin-flip events coming from every scattering \cite{RevModPhys.76.323}.

\subsubsection{The Dyakonov-Perel mechanism}

In systems where inversion symmetry is broken, such as graphene supported on a substrate or subjected to a transverse electric field, the presence of Rashba spin-orbit coupling produces a momentum-dependent effective magnetic field acting on the electron spins. In this regime the spin does not behave as a single eigenstate of the Hamiltonian and instead precesses around the effective field $\boldsymbol{\Omega}(\mathbf{k})$ between successive scattering events. This precession is the defining feature of the Dyakonov-Perel (DP) mechanism.

For graphene with Rashba SOC, the effective field takes the symmetry-allowed form
\begin{equation}
    \boldsymbol{\Omega}(\mathbf{k}) = \frac{2\lambda_R}{\hbar}
    \left( -\sin\theta_{\mathbf{k}}, \; \cos\theta_{\mathbf{k}}, \; 0 \right),
\end{equation}
where $\theta_{\mathbf{k}}$ is the polar angle of the crystal momentum. Therefore, the spin lies predominantly in the graphene plane, orthogonal to~$\mathbf{k}$, in agreement with the characteristic in-plane spin texture shown in the bottom left panel of  Fig.\ \ref{fig:SOC graphene}. In the presence of disorder, each scattering event randomizes the direction of $\mathbf{k}$ and, therefore, the precession axis.

\begin{figure}[h!]
    \centering
    \includegraphics[width=0.5\linewidth]{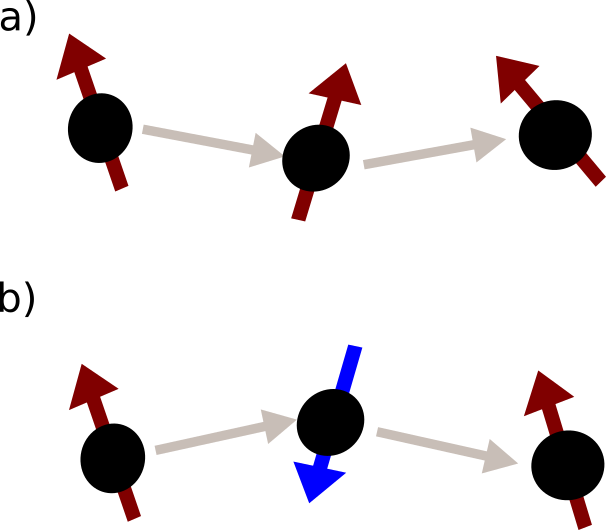}
    \caption{Schematic illustrations for Dyakonov-Perel (a) and Elliott-Yafet (b) mechanisms. }
    \label{fig:placeholder}
\end{figure}

In general, the presence of Rashba SOC can lead to two different regimes of spin dynamics depending on the competition between the spin precession time $\tau_\Omega = \Omega^{-1}$ and the momentum-scattering time $\tau_p$ \cite{Dyakonov-Perel}.  

(i) When spin precession is fast compared to momentum scattering ($\tau_p \gg \tau_\Omega$), the spin can execute many rotations between collisions, leading to a precession-dominated regime. Here, more scattering events imply a larger modification of the precession angle and thus $\tau_s$ scales with $\tau_p$.

(ii) Conversely, when momentum scattering is fast compared to spin precession ($\tau_p \ll \tau_\Omega$), the system enters the motional-narrowing limit: frequent scattering events constantly reorient the precession axis, suppressing coherent rotation of the spin. This means that when many scattering events occur, the precession of the spin is averaged, and it relaxes more slowly. This is known as the Dyakonov-Perel (DP) relaxation mechanism, where the relaxation rate takes the characteristic form
\begin{equation}\label{eq:Dyakonov-perel}
    \frac{1}{\tau_s^{\mathrm{DP}}} \propto \Omega^{2}\,\tau_p.
\end{equation}

However, for monolayer graphene, the weak SOC frequently ensures that $\tau_\Omega$ is much longer than $\tau_p$ under realistic conditions (room temperature, typical substrates, or curvature-induced SOC). Thus, graphene is firmly in the motional-narrowing DP limit \cite{Fabian_2007,Cummings_2017}. For the pristine case, the situation is even less favorable for the DP mechanism, as the extreme weakness of the Rashba SOC frequently makes the DP contributions to spin relaxation negligible. Thus, the DP mechanism is frequently considered relevant only when Rashba SOC is enhanced, either by a strong substrate potential, by adatoms, or by externally applied electric fields \cite{Ertler2009,Cummings_2017}.

\subsubsection{The Elliott-Yafet mechanism}

The Elliott-Yafet (EY) mechanism describes spin relaxation originating from the fact that, in the presence of spin-orbit coupling, Bloch eigenstates are not pure spin states. Instead, each nominal spin-up eigenstate contains a small admixture of spin-down character and vice versa. This spin mixing implies that every momentum-scattering event produced by impurities, phonons, or corrugations carries an associated probability of flipping the electron spin. This mechanism was first identified in elemental metals by Elliott \cite{PhysRev.96.266} and later formalized by Yafet within a microscopic scattering framework \cite{YAFET19631}.

Let $\ket{\psi_{\mathbf{k}}^{\uparrow}}$ denote an eigenstate of \eqref{eq:H_eff_Rashba} whose dominant spin projection is $\uparrow$, we can expand this state in the spin space as
\begin{equation}
   \ket{\psi_{\mathbf{k}}^{\uparrow}} 
   = a_{\mathbf{k}} \ket{\uparrow} + b_{\mathbf{k}} \ket{\downarrow},
\end{equation}
the quantity $|b_{\mathbf{k}}|^2$ is called the spin mixing parameter. A momentum-scattering event that changes $\mathbf{k}\!\to\!\mathbf{k}'$ also transforms the admixture amplitudes, giving the spin a finite probability of flipping during scattering. As a consequence, the EY spin-relaxation rate will scale with the scattering time, following the proportionality
\begin{equation}
   \frac{1}{\tau_s^{\mathrm{EY}}} \simeq \frac{\langle |b_{\mathbf{k}}|^2 \rangle}{\tau_p},
\end{equation}
where $\tau_p$ is the momentum-relaxation time. This opposite scaling, $\tau_s^{\mathrm{EY}} \propto \tau_p$, is experimentally one of the clearest signatures distinguishing EY from DP relaxation in graphene.

In graphene, the spin-mixing parameter $|b_{\mathbf{k}}|$ is mainly governed by the strength of the spin-orbit coupling. For pristine graphene, the intrinsic SOC is extremely small (of order $10~\mu$eV), producing negligible mixing \cite{PhysRevB.92.155403, Yazyev_2010}. However, the EY mechanism has been predicted to become relevant once the spin-orbit coupling is enhanced by extrinsic sources such as substrate-induced Rashba fields or curvature-induced hybridization \cite{PhysRevLett.103.026804, Cummings_2017}. These perturbations increase the SOC matrix elements, amplifying the spin-mixing parameter.

\section{A model for corrugations}
\label{sec:corrugations}

\subsubsection{Flat graphene}

To quantify the influence of corrugations on spin transport, we require an atomistic model that captures how out-of-plane deformations modify the electronic structure and induce effective spin-orbit coupling (SOC). In this subsection, we introduce the tight-binding formalism used throughout this chapter. 

\begin{figure}
    \centering
    \includegraphics[width=.5\linewidth]{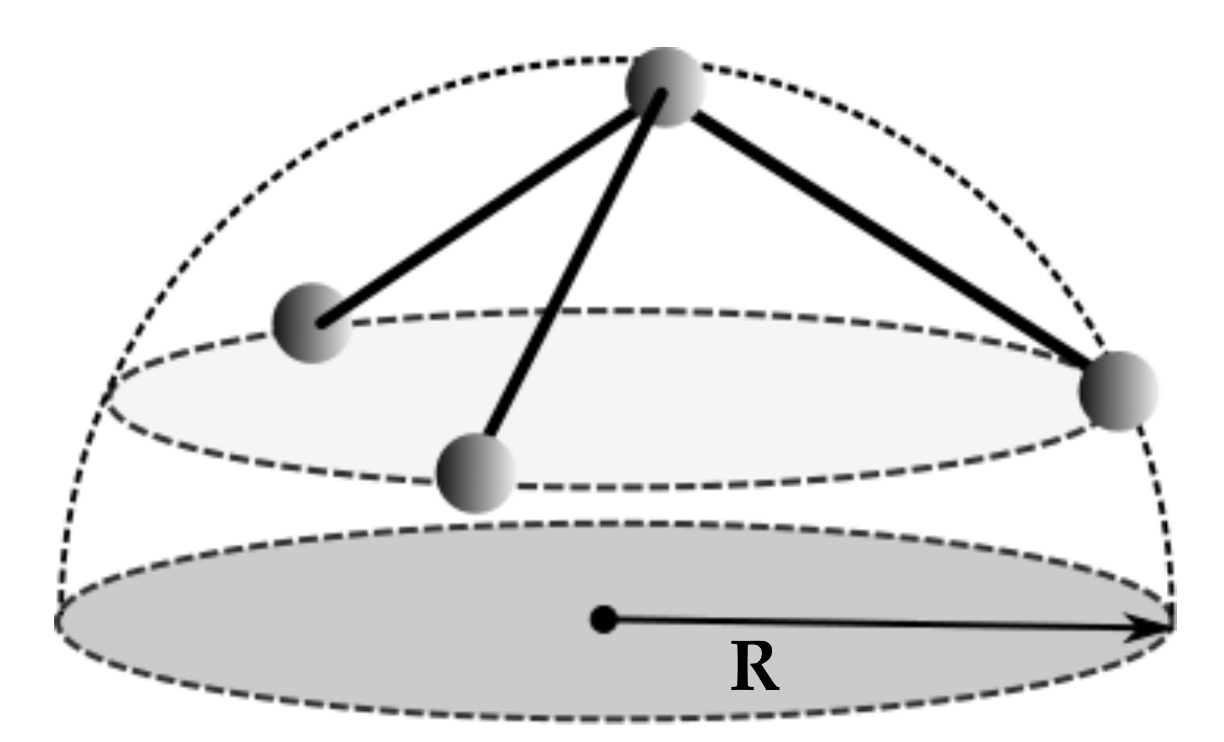}
    \caption{Example of the curvature modeled for Sec.\ \ref{sec:corrugations}.}
    \label{fig:Curvature}
\end{figure}

The key idea is that curvature locally breaks mirror symmetry (Fig.\ \ref{fig:Curvature}), generating short-range SOC fields. These fields fluctuate on atomic length scales and ultimately limit spin relaxation in suspended graphene. We describe the electronic states starting from the second-nearest-neighbor version of equation\ \eqref{tight binding} and including the Kane-Mele term \eqref{eq:HI_graphene} we arrive at the expression of our base Hamiltonian in flat graphene as

\begin{equation}
\hat{H}_0 
= \sum_{\langle i,j\rangle} t^{n}_{ij}\, c^\dagger_i c_j
+ \sum_{\langle\!\langle i,j\rangle\!\rangle} t^{nn}_{ij}\, c^\dagger_i c_j
+ i\frac{\lambda_I}{3\sqrt{3}}
  \sum_{\langle\!\langle i,j\rangle\!\rangle}
  \eta_{ij}\, c^\dagger_i s_z c_j ,
\label{eq:H0_flat}
\end{equation}
where $t^n_{ij}$ and $t^{nn}_{ij}$ are nearest- and next-nearest-neighbor hoppings, $\lambda_I$ is the intrinsic Kane--Mele SOC, $\eta_{ij}=\pm 1$ distinguishes clockwise/counterclockwise paths, and $c^\dagger_i$ creates a $p_z$ electron on site $i$. This Hamiltonian is frequently used to reproduce the Dirac cones and the extremely weak intrinsic SOC of pristine graphene \cite{PhysRevB.80.235431,PhysRevB.82.245412}. The tight-binding parameters have been adapted to the corrugations in the same way as we did in \eqref{eq:decay-Magic-angle-hoppings}. For this case an exponential decay of the hopping has been used as \cite{Cummings_2025,Cummings2019}
\begin{equation}\label{eq:dubois model}
t^{n,nn}_{ij}=t_0^{n,nn}\cdot \exp\left[-\beta_{n,nn}\left(\frac{r_{ij}}{a_{n,nn}}-1\right)\right],
\end{equation}
where $t_0^{nm}$ are the average carbon-carbon hoppings and $r_{ij}$ the distance to the (next) nearest neighbor.

\subsubsection*{Local description of curvature}

Out-of-plane deformations are described by a local curvature field $d_i$ computed at each lattice site. Operationally, $d_i$ is defined as the inverse radius of curvature of the sphere that passes through atom $i$ and its three nearest neighbors ($R$ of Fig.\ \ref{fig:Curvature}). Positive and negative curvatures represent locally convex or concave patches and are distinguished by a sign factor $\sigma_i=\pm 1$, so that each site carries a deformation vector
\begin{equation}\label{eq:curvature}
    \mathbf{d}_i = d_i \sigma_i\, \hat{\mathbf{z}} .
\end{equation}

Curvature modifies the electronic structure in three distinct ways:

\begin{enumerate}
\item \textbf{Scalar potential term} (local doping):
\begin{equation}\label{eq: scalar_potential}
    \hat{H}_{\mu} = 
    \mu_d \sum_i d_i\, c^\dagger_i c_i \;,
\end{equation}
where $\mu_d$ captures the charge redistribution due to corrugation.

\item \textbf{Hopping renormalization}:
\begin{equation}\label{eq: bond renormalization}
    \hat{H}_{t} =
    \beta^{d}_n \sum_{\langle i,j\rangle} d^{n}_{ij}\, c^\dagger_i c_j
    + \beta^{d}_{nn}\sum_{\langle\!\langle i,j\rangle\!\rangle} d^{nn}_{ij}\, c^\dagger_i c_j \;,
\end{equation}
with $\beta^d_{n,nn}$ accounting for the modifications in the hoppings induced by curvature.

\item \textbf{Curvature-induced SOC terms}:
\begin{align}
\hat{H}_{\mathrm{SOC}}^{\mathrm{curv}} &=
 i\lambda_R^{d}
 \sum_{\langle i,j\rangle}
 c^\dagger_i
 \big(\mathbf{d}^{n}_{ij}\times \mathbf{n}_{ij}\big)\!\cdot\!\mathbf{s}\,
 c_j  
 \nonumber\\
&\quad
 + i\lambda_I^{d}
 \sum_{\langle\!\langle i,j\rangle\!\rangle}
 c^\dagger_i
 \eta_{ij}\,(d^{nn}_{ij})^2\, s_z\, c_j 
 \nonumber\\
&\quad
 + i\lambda_{\mathrm{PIA}}^{d}
 \sum_{\langle\!\langle i,j\rangle\!\rangle}
 c^\dagger_i
 \big(\mathbf{d}^{nn}_{ij}\times \mathbf{n}_{ij}\big)\!\cdot\!\mathbf{s}\,
 c_j\;.
\label{eq:H_SOC_curv}
\end{align}
Here $\lambda_R^{d}$ emerges from the breaking of the local inversion symmetry, generating a Rashba SOC induced by local curvature, $\lambda_I^{d}$ is a correction induced by curvature to the intrinsic SOC of graphene, and $\lambda_{\mathrm{PIA}}^{d}$, whose effect is the smallest, can be understood as a local version for the pseudospin-inversion-asymmetry (PIA) SOC as shown in \cite{PhysRevB.80.235431}.
\end{enumerate}

In the previous equations, we use $d^{n}_{ij}$ and $d^{nn}_{ij}$ for the curvature fields of the nearest and next-nearest neighbors, respectively. Here, $d^n_{ij}$ is computed as the arithmetic mean of the curvature computed as in \eqref{eq:curvature} for the nearest-neighbor positions $i,j$. The second-nearest-neighbor curvature field $d^{nn}_{ij}$ where $i,j$ correspond to atoms that are second nearest neighbors, can be computed as the value of $d_k$ where $k$ is the common neighbor between sites $i$ and $j$.

The addition of the three contributions results in a full curvature Hamiltonian that takes the compact form
\begin{equation}\label{eq:corrugation-full-hamiltonian}
    \hat{H}_{\mathrm{curv}}
    = \hat{H}_{\mu} + \hat{H}_{t} + \hat{H}_{\mathrm{SOC}}^{\mathrm{curv}}\; . 
\end{equation}

All the parameters mentioned in equations  \eqref{eq: scalar_potential}, \eqref{eq: bond renormalization}, and \eqref{eq:H_SOC_curv} were fitted from bands obtained by first-principles calculations done by collaborators at UCLouvain on flat and corrugated graphene samples \cite{Cummings_2025}. They were made with the all-electron full-potential linearized augmented plane wave (FP-LAPW) method implemented in the Elk code \cite{elk_code}. Self-consistent calculations with SOC were carried out within the local density approximation with a muffin tin radius of 1.316 Bohr for carbon atoms and an APW cutoff of 5.32 Bohr$^{-1}$. A 33 $\times$ 33 k-point mesh was used to sample the first Brillouin zone of pristine graphene, and an equivalent k-point density was used for the 2 $\times$ 2 supercell of graphene. The fitting of the tight-binding model to the first-principles results was limited to the $[-1, 1]$ eV energy window around the Fermi level; a similar fitting procedure has previously been used in \cite{Cummings2019} for flat graphene. The values of the fitted parameters are given in Table \ref{tab_params_upper}.

\begin{table}[t]
\begin{tabular*}{\columnwidth}{@{\extracolsep{\fill} } c c c}
\hline
Parameter & Description & Value \\
\hline\hline
$t_0^n$ & 1st-neighbor & $-2.51$ eV \\
$a_n$ & hopping & $1.418$ \AA \\
$\beta_n$ &  & $2.62$ \\
$t_0^{nn}$ & 2nd-neighbor & $-0.17$ eV \\
$a_{nn}$ & hopping & $2.456$ \AA \\
$\beta_{nn}$ &  & $7.57$ \\
$\lambda_\mathrm{I}$ & intrinsic SOC& $13.4$ $\upmu$ eV \\
\hline
$\mu_d$ & local doping & $18.6$ meV$\cdot$nm \\
$\beta_n^d$ & local hopping & $-4.33$ meV$\cdot$nm \\
$\beta_{nn}^d$ & & $-27.8$ meV$\cdot$nm \\
$\lambda_\mathrm{R}^d$ & local Rashba& $-9.69$ meV$\cdot$nm \\
$\lambda_\mathrm{PIA}^d$ & & $-0.322$ meV$\cdot$nm \\
$\lambda_\mathrm{I}^d$ & local intrinsic& $-4.44$ $\upmu$ eV$\cdot$nm$^2$ \\
\hline
\end{tabular*}
\caption{Tight-binding parameters for the corrugated graphene Hamiltonian given in Eqs.\ \eqref{eq: scalar_potential}, \eqref{eq: bond renormalization}, and \eqref{eq:H_SOC_curv}, extracted from \cite{Cummings_2025}.}
\label{tab_params_upper}
\end{table}

\subsubsection{Magnitude of the corrugations}

\begin{figure}
    \centering
    \includegraphics[width=\linewidth]{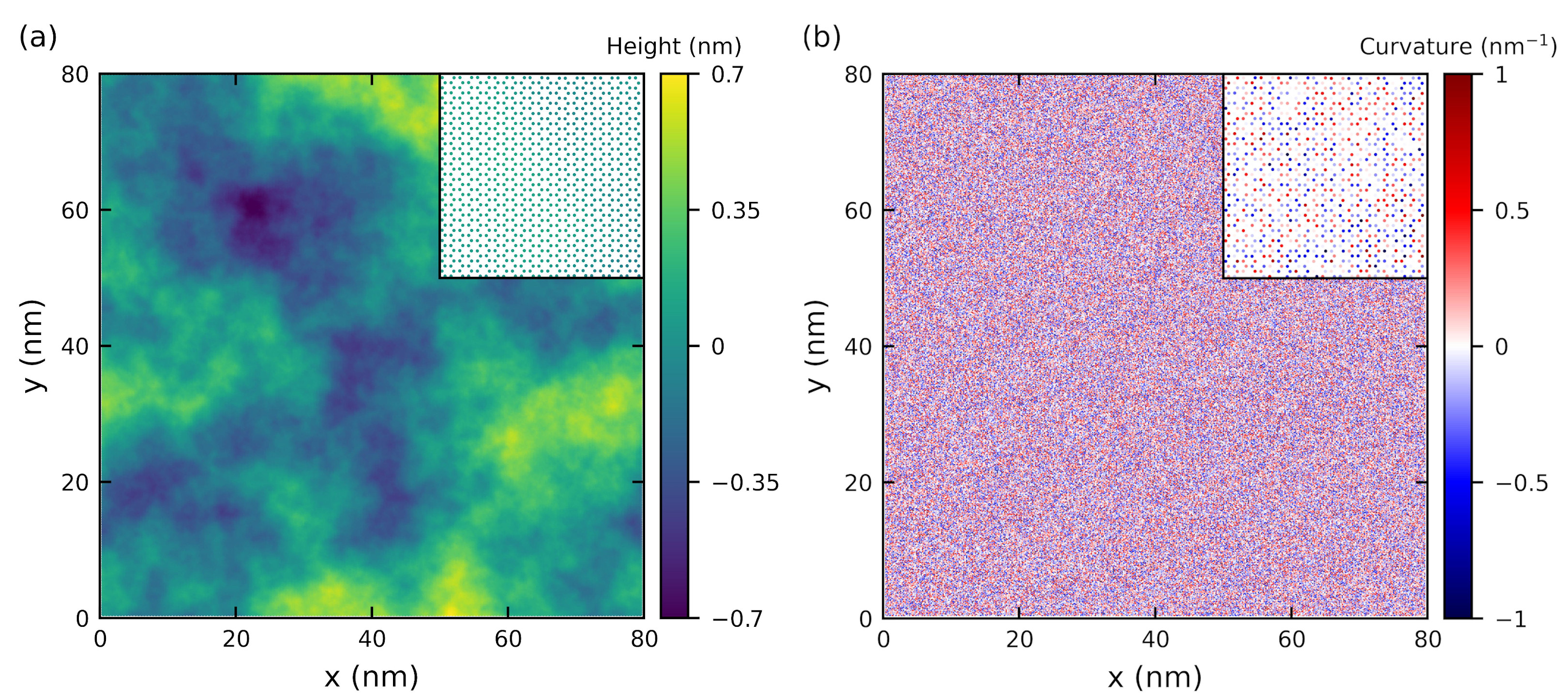}
    \caption{Height (a) and local curvature (b) real-space maps for a thermally corrugated graphene sample thermalized at $300\;\mathrm K$ the insets show a 5nm$\times$ 5nm zoom of the lower-left corner. Figures extracted from \cite{Cummings_2025}.}
    \label{fig:corrugated-graphene-samples}
\end{figure}

To obtain realistic corrugation profiles appropriate for suspended monolayer graphene, we generate atomic configurations by means of classical molecular-dynamics simulations. The system is evolved using the LAMMPS package under an optimized Tersoff-Brenner description of the carbon-carbon interactions, which provides an accurate representation of the bending rigidity and thermal fluctuations of graphene sheets. Starting from an initially flat sample, the structure is first relaxed at $T=0$\,K and subsequently thermalized at the target temperature in an NPT ensemble with zero external pressure. After an equilibration stage of $\sim 100$\,ps, additional simulation time is used to extract statistically independent snapshots of the atomic positions. In Fig.\ \ref{fig:corrugated-graphene-samples} we can see the local maps of the height and the temperature for one snapshot of the $300\;\mathrm{K}$ simulation. Here we can observe a thickness of corrugated graphene of $\sim 1.4\;\mathrm{nm}$, similar to those measured in suspended graphene samples \cite{Locatelli2010,Meyer2007,PhysRevB.84.235417}. 

In Fig.\ \ref{fig:corrugated-graphene-samples} b, we observe that the local corrugations $d_i$ are dominated by short-range fluctuations, with no apparent relation with the height profile. Following equation\ \eqref{eq:H_SOC_curv}, the local Rashba SOC in this sample is going to be generated from these curvatures and proportional to $d_i$. From Table \ref{tab_params_upper} and Fig.\ \ref{fig:corrugated-graphene-samples} b, we see that the local value for Rashba SOC in this sample can reach values as large as $10\ \mathrm{meV}$, approximately three orders of magnitude larger than the usual Rashba SOC obtained in monolayer graphene deposited on $\mathrm{SiO_2}$, implying that even in the absence of external substrates or impurities, suspended graphene can sustain SOC fields strong enough to drive efficient Dyakonov-Perel spin relaxation
\cite{Cummings_2025}.

\subsection{Charge transport in corrugated graphene}

Before addressing the spin dynamics, it is useful to characterize how thermally induced corrugations affect charge transport. Using the tight-binding Hamiltonian defined in equation\ \eqref{eq:corrugation-full-hamiltonian}, we compute the time evolution of an electronic wave packet and evaluate the mean-squared displacement (MSD) and the corresponding diffusion coefficient using the Kubo-Greenwood formalism described in Sec. \ref{sec:Mean Squared Displacement-formulas}. For each molecular-dynamics snapshot, we compute the diffusion coefficient using \eqref{eq:diffusion coefficient definition} and \eqref{eq:MSD LSQUANT}.

To extract an effective momentum-relaxation time $\tau_p(E)$ and mean free path $\ell(E)$, we will follow a slightly modified procedure from the one in chapter\ \ref{Chapter: Transport in twisted multilayer graphene}. There, we extracted $\tau_p$ and $\ell$ using the equations \eqref{eq:deffinition_momentum_scattering_time} and \eqref{eq:deffinition mfp}. For our case in corrugated graphene, the mean free paths are not small enough for us to capture the full ballistic-to-diffusive transition within our simulations. Instead, we will make use of 

\begin{equation}\label{eq:mfp-corrugated-graphene}
    \langle \Delta X^2\rangle=2\ell^2\left(t/\tau_p-1+e^{-t/\tau_p}\right).
\end{equation}
We can see the fits compared to our calculations in the inset of Fig.\ \ref{fig:corrugated-graphene-charge-transport} where the solid lines represent the simulations obtained by \eqref{eq:MSD LSQUANT} and the dashed lines the fits to \eqref{eq:mfp-corrugated-graphene}.

\fullpagefigure[1]{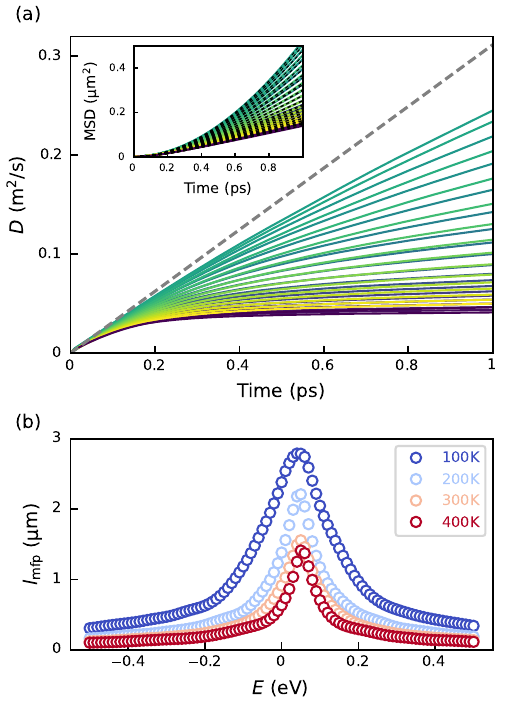}
{a: Diffusion coefficient (main) and mean-squared displacement (inset) as a function of time at 300 K for different values of the chemical potential $E\in\left[-0.5,0.5\right]$ eV where green curves represent $E=0$ and purple and yellow colors increase both for negative and positive values respectively. The dashed lines in the inset represent the fits of \eqref{eq:mfp-corrugated-graphene} b: mean free paths as a function of chemical potential for different values of the temperature. Figures extracted from \cite{Cummings_2025}.}{fig:corrugated-graphene-charge-transport}


In Fig.\ \ref{fig:corrugated-graphene-charge-transport} we can see the mean free paths reached from our corrugated graphene samples for different values of temperature. Despite the presence of strong atomic-scale curvature (Fig. \ref{fig:corrugated-graphene-samples}), the resulting charge transport remains remarkably close to the pristine limit. During a broad energy window around the Dirac point, the mean free paths extracted reach several hundred nanometers, which is consistent with the high mobilities measured experimentally in suspended graphene devices \cite{Bolotin_2008}.

To understand this, it is important to first note that our calculations obtain similar results of $\tau_p$ when we set the hopping renormalizations of equation\ \eqref{eq: bond renormalization} to zero, proving that the main cause of momentum scattering is the local doping established in equation \eqref{eq: scalar_potential}.

In addition, this can also explain the long mean free paths. Following from \eqref{eq: scalar_potential} and Table\ \ref{tab_params_upper}, where we can see that the local shifts in energy are $\sim 20\ \mathrm{meV}\approx(1/135) t_{ij}^n$, generating an equivalent source of disorder as the one shown in \eqref{eq: Anderson disorder} with a much smaller magnitude, which explains the long relaxation lengths. 
Overall, the corrugations considered here, although substantial at the sub-nanometer scale, do not significantly degrade the electronic mobility. This shows that curvature has only a minor effect on charge transport. In the next section, we will see that the corrugations have a strong impact on spin relaxation and we will analyze and quantify this effect.

\section{Spin lifetime}

Having established that thermally induced corrugations have only a minor impact on charge transport, we now focus on their influence on spin dynamics. In contrast to the long mean free paths shown in Fig. \ref{fig:corrugated-graphene-charge-transport}, the curvature-induced spin-orbit fields generate efficient spin relaxation even in clean suspended graphene.

To quantify the spin lifetime $\tau_s(E)$, we compute the quantity
\begin{equation}\label{eq: time-evolution-trace-spin-relax}
\langle S_\alpha(E,t)\rangle = \mathrm{Tr}\left[\rho_{\mathrm{eq}}(E)\,
U^\dagger(t)\, s_\alpha\, U(t)\right]_\uparrow,\quad \alpha = \lbrace x,y,z\rbrace\;,
\end{equation}
where $\mathrm{Tr\left[\,\right]}_\uparrow$ is the trace in the subspace of up $s_\alpha$. We can draw an analogy with this equation and equation\ \eqref{eq:lsquant:time-evolution} and repeat the process in Sec.\ \ref{Sec:Time evolution} choosing the random initial states $\ket{\phi_\uparrow}$, as states completely localized in the target spin subspace. Thus, following similar arguments as in Sec.\ \ref{sec:stochastic trace}, we can express the trace of \eqref{eq: time-evolution-trace-spin-relax} as the stochastic limit of

\begin{equation}\label{eq:time-evolution-spin-transport-random-phase}
\langle S_\alpha(E,t)\rangle = \bra{\psi_\uparrow}\rho_{\mathrm{eq}}(E)\,U^\dagger(t)\, s_\alpha\, U(t)\ket{\psi_\uparrow},\quad \alpha = \lbrace x,y,z\rbrace\;,
\end{equation}
where $\ket{\psi_\uparrow}$ are the random phase functions similar to those of \eqref{eq: random phase state definition} but completely localized in the spin-up state.

For all temperatures and energies considered, the spin signal exhibits a single-exponential envelope,
\begin{equation}
\langle S_\alpha(E,t) \rangle\simeq 
\langle S_\alpha(E,0)\rangle\, e^{-t/\tau_s(E)},
\end{equation}
consistent with the predominance of a Dyakonov-Perel dephasing in the motional-narrowing regime.



Fig.\ \ref{fig:corrugated-graphene-spin-relaxation}a shows the extracted $\tau_s(E)$ for different thermalization 
temperatures. At room temperature, the spin lifetime lies in the range 
$\tau_s \sim 1$--$10\,\mathrm{ns}$ over a wide energy window around the Dirac point. 

Fig.\ \ref{fig:corrugated-graphene-spin-relaxation} b shows the spin-lifetime anisotropy $\tau_z/\tau_x$. Although Rashba fields are theoretically predicted to generate a 1/2 anisotropy for graphene on substrates \cite{Fabian_2007}, experiments consistently measure an anisotropy of 1 in suspended samples \cite{Raes2016,PhysRevB.95.085403,PhysRevB.97.205439}. This discrepancy is consistent for many numerical simulations made for suspended graphene, and we cannot correlate it between theory and experiments to this curvature-induced mechanism for spin-relaxation. 

\fullpagefigure[1]{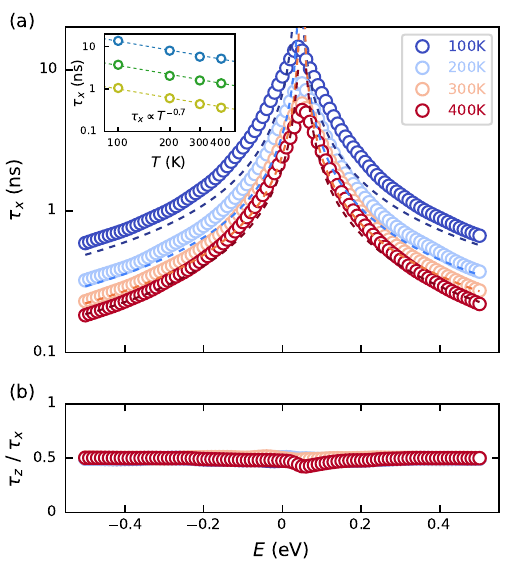}
{a: Spin relaxation time in $x$ of corrugated graphene as a function of chemical potential at different values for the thermalization temperature. Dashed lines correspond to predictions from fluctuating SOC fields, see \eqref{eq: taus random-SOC}. The inset shows the scaling of $\tau_s$ with temperature at the surroundings of the Dirac point with symbols corresponding to simulations and dashed lines the evolution $t\propto T^{-0.7}$. Here blue, green and yellow colors correspond to the Dirac point, $E=0.15\mathrm{eV}$ and $E=0.35\mathrm{eV}$ respectively. b: Spin lifetime anisotropy as a function of chemical potential at different temperatures.}{fig:corrugated-graphene-spin-relaxation}

\section{The relaxation mechanism}

In Sec.\ \ref{Section:Main-relaxation mechanisms} we saw the main mechanisms for spin relaxation in graphene and showed that for pristine graphene, the EY mechanism is often considered dominant with respect to DP due to the weak Rashba SOC fields. On the other hand, in Sec.\ \ref{sec:corrugations} we saw that the local Rashba field $\lambda_R^d$ presents extremely large values $\sim 10\,\mathrm{meV}$ but with a short correlation length of a few lattice constants. This raises the question of whether these huge local SOC fields are strong enough to produce a DP-type relaxation or not.

To do so, we will follow the work of \cite{Cummings_2025,Dugaev2011}, where the relaxation lifetime is studied as a result of a randomly fluctuating Rashba SOC. Here, the relevant magnitudes are the mean free path $\ell$, the spin precession length generated by the SOC $l_{SOC}\sim\hbar v_F/\lambda_{RMS}$ with $v_F$ the Fermi velocity and $\lambda_{RMS}$ the root mean square of the SOC, and the Fermi wavelength $l_F=2\pi/k_F$ with $k_F$ the Fermi wave number and the length scale of the SOC fluctuations $R$. 

There, relaxation takes the shape of

\begin{equation}\label{eq: taus random-SOC}
    \tau_x^{-1}=\pi\left(\frac{\lambda_{RMS}}{\hbar}\right)\left(\frac{R}{v_F}\right)\pi k_F R.
\end{equation}

This is reminiscent of the standard DP relation in the motional-narrowing regime, guided by \eqref{eq:Dyakonov-perel} where now the fluctuations of the SOC field produce the variations in the SOC angle instead of the momentum scattering via the ratio $R/v_F$.

\begin{figure}[h]
    \centering
    \includegraphics[width=\linewidth]{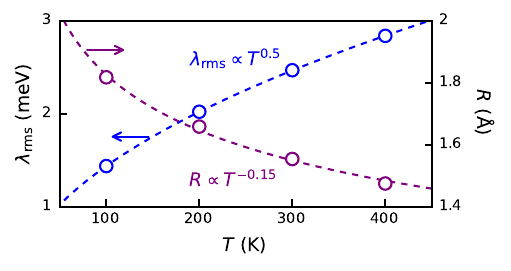}
    \caption{Temperature scaling of the root-mean-squared $\lambda_{RMS}$ and the length scale of the fluctuations $R$. Symbols represent simulations and dashed lines numerical scaling.}
    \label{fig:corrugated-graphene-spin-evolution}
\end{figure}

We can easily obtain $\lambda_{RMS}$ taking the root-mean-square of the spatial distribution of Rashba in our corrugations, shown in Fig.\ \ref{fig:corrugated-graphene-samples} b, and $R$ if we calculate the autocorrelation function.

We find $R$ by calculating the autocorrelation of the SOC field \cite{Dugaev2011},
\begin{equation}
C(\mathbf{r}-\mathbf{r}')=\langle \lambda(\mathbf{r})\lambda(\mathbf{r}')\rangle
   =\mathcal{F}^{-1}\{\mathcal{F}[\lambda(\mathbf{r})]\,
        \mathcal{F}^\ast[\lambda(\mathbf{r})]\}.
\end{equation}
Here, $\mathcal{F}\{\,\}$ denotes the Fourier transform.  
For each temperature considered, the autocorrelation function is well described by a Lorentzian profile,
\begin{equation}
C(\mathbf{r}-\mathbf{r}')=\lambda_{\mathrm{RMS}}^{2}
\frac{R^{2}}{|\mathbf{r}-\mathbf{r}'|^{2}+R^{2}},
\end{equation}
allowing us to extract the characteristic correlation length $R$ by fitting this expression to the numerical data.

If we perform the calculations, we see that the picture described in \eqref{eq: taus random-SOC} is fully consistent with our numerical results, as can be seen from the dashed lines in Fig.\ \ref{fig:corrugated-graphene-spin-relaxation}. Here we can see a direct comparison between the results obtained by this theory (dashed lines) and the results obtained by our quantum transport simulations (markers).

Figure\ \ref{fig:corrugated-graphene-spin-evolution} displays the temperature dependence of $\lambda_{\mathrm{RMS}}$ and $R$.  
We find that the correlation length scales weakly with temperature as $R\propto T^{-0.15}$. These values are only slightly larger than the carbon-carbon nearest-neighbor distance, highlighting that the curvature-induced SOC originates from truly atomic-scale deformations.

In contrast, the SOC amplitude follows a much stronger temperature dependence, $\lambda_{\mathrm{RMS}}\propto T^{0.5}$, reaching values in the range $1$-$3$ meV, exceptionally large for monolayer graphene.

Combining these scalings with equation\,(7.19) leads directly to
\begin{equation}
\tau_x \propto \frac{1}{(\lambda_{\mathrm{RMS}} R)^{2}}
         \propto T^{-0.7},
\end{equation}
in quantitative agreement with the behavior extracted from the inset in Fig.\ \ref{fig:corrugated-graphene-spin-relaxation}.  

This proves unambiguously that the curvature-induced Rashba SOC is the dominant contribution to spin relaxation in our suspended graphene model. In addition, when the intrinsic SOC terms induced by PIA and curvature are artificially set to zero, $\tau_s$ remains unchanged.

\chapter{Conclusions and outlook}
\label{ch:conclusions}

The central message of this thesis is that transport in graphene-based structures with long-wavelength modulations is governed by the way in which several length scales and scattering channels compete. When the electronic structure is reshaped by a superlattice, a moir\'e pattern, or a quasicrystalline arrangement, the disorder competes with the modulations in real space, thus generating new transport effects. Our real-space time-resolved approach makes these qualifications explicit: it reveals where the dynamics are controlled by geometry, and where they are controlled by genuinely random scattering.

A second conclusion is methodological. For these systems, static indicators alone are rarely sufficient. The density of states peaks, velocity renormalization, and real-space patterns in the local density of states are informative, but do not uniquely determine whether carriers spread ballistically, diffusively, or in an anomalous way over experimentally relevant scales. Following wave-packet evolution and extracting diffusion-related quantities provides a direct bridge between band-structure features and measurable transport trends, and it makes it possible to compare very different platforms using the same language.

\section{Superimposed potentials}

The results obtained for superimposed periodic and quasiperiodic potentials serve as a tool to understand which effects are purely due to geometry. A strictly superperiodic modulation produces transport signatures that remain compatible with ordinary Bloch dynamics, even when velocities are strongly renormalized: the wave packet may slow down, but its spreading keeps the characteristic structure of a periodic medium once the relevant time scales are reached. By contrast, below the superperiodic length, the structures introduce an important intermediate regime where transport does not simply emerge from the two periodicities. Instead, it develops sharp renormalizations due to the scattering at spreads related to the superperiodic lengths that are visible in time-dependent diffusion and can be seen as indicators of superperiodicity in the time/length dependent transport at very short time and length scales.

Within that controlled framework, truly quasiperiodic modulations stand out by producing transport that does not settle into the same periodic expectations. The key point is not simply that quasiperiodicity can suppress transport but that it can do so in a way that is not captured by a single renormalized velocity or a single mean free path. The dynamics reflect the absence of translational invariance and the resulting hierarchy of scattering processes. The most useful conclusion here is conceptual: quasiperiodicity is best understood as generating a structured, deterministic form of complexity that can act as an alternative to randomness in producing nontrivial spreading laws. That distinction becomes crucial later, when comparing quasicrystalline bilayers with disordered moir\'e systems, because it separates effects that are robust to additional scattering from effects that are intrinsically fragile.

\section{Moir\'e systems}

For magic-angle twisted bilayer graphene, the analysis supports a nuanced picture of how disorder reshapes flat-band transport. In a finite and physically relevant disorder window, where the flat-band spectral features remain identifiable, disorder can increase the mean free path within the flat-band energy range. The transport interpretation is that moderate disorder broadens the sharp spectral structures associated with the flat bands and reduces the effectiveness of elastic scattering, producing what is best described as disorder-induced delocalization at the single-particle level.

This is not a semantic point, it changes how one should think about ``clean'' versus ``dirty'' regimes in moir\'e materials. The onset of disorder does not necessarily push the system immediately toward stronger localization within the flat-band window. The picture that emerges is closer to a competition between two tendencies: band flattening promotes localization by suppressing velocities and enhancing sensitivity to scattering, while moderate broadening can restore connectivity in energy space and, with it, more extended real-space dynamics. Once disorder becomes too strong, the conventional trend takes over, and transport is suppressed across energies.

The link between transport and quantum geometry strengthens this conclusion. The quantum metric can track changes in the effective spatial extent of the electronic states and connects with other relevant properties such as superfluid weight or optical transport. This makes the disorder-driven evolution not only a kinetic detail but a real change in the character of the ground-state manifold. In that sense, transport becomes a practical diagnostic of how flat-band-like the system remains as inhomogeneity increases, not only spectrally but also in terms of real-space structure, and predicts an impact on the thermal or disorder vanishing of correlated phases.

 Quasicrystalline bilayers host prominent high-energy resonant states and striking spatial patterns that carry a self-similar character. Transport in that regime shows deviations from simple ballistic expectations, consistent with the idea that quasiperiodic order can generate unconventional dynamics. However, the decisive conclusion is fragility. Both disorder and proximity to an additional conducting graphene layer are sufficient to wash out those quasicrystalline fingerprints. In the trilayer stack, the presence of the magic-angle interface provides additional hybridization backgrounds and scattering pathways that strongly degrade the high-energy resonances, even before disorder is introduced. When disorder is added, the dynamics approach more standard diffusion.

This contrast between low-energy robustness and high-energy fragility is instructive. It suggests that not all spectacular real-space patterns translate into equally robust transport signatures once the environment is made more realistic. It also suggests a design rule: adding layers is not a neutral modification. Even if the additional layer is only weakly coupled at low energies, it can still destroy the high-energy structure by opening new channels that effectively act as an internal bath for the states that carry the quasicrystalline resonances.

The magnetism discussion fits naturally into this broader transport-centered picture. Even without completing a fully quantitative KPM-mean-field program, the review of experimental magnetic and topological phases in magic-angle systems highlights the subtleties of their effects even in the mean field in comparison to other structures (such as the zigzag graphene nanoribbons) and suggests their weakness against disorder. Any successful microscopic account of magnetism and topology in these systems must coexist with disorder-sensitive transport phenomenology, including the possibility that moderate disorder can partially restore delocalization in the flat-band window while still allowing strong interaction effects. This is an important outlook for this work and motivates work in having a more precise mean-field description for the flat-band states against disorder.

\section{Spin transport}

The spin-transport results provide a complementary conclusion for monolayer graphene. In suspended graphene, atomic-scale corrugations generate short-range fluctuating spin-orbit fields, particularly of Rashba type, through curvature-induced terms. These fields can dominate spin relaxation even when charge transport remains close to the ballistic limit. The important implication is that high mobility does not guarantee long spin lifetimes. The relevant disorder for spin dynamics can be encoded in structural fluctuations on length scales that hardly affect charge conduction.

This resolves a persistent tension between idealized expectations and experiments. Early theoretical estimates of exceptionally long spin lifetimes relied on nearly flat, perfectly clean graphene with weak intrinsic spin-orbit coupling. The present results indicate that a realistic structural landscape introduces an efficient relaxation channel that can reduce spin lifetimes to the nanosecond scale without requiring extrinsic magnetic impurities or strong substrate-induced coupling. In practical terms, corrugations are not a minor correction. They are a systematic, unavoidable feature in suspended samples and a plausible limiting factor even in devices where other disorder sources have been minimized.

The broader lesson is that spin transport is an unusually sensitive probe of atomic-scale structure. Although charge transport averages over many microscopic details, spin relaxation can respond strongly to local curvature and its spatial fluctuations. 

\section{Final remarks}
Bringing these strands together, the conclusions of this thesis can be summarized succinctly. First, long-wavelength modulations create transport regimes that are governed by crossovers and competing scales, so interpreting experiments requires identifying which regime is being probed. Second, in moir\'e systems, the flat-band window can host counterintuitive disorder trends, including a regime where moderate disorder increases the mean free path, and this behavior naturally connects to changes in the extent and geometry of the electronic states in real space. Third, quasicrystalline signatures, especially at high energies, are intrinsically fragile to both disorder-induced and proximity-induced scattering channels, which limits how directly they can be transferred from ideal models to realistic stacks. Fourth, spin transport is constrained by structural complexity in a way that is largely decoupled from charge transport, and atomic corrugations provide a concrete microscopic mechanism for the observed spin-lifetime limits.

The outlook that follows from these conclusions is clear. On the moir\'e side, the next step is to incorporate interactions into our framework to understand the effects of disorder on electron-electron interactions. Then, similar disorder-dependent transport metrics can be established in the interacting case instead of treating disorder and correlations as separate problems. On the quasicrystal side, progress requires larger-scale and more genuinely aperiodic treatments, together with disorder models that reflect realistic inhomogeneity beyond onsite randomness. On the spin side, the most direct route is to extend the corrugation-based description to supported and encapsulated devices and to multilayers, where the corrugations are reduced but additional proximity mechanisms may appear. Across all cases, the most reliable path forward is to keep transport, geometry, and structure in the same quantitative framework and provide new metrics for robustness of these interesting highly-correlated new physics.

\appendix

\chapter{Green's function solution to LDoS in AB bilayer graphene}\label{Appendix: green's function}

This appendix outlines the Green's-function route to compute the density of states (DoS) and local density of states (LDoS), which is the approach used in the thesis when estimating DoS through
Green's functions. 

\section{Retarded Green's function and spectral representation}

Green's functions are response (or propagation) operators: in a single-particle quantum problem they quantify how an excitation created at a given orbital propagates through the system. The Green's function operator can be defined as \cite{economou},
\begin{equation}
\hat G(z)=(\lambda-\hat H)^{-1},
\end{equation}
where $\lambda\in\mathbb{C}$. These functions are defined in the same as in \eqref{eq:green_functions}

As the operator $E-\hat H$ is singular in the eigenvalues of the system, the retarded (advanced) Green's function is often used when dealing with propagation, obtained by approaching the real axis from the upper (lower) imaginary half-plane,
\begin{equation}
\hat G^{R(A)}(E)=\lim_{\eta\to 0^{+}}(\lambda\pm i\eta-\hat H)^{-1}.
\label{eq:app_GR_def}
\end{equation}
Here, the positive (negative) sign is taken for the retarded (advanced) function. Because of its physical relevance, we will derive this analysis making use of the retarded Green's function. There, the infinitesimal $\eta>0$ enforces causality (retarded boundary condition) and selects the correct analytic continuation for matrix inversion after the limit $\eta\to 0$ is taken. In numerical calculations, it is common to keep a small finite $\eta>0$, which produces a controlled energy broadening, similar to the broadening induced by the KPM formalism explained in Sec.\ \ref{Sec: KPM-intro}.

If we make use of the expansion of operators in our Hamiltonian eigenbasis, from \eqref{eq:Chebyshev expansion delta} we have
\begin{equation}
\hat G^{R}(E)=\sum_n \frac{| n\rangle\langle n|}{E-E_n+i\eta}\approx\left[E-H+i\eta\right]^{-1},
\label{eq:app_GR_spectral}
\end{equation}
where we have assumed that the operator $\left[E-H+i\eta\right]$ is invertible due to the small broadening $\eta$.

\section{DoS from the imaginary part of the Green's function}
The DoS is defined as
\begin{equation}
\rho(E)=\mathrm{Tr} \left[\delta(E-\hat H)\right]=\sum_n \delta(E-E_n).
\label{eq:app_dos_def}
\end{equation}
Using equation~\eqref{eq:app_GR_spectral} and approximating $\delta$ as
\begin{equation}
\delta(x)=\lim_{\eta\to 0^{+}} \frac{1}{\pi}\frac{\eta}{x^2+\eta^2},
\end{equation}
one obtains the key operator relation;
\begin{equation}
\delta(E-\hat H)=-\frac{1}{\pi}\,\lim_{\eta\to 0^{+}}\mathrm{Im}\,\hat G^{R}(E),
\label{eq:app_delta_from_GR}
\end{equation}
and therefore the central formula
\begin{equation}
\rho(E)=-\frac{1}{\pi}\mathrm{Im}\,\mathrm{Tr}\left[\hat G^{R}(E)\right]=-\frac{1}{\pi}\mathrm{Im}\,\mathrm{Tr}\left[\left(H-E+i\eta\right)^{-1}\right].
\label{eq:app_dos_from_GR}
\end{equation}
For finite $\eta$, $\rho(E)$ becomes a broadened DoS where each discrete level is represented by a Lorentzian of width $\eta$.

\section{Local density of states}

The LDoS at the site $i$ can now be computed by direct projection onto the basis where
\begin{equation}
\rho_i(E)=\langle i| \delta(E-\hat H)| i\rangle
=-\frac{1}{\pi}\,\mathrm{Im}\,G^{R}_{ii}(E),
\label{eq:app_ldos}
\end{equation}

In practice, it is going to be useful to take the Fourier transform of the Green's function onto the reciprocal space, where now we can make

\begin{equation}
    \rho_i(E)=-\frac{1}{\pi}\mathrm{Im}G_{ii}^R(E)=-\sum_k\mathrm{Im}\left[(E-H_k+i\eta)^{-1}_{ii}\right].
\end{equation}

Summing over a finite set of orbitals, we recover the projected DoS onto  a region of space $\Omega$: $\rho_\Omega(E)=\sum_i \rho_i(E)\quad\forall i \in \Omega$.

After obtaining the $\rho_i$ at each site, the zero temperature occupations can be computed as usual as 
\begin{equation}
    n_i(E)=\int_{-\infty}^{E} \rho(E')\mathrm d E'.
\end{equation}

\section{The case of the AB graphene}

For the solution to equation\ \eqref{eq:AB_graphene_hamiltonian}, we can analytically derive the projected density of states at each of the layers following this procedure as

\begin{equation}
\begin{array}{l}
\mathrm{\rho}_{\mathrm{layer}}(V_\pi,V_\sigma,\Gamma,z,\Delta V)
=\\\qquad
\sum_k\mathrm {Im} \left[
\frac{
2 z\left(2\Delta V^{2}+2\Gamma^2 V_\pi^{2}+V_\sigma^{2}-2 z^{2}\right)
\pm\left(2\Delta V^{3}+\Delta V V_\sigma^{2}-2\Delta V z^{2}-2\Delta V \Gamma^2 V_\pi^{2}\right)
}{
\pi\Big(    
\Delta V^{4}
-2\Delta V^{2}\Gamma^2 V_\pi^{2}
+\Delta V^{2}V_\sigma^{2}
-2\Delta V^{2}z^{2}
+\Gamma^4V_\pi^{4}
-2\Gamma^2 V_\pi^{2}z^{2}
-V_\sigma^{2}z^{2}
+z^{4}
\Big)
}
\right]\\\\
z=V_{avg}+i\eta\\\\
\Gamma=\left|\sum_{p} e^{ik\mathbf{r}_{neigh,p}}\right|,
\end{array}
\end{equation}
where the + (-) sign is chosen for the top (bottom) layer. Here $V_{\pi}, V_{\sigma}$, and $\Delta V$, $V_{avg}$ are the hopping integrals, interlayer asymmetry, and average potential used in \eqref{eq:AB_graphene_hamiltonian}, $\eta$ is the broadening and $\mathbf r_{neigh}$ are the distance vectors to the nearest neighbors in the direction $A$ to $B$ where

\begin{equation}
    \mathbf r_{neigh,1}=(1,0)\qquad 
    \mathbf r_{neigh,2}=\left(-\frac{1}{2},\frac{\sqrt 3}{2}\right)\qquad
    \mathbf r_{neigh,3}=\left(-\frac{1}{2},-\frac{\sqrt 3}{2}\right) .
\end{equation}

Using that function, we can achieve the description of the density of states projected on the layer as the one plotted in Fig.\ \ref{fig:AB_DOS}.

\chapter{Solution to the Poisson Equation for a heterostack}\label{appendix poisson_solution}

In this appendix, we will explain how to efficiently solve the Poisson equation in a periodic two-dimensional material stack.
Let us start from \eqref{eq: The Poisson equation}. Let us choose the $z$ direction for the 2D-materials stacking. Here, our system is periodic in the $x$ and $y$ directions. First, we will assume that the dielectric tensor $\epsilon$ is diagonal with its only non-zero values being $\epsilon_{xx},\epsilon_{yy}, \epsilon_{zz}$. 

Let us also assume that the only dependence in the $\epsilon_{ii}(z)$ is in the stacking direction $z$, this assumption emerges naturally from assuming homogeneity of the dielectric potential along each layer of the stack. With that, we can expand \eqref{eq: The Poisson equation} as:

\begin{equation}\label{eq:poisson separated}
    \epsilon_{xx}\frac{\partial^2}{\partial x^2}V(\mathbf{r})+
    \epsilon_{yy}\frac{\partial^2}{\partial y^2}V(\mathbf{r})+
    \frac{\partial}{\partial z}\left(\epsilon_{zz}\frac{\partial}{\partial z}V(\mathbf{r})\right)=en(\mathbf{r}) .
\end{equation}

Here, we will take advantage of the periodicity of the potential along $x,y$, assuming that the potential and the electronic density are invariant under translations in the $x,y$ plane of the shape:
\begin{equation}
    V(x+\alpha L_x,y+\beta L_y)\qquad\forall\;\alpha,\beta\in\mathbb{Z} \;.
\end{equation}

For computational reasons, we will take a discretization of the potential unit cell to a square grid of dimensions $L_x\times L_y$ onto the set

\begin{equation}
    \lbrace x_n\rbrace=n\frac{L_x}{N}\qquad \lbrace y_m\rbrace=m\frac{L_y}{M}\;.
\end{equation}

On this grid, we will take the discrete Fourier transform, making

\begin{equation}
    V(x_n,y_m,z)=\frac{1}{NM}\sum_{k_x,k_y} \hat V_{k_x,k_y}(z)e^{i2\pi \mathbf{k}\cdot \mathbf{r}_{xy}}\;,
\end{equation}
where the integral is taken over the whole reciprocal space.  $\mathbf{r}_{xy}=(n/N,m/M)$ is the dimensionless in-plane position vector,  $\mathbf{k}=(k_x,k_y)$ the in-plane reciprocal space vector and $\hat V_{k_xk_y}(z)$ the $z$ dependent Fourier components of the potential at the reciprocal space sites $k_x,k_y$. Including this potential in \eqref{eq:poisson separated} and taking the Fourier transform of the electron density\\ $
    n(x_n,y_m,z)=\frac{1}{NM}\sum_{k_x,k_y} \hat n_{k_x,k_y}(z)e^{i 2\pi \mathbf{k}\cdot \mathbf{r}_{xy}}$, we achieve the 1-D version of the Poisson equation for a heterostack \eqref{eq:1-d-version-poisson}
These transformations leave \eqref{eq: The Poisson equation} as a 1-dimensional differential equation in $z$ for each value of $\mathbf{k}$.

\section{Solving the finite-difference 1D equation}

To solve the differential equation, we will now discretize the stacking coordinate $z$ into the set $\lbrace z_p\rbrace=p\frac{L_z}{P}$, where we can replace the derivatives by their discretized version \cite{EYMARD2000713}.

\begin{equation}
    \frac{\partial}{\partial z} f(z)\approx P\frac{f(z_{p+\frac{1}{2}})-f(z_{p-\frac{1}{2}})}{L_z}
\end{equation}

By substitution in \eqref{eq:1-d-version-poisson} we achieve
\begin{equation}\label{eq: poisson linar system}
    M\mathbf{\hat V}_{k_xk_y}=e\mathbf{\hat n}_{k_xk_y}\;,
\end{equation}
where
$$
\mathrm{\hat V}_{k_xk_y}=
\left(
\begin{array}{c}
     V_{k_xk_y}(0)  \\
     \vdots  \\
     V_{k_xk_y}(p\frac{L_z}{P})  \\
     \vdots  \\
     V_{k_xk_y}(L_z)  \\
     
\end{array}
\right)\qquad\qquad
\mathrm{\hat n}_{k_xk_y}=
\left(
\begin{array}{c}
     n_{k_xk_y}(0)  \\
     \vdots  \\
     n_{k_xk_y}(p\frac{L_z}{P})  \\
     \vdots  \\
     n_{k_xk_y}(L_z)  \\
     
\end{array}
\right)
$$
and $M$ is a tridiagonal matrix defined as
\begin{equation}\label{eq:M_matrix_tridiagonal}
    M=\left(
    \begin{array}{ccccccccc}
         \alpha_0-\beta_0&
         \epsilon_{0+}&&&&&&&\\
         \epsilon_{1-}&\alpha_1-\beta_1&\epsilon_{1+}&&&&&\\
         &\epsilon_{2-}&\alpha_2-\beta_2&\epsilon_{2+}&&&&&\\
         &&\ddots&\ddots&\ddots&&&&\\
         &&&\ddots&\ddots&\ddots&&&\\
         &&&&\epsilon_{P-}&\alpha_P-\beta_P&\epsilon_{P+}\\
         
    \end{array}
    \right).
\end{equation}
Here $\alpha_p$ and $\beta_p$ are defined as
\begin{equation}
    \alpha_p=4\pi^2\left(\frac{k_x^2}{L_x^2}\epsilon_{xx}(z_p)+\frac{k_y^2}{L_y^2}\epsilon_{yy}(z_p)\right)
    \qquad\beta_{p}=P\frac{\epsilon_{p+}+\epsilon_{p-}}{L_z}\;,
\end{equation}
and $\epsilon_{p\pm}=\epsilon_{zz}(z_{i}\pm\frac{pL_z}{2P})$. These are values that we can analytically compute from the analytical expression of $\epsilon_{zz}(z)$ or obtain by linear interpolation.

With that we have \eqref{eq: poisson linar system} as a tridiagonal linear system of equations that we can efficiently solve by using the Thomas algorithm \cite{Ames,Quarteroni}.

\chapter{Code Availability}

The numerical codes developed and used throughout this thesis are openly available to ensure transparency, reproducibility, and reuse.

The implementation of the self-consistent Poisson-Schr\"odinger solver, employed to compute electrostatic profiles and charge redistribution, as well as the mean-field routines used to model interaction effects, have been implemented by the author and are publicly accessible.

All source code can be found in dedicated Git repositories maintained by the author\footnote{\href{https://gitlab.com/palcazar/poisson_solver_dots.git}{https://gitlab.com/palcazar/poisson\_solver\_dots.git}}\footnote{\href{https://gitlab.com/palcazar/humema/-/tree/c1dc0ec01185c0cb9d7e1cb1c04a4942e600cca1/}{https://gitlab.com/palcazar/humema/-/tree/c1dc0ec01185c0cb9d7e1cb1c04a4942e600cca1/}} (see footnotes). These repositories include the full workflow required to reproduce the numerical results presented in this thesis, subject to the computational resources described in the corresponding chapters.

\bibliography{exemple_bibtex}

\backmatter
\printindex

\end{document}